\documentclass[12pt]{article}

\usepackage{amsfonts,amssymb,bm,cite}
\usepackage[dvips]{graphicx}

\newcommand {\beq}{\begin{equation}}
\newcommand {\eeq}{\end{equation}}
\newcommand {\beqa}{\begin{eqnarray}}
\newcommand {\eeqa}{\end{eqnarray}}
\newcommand {\n}{\nonumber \\}

\renewcommand{\theequation}{\thesection.\arabic{equation}}

\begin{document}
\setlength{\oddsidemargin}{0cm}
\setlength{\baselineskip}{7mm}

\begin{titlepage}
\renewcommand{\thefootnote}{\fnsymbol{footnote}}
\begin{normalsize}
\begin{flushright}
\begin{tabular}{l}
OU-HET 560\\
May 2006
\end{tabular}
\end{flushright}
  \end{normalsize}

~~\\

\vspace*{0cm}
    \begin{Large}
       \begin{center}
         {${\cal N}=4$ SYM on $R \times S^3$ and Theories with 16 Supercharges}
       \end{center}
    \end{Large}
\vspace{1cm}

\begin{center}
           Goro I{\sc shiki}\footnote
            {
e-mail address : 
ishiki@het.phys.sci.osaka-u.ac.jp},
           Yastoshi T{\sc akayama}\footnote
            {
e-mail address : 
takayama@het.phys.sci.osaka-u.ac.jp}
           {\sc and}
           Asato T{\sc suchiya}\footnote
           {
e-mail address : tsuchiya@het.phys.sci.osaka-u.ac.jp}\\
      \vspace{1cm}
                    
               {\it Department of Physics, Graduate School of  
                     Science}\\
               {\it Osaka University, Toyonaka, Osaka 560-0043, Japan}
\end{center}

\vspace{3cm}

\begin{abstract}
\noindent
We study ${\cal N}=4$ SYM on $R \times S^3$ and theories with 16 supercharges
arising as its consistent truncations. These theories
include the plane wave matrix model, ${\cal N}=4$ SYM on $R\times S^2$ and
${\cal N}=4$ SYM on $R\times S^3/Z_k$, and their gravity duals 
were studied by Lin and Maldacena. We make a harmonic 
expansion of the original 
${\cal N}=4$ SYM on $R \times S^3$ and obtain each of the truncated
theories by keeping a part of the Kaluza-Klein modes. This enables 
us to analyze all the theories in a unified way. We explicitly
construct some nontrivial
vacua of ${\cal N}=4$ SYM on $R\times S^2$. 
We perform 1-loop analysis of the original and truncated theories.
In particular, we examine states
regarded as the integrable $SO(6)$ 
spin chain and a time-dependent BPS solution, which is considered to
correspond to the AdS giant graviton in the original theory.
\end{abstract}
\vfill
\end{titlepage}
\vfil\eject

\setcounter{footnote}{0}

\tableofcontents

\section{Introduction}
\setcounter{equation}{0}
\renewcommand{\thefootnote}{\arabic{footnote}} 
It is important to collect various examples of the gauge/gravity correspondence
in order to elucidate how universal this phenomena is. 
Recently this direction has 
been pursued successfully by Lin and Maldacena \cite{Lin:2005nh}. 
They gave a general method for
constructing the gravity solutions dual to a family of
theories with 16 supercharges. All these theories share the common feature
that they have a mass gap, a discrete spectrum of excitations and a 
dimensionless parameter, which connect weak and strong coupling regions.
This method is an extension of the so-called
bubbling AdS geometries \cite{LLM,Berenstein,Jevicki}. 
The symmetry algebra of some of the theories
is $SU(2|4)$ supergroup, while the other theories
have $SO(4)\times SO(4)$ symmetry.
The theories with the $SU(2|4)$ symmetry arise as consistent truncations of
${\cal N}=4$ super Yang Mills (SYM) on $R\times S^3$ as explained below.
They include the plane wave matrix model \cite{Berenstein:2002jq},
${\cal N}=4$ SYM on $R\times S^2$ \cite{Maldacena:2002rb} 
and ${\cal N}=4$ SYM on $R\times S^3/Z_k$.

${\cal N}=4$ SYM on $R\times S^3$ has the superconformal symmetry
$SU(2,2|4)$, whose bosonic subgroup is $SO(2,4)\times SO(6)$, where
$SO(2,4)$ is the conformal group in 4 dimensions and $SO(6)$ is the R-symmetry.
$SO(2,4)$ has a subgroup $SO(4)$ that is the isometry of the $S^3$ on which
the theory is defined. $SO(4)$ is identified with $SU(2)\times \tilde{SU}(2)$,
where we marked one of two $SU(2)$'s with a tilde to focus on it.
By quotienting the original ${\cal N}=4$ SYM on $R\times S^3$ by various
subgroups of $\tilde{SU}(2)$, one obtains the above mentioned theories whose
symmetry algebra is $SU(2|4)$.
Quotienting by full $\tilde{SU}(2)$, $U(1)$ and $Z_k$ give rise to
the plane wave matrix model,
${\cal N}=4$ SYM on $R\times S^2$ and ${\cal N}=4$ SYM on $R\times S^3/Z_k$,
respectively.
Indeed, the consistent truncation to the plane wave matrix model was first found in \cite{Kim:2003rz}.
The original ${\cal N}=4$ SYM on $R\times S^3$ has a unique vacuum, 
while the truncated theories have many vacua.
The method by Lin and Maldacena give in principle
gravity solutions that describe these vacua and fluctuations around them, 
and they indeed obtained a few explicit solutions \cite{Lin:2005nh}.

It is obviously relevant to study the dynamics of
the above truncated theories and compare the results 
with those obtained on the 
gravity side. Indeed, some studies on the dynamics of
the plane wave matrix model have already been carried out 
\cite{Maldacena:2002rb}$\sim$\cite{Beisert:2005wv}. 
It should also be worthwhile to study the original 
${\cal N}=4$ SYM on $R\times S^3$ itself 
\cite{Breitenlohner:1982jf}$\sim$\cite{Okuyama:2002zn}, 
although it is believed to be equivalent to 
${\cal N}=4$ SYM on $R^4$ at conformal point, which is much easier
to analyze. The reasons are as follows.
First, the pp-wave limit on the gravity
side is taken for $AdS_5\times S^5$ in the global coordinates, and the 
boundary of $AdS_5$ is $R\times S^3$. 
The holography in the pp-wave limit could, therefore, be well understood in 
${\cal N}=4$ SYM on $R\times S^3$. Next, the original theory has a classical
time-dependent BPS solution, which is considered to correspond to the AdS
giant graviton \cite{Hashimoto,Jevicki}. 
The quantum dynamics of the AdS giant graviton is expected to
be understood by examining the quantum fluctuation around this classical
solution. The classical solution is, however, mapped to a classical vacuum
solution of ${\cal N}=4$ SYM on $R^4$ that breaks the conformal symmetry,
so that the equivalence between ${\cal N}=4$ SYM on $R\times S^3$ and $R^4$
does not seem to hold in this case.
Third, one can consider ${\cal N}=4$ SYM on $S^1\times S^3$, 
which is the finite temperature 
version of ${\cal N}=4$ SYM on $R\times S^3$ and is not equivalent to
${\cal N}=4$ SYM on $R^4$. This theory is known to show
a phase transition \cite{Witten:1998zw,Sundborg,Aharony}, 
which should correspond to the thermal phase transition
between the AdS space and the AdS black hole \cite{Hawking-Page}. The study of 
${\cal N}=4$ SYM on $R\times S^3$ serves as a preparation for that of 
this theory.

In this paper, we study the dynamics of
the original ${\cal N}=4$ SYM on $R\times S^3$ and the truncated theories,
by making a harmonic expansion of the original theory on $S^3$. 
We obtain each of the truncated theories by keeping a part of the Kaluza-Klein
(KK) modes of the original theory. This enables 
us to analyze all of the original and truncated theories in a unified way.

In section 2, we review basic properties of ${\cal N}=4$ SYM on $R\times S^3$.
In section 3, we develop the harmonic expansion on $S^3$. In particular,
we obtain a new formula for the integral of the product of three harmonics,
which is used in the following sections. In section 4, by applying
the results of section 3, we carry out a harmonic expansion of ${\cal N}=4$ SYM on $R\times S^3$ including all interaction terms.
The result in this section is an extension of the work\cite{Kim:2003rz}, where the authors carried out the mode expansion of the free part in detail and analyzed interactions between the lowest modes needed for the truncation to the plane wave matrix model.

In section 5, we describe the consistent truncations of the original
${\cal N}=4$ SYM on $R\times S^3$ to the theories with $SU(2|4)$ symmetry.
We realize each quotienting by keeping a part of the KK
modes of the original theory. We verify that
quotienting by $U(1)$ indeed yields 
${\cal N}=4$ SYM on $R\times S^2$ by comparing the KK modes we kept
with the KK modes of ${\cal N}=4$ SYM on $R\times S^2$.
We explicitly construct some of the nontrivial vacua
of ${\cal N}=4$ SYM on $R\times S^2$ in terms of the KK modes.

In section 6, we first calculate 
1-loop diagrams in the original theory. 
We introduce cut-offs for loop angular momenta and
see that this cut-off scheme yield correct coefficients of
logarithmic divergences, which are consistent with the Ward identities and the 
vanishing of the beta function. We next determine some counter terms in the
original theory and the truncated theories in the trivial vacuum by 
using the non-renormalization of energy of the BPS states. This reveals that
the states built by the sequence of the scalars in both the original theory and
the truncated theories in the trivial vacuum are mapped to the same
integrable $SO(6)$ spin chain. 

In section 7, we examine the time-independent
BPS solution in the original and truncated theories, which is considered to
correspond to the AdS giant graviton in the original theory. We see that
the 1-loop effective action around this solution vanishes. 

Section 8 is
devoted to summary and discussion. In appendix A, 
we gather some formulae concerning the representation of $SU(2)$.
In appendix B, we describe 
the vertex coefficients which are
used in representing the interaction terms by the modes.
In appendix C, we describe some properties of the spherical harmonics
on $S^2$, which are used in section 5. In appendix D, we list the 1-loop
diagrams and the divergent parts of those diagrams.
In appendix E, we give the expressions for the 1-loop effective action
around the time dependent BPS solution in the truncated theories.

\section{Basic properties of ${\cal N}=4$ SYM on $R\times S^3$}
\setcounter{equation}{0}
In this section, we review the basic properties
of ${\cal N}=4$ SYM on $R\times S^3$ 
\cite{Breitenlohner:1982jf}$\sim$\cite{Okuyama:2002zn}.
We restrict ourselves to the $U(N)$ gauge group and the 't Hooft limit
throughout this paper.
However, the generalization to other gauge groups that
allow the 't Hooft limit is easy.
We follow the notation of \cite{Okuyama:2002zn} with slight modification. 
We set the radius of $S^3$ at one.
Borrowing the ten-dimensional notation, we can write down the action 
as follows:
\beqa
&&S=\frac{1}{g_{YM}^2}\int d^4x \:e\: \mbox{Tr}\left(
-\frac{1}{4}F_{ab}F^{ab}-\frac{1}{2}D_aX_mD^aX_m-\frac{1}{12}RX_m^2 \right.\n
&&\hspace{4cm}\left.-\frac{i}{2}\bar{\lambda}\Gamma^aD_a\lambda
-\frac{1}{2}\bar{\lambda}\Gamma^m[X_m,\lambda]+\frac{1}{4}[X_m,X_n]^2\right),
\label{action}
\eeqa
where $a$ and $b$ are local Lorentz indices and run from $0$ to $3$,
and $m$ runs from $4$ to $9$.
$\Gamma^a$ and $\Gamma^m$ are the 10-dimensional gamma matrices, which satisfy
\beqa
\{\Gamma^a,\Gamma^b\}=2\eta^{ab},\;\;\;\;\; \{\Gamma^m,\Gamma^n\}=2\delta^{mn},
\eeqa 
where $\eta^{ab}=\mbox{diag}(-1,1,1,1)$.
$\lambda$ is the Majorana-Weyl spinor in 10 dimensions.
$e$ is the determinant of the vierbein $e_{\mu}^a$ on $R\times S^3$.
$R$ is the scalar curvature of $S^3$ which is equal to $6$.
The field strength and the covariant derivatives take the form
\beqa
&&F_{ab}=\nabla_aA_b-\nabla_bA_a-i[A_a,A_b]=e_a^{\mu}e_b^{\nu}F_{\mu\nu}, \n
&&D_aX_m=\nabla_a X_m-i[A_a,X_m], \n
&&D_a\lambda=\nabla_a\lambda-i[A_a,\lambda],
\eeqa
where 
\beqa
\nabla_aA_b=e^{\mu}_a(\partial_{\mu}A_b+\omega_{\mu b}^{\;\;\;\;\;c}A_c),\;\;\;
\nabla_aX_m=e^{\mu}_a\partial_{\mu}X_m, \;\;\;
\nabla_a\lambda=e^{\mu}_a(\partial_{\mu}\lambda+\frac{1}{4}\omega_{\mu}^{bc}
\Gamma_{bc}\lambda),
\eeqa
and $\omega_{\mu}^{ab}$ is the spin connection on $R\times S^3$ determined by
$de^a+\omega^a_{\;b}\wedge e^b=0$. 

The classical action (\ref{action}) with arbitrary 
gauge group has the superconformal symmetry
$SU(2,2|4)$. This symmetry is preserved at the quantum level.
This is ensured by the following two facts.
One is that
the Weyl anomaly for the $g_{YM}=0$ was shown to vanish 
on $R\times S^3$ \cite{Cappelli:1988vw}.
The other is that
the beta function vanishes for arbitrary $g_{YM}$ because it only reflects
the short distance structure of the theory and indeed vanishes on $R^4$.
In what follows, we describe the transformation laws of the fields
under each element of 
$SU(2,2|4)$ and see that the action (\ref{action}) 
is invariant under such transformations.

First, let us see the conformal invariance of the action.
If the metric and the vierbein were allowed to vary, 
the action would possess the Weyl invariance,
\beqa
\delta_W A_a=-\alpha A_a, \;\;\; \delta_WX_m=-\alpha X_m,\;\;\; 
\delta_W \lambda=-\frac{3}{2}\alpha \lambda,\;\;\; 
\delta_W e_{\mu}^a=\alpha e_{\mu}^a,
\eeqa
the diffeomorphism invariance,
\beqa
&&\delta_{\xi}A_a=\xi^{\mu}\partial_{\mu}A_a, \;\;\; 
\delta_{\xi}X_m=\xi^{\mu}\partial_{\mu}X_m, \;\;\;
\delta_{\xi}\lambda=\xi^{\mu}\partial_{\mu}\lambda, \n
&&\delta_{\xi}e_{\mu}^a=\xi^{\nu}\nabla_{\nu}e_{\mu}^a
+\nabla_{\mu}\xi^{\nu}e_{\nu}^a.
\eeqa
and the local Lorentz invariance,
\beqa
\delta_LA_a=\varepsilon_a^{\;b}A_b,\;\;\; \delta_LX_m=0,\;\;\; 
\delta_L\lambda=\frac{1}{4}\varepsilon_{ab}\Gamma^{ab}\lambda,\;\;\;
\delta_Le_{\mu}^a=\varepsilon^a_{\;b}e_{\mu}^b.
\eeqa
Let $\xi$ be a conformal Killing vector satisfying
\beqa
\nabla_a \xi_b+\nabla_b \xi_a=\frac{1}{2}\nabla_c\xi^c \eta_{ab},
\label{conformalKillingequation}
\eeqa
and set $\alpha=-\frac{1}{4}\nabla_a\xi^a$ and 
$\varepsilon_{ab}=\xi^{\mu}\omega_{\mu ab}
+\frac{1}{2}(\nabla_a\xi_b-\nabla_b\xi_a)$.
Then, 
\beqa
(\delta_{\xi}+\delta_W+\delta_L)e_{\mu}^a=0.
\eeqa
The action is, therefore, invariant under the conformal transformation
$\delta_c=\delta_{\xi}+\delta_W+\delta_L$, where the metric and the vierbein
are fixed. The conformal transformation act on each field as follows:
\beqa
&&\delta_cA_a=\xi^b\nabla_bA_a+\nabla_a\xi^bA_b,\n
&&\delta_cX_m=\xi^a\nabla_aX_m+\frac{1}{4}\nabla_a\xi^aX_m,\n
&&\delta_c\lambda=\xi^a\nabla_a\lambda+\frac{1}{4}\nabla_a\xi_b\Gamma^{ab}\lambda
+\frac{3}{8}\nabla_a\xi^a\lambda.
\eeqa

It is often convenient
to rewrite the action in the the $SU(4)$ symmetric form.  
The 10-dimensional Lorentz group has been decomposed as
$SO(9,1) \supset SO(3,1)\times SO(6)$. We identify $SO(6)$ with $SU(4)$.
We use $A,B=1,2,3,4$ as the indices of $\mbox{\boldmath $4$}$
in $SU(4)$ while we have used $m,n=4,\cdots,9$ as the indices of
$\mbox{\boldmath $6$}$ in $SO(6)$. The $SO(6)$ vector, $\mbox{\boldmath $6$}$,
corresponds to the antisymmetric tensor of $\mbox{\boldmath $4$}$ in $SU(4)$.
The $SO(6)$ and $SU(4)$ basis are related as
\beqa
&&X_{i4}=\frac{1}{2}(X_{i+3}+iX_{i+6}) \;\;\; (i=1,2,3), \n
&&X_{AB}=-X_{BA},\;\;\;X^{AB}=-X^{BA}=X_{AB}^{\dagger},\;\;\;
X^{AB}=\frac{1}{2}\epsilon^{ABCD}X_{CD},
\eeqa
Similar identities hold for the gamma matrices:
\beqa
\Gamma^{i4}=\frac{1}{2}(\Gamma^{i+3}-i\Gamma^{i+6}), \;\;\;\mbox{etc.}
\eeqa
The 10-dimensional gamma matrices are decomposed as
\beqa
\Gamma^a=\gamma^a\otimes 1_8,\;\;\;
\Gamma^{AB}=\gamma_5\otimes \left( \begin{array}{cc}
                                   0          &  -\tilde{\rho}^{AB} \\
                                   \rho^{AB}  &  0
                                   \end{array}  \right)
=-\Gamma^{BA},
\eeqa
where $\gamma^a$ is the 4-dimensional gamma matrix, satisfying
$\{\gamma^a,\gamma^b\}=2\eta^{ab}$, and 
$\gamma_5=i\gamma^0\gamma^1\gamma^2\gamma^3$. $\Gamma^{AB}$ satisfies
$\{\Gamma^{AB},\Gamma^{CD}\}=\epsilon^{ABCD}$, and $\rho^{AB}$ and
$\tilde{\rho}^{AB}$ are defined by
\beqa
(\rho^{AB})_{CD}=\delta^A_C\delta^B_D-\delta^A_D\delta^B_C,\;\;\;
(\tilde{\rho}^{AB})^{CD}=\epsilon^{ABCD}.
\eeqa
The charge conjugation matrix and the chirality matrix are given by
\beqa
C_{10}=C_4 \otimes \left( \begin{array}{cc}
                          0   &  1_4 \\
                          1_4 &  0   
                          \end{array} \right), \;\;\;\;\;
\Gamma^{11}=\Gamma^0\cdots\Gamma^9=\gamma_5\otimes 
            \left( \begin{array}{cc}
                     1_4   &  0    \\
                     0     &  -1_4    
                   \end{array} \right),
\eeqa
where $(\Gamma^{a,m})^T=-C_{10}^{-1}\Gamma^{a,m}C_{10}$ and 
$C_4$ is the charge conjugation matrix in 4 dimensions.
The Majorana-Weyl spinor in 10 dimensions is decomposed as
\beqa
\lambda=\Gamma_{11}\lambda
=\left(\begin{array}{c} \lambda_+^A \\ \lambda_{-A} \end{array}\right),
\label{10to4}
\eeqa
where $\lambda_{-A}$ is the charge conjugation of $\lambda_+^A$:
\beqa
\lambda_{-A}=(\lambda_+^A)^c=C_4(\bar{\lambda}_{+A})^T,\;\;\;\;\;
\gamma_5\lambda_{\pm}=\pm\lambda_{\pm}.
\eeqa
The action is rewritten in terms of $SU(4)$ symmetric notation as follows:
\beqa
&&S=\frac{1}{g_{YM}^2}\int d^4x \:e\: \mbox{Tr}\left(
-\frac{1}{4}F_{ab}F^{ab}-\frac{1}{2}D_aX_{AB}D^aX^{AB}
-\frac{1}{2}X_{AB}X^{AB}-i\bar{\lambda}_{+A}\gamma^aD_a\lambda_+^A\right.\n
&&\hspace{4cm}\left.
-\bar{\lambda}_{+A}[X^{AB},\lambda_{-B}]-\bar{\lambda}_-^A[X_{AB},\lambda_+^B]
+\frac{1}{4}[X_{AB},X_{CD}][X^{AB},X^{CD}]\right), \n
\label{actionSU(4)symmetric}
\eeqa
It is easy to see that the action (\ref{actionSU(4)symmetric}) 
is invariant under the $SU(4)$ R-symmetry
\beqa
\delta_R X^{AB}=iT^A_{\;\;C}X^{CB}+iT^B_{\;\;C}X^{AC},\;\;\;
\delta_R\lambda_+^A=iT^A_{\;\;B}\lambda_+^B,\;\;\; 
\delta_R\bar{\lambda}_{-A}=-i\bar{\lambda}_{-B}T^B_{\;\;A},
\eeqa
where $T^A_{\;\;B}$ is a hermitian traceless matrix.

Finally, we consider the superconformal symmetry.
The conformal Killing spinor equation on $R\times S^3$ takes the form
\beqa
\nabla_a \epsilon_+=\pm \frac{i}{2}\gamma_a \gamma^0 \epsilon_+,\;\;\;\;\;
\gamma_5 \epsilon_+=\epsilon_+.
\label{conformalKillingequation}
\eeqa
A general solution to (\ref{conformalKillingequation}) for each sign
includes arbitrary constant Weyl spinor and is obtained by projecting
the Killing spinor
on $AdS_5$ on the boundary \cite{Breitenlohner:1982jf,Bergshoeff:1987dh}. 
We construct a 10-dimensional Majorana-Weyl spinor as
\beqa
\epsilon=\left(\begin{array}{c}\epsilon_+^A \\ \epsilon_{-A}
               \end{array}\right),
\label{epsilon}
\eeqa
where $\epsilon_+^A$ satisfies (\ref{conformalKillingequation}) and 
$\epsilon_{-A}$ is the charge conjugation of $\epsilon_+^A$ and satisfies
\beqa
\nabla_a \epsilon_{-A}=\mp \frac{i}{2}\gamma_a \gamma^0 \epsilon_{-A},
\;\;\;\;\;
\gamma_5 \epsilon_{-A}=-\epsilon_{-A}.
\eeqa
The action (\ref{action}) is invariant under the superconformal transformation
\beqa
&&\delta_{\epsilon}A_a=i\bar{\lambda}\Gamma_a\epsilon,\;\;\;
\delta_{\epsilon}X_m=i\bar{\lambda}\Gamma_m\epsilon, \n
&&\delta_{\epsilon}\lambda=\left[\frac{1}{2}F_{ab}\Gamma^{ab}
+D_aX_m\Gamma^a\Gamma^m-\frac{1}{2}X_m\Gamma^m\Gamma^a\nabla_a
-\frac{i}{2}[X_m,X_n]\Gamma^{mn}\right]\epsilon.
\label{superconformaltransformation}
\eeqa
$\epsilon_+$ in (\ref{conformalKillingequation}) includes 
four real degrees of freedom for each sign as mentioned above
and there are four
$SU(4)$ indices, so that $\epsilon$ in (\ref{epsilon}) possess 32 real
degrees of freedom. Namely, the superconformal symmetry 
(\ref{superconformaltransformation}) has 32 real supercharges.
In the $SU(4)$ symmetric notation, the transformation 
(\ref{superconformaltransformation}) is written as
\beqa
&&\delta_{\epsilon}A_a=i(\bar{\lambda}_{+A}\gamma_a\epsilon_+^A
-\bar{\epsilon}_{+A}\gamma_a\lambda_+^A), \n
&&\delta_{\epsilon}X^{AB}=i(-\bar{\epsilon}_-^A\lambda_+^B
+\bar{\epsilon}_-^B\lambda_+^A+\epsilon^{ABCD}\bar{\lambda}_{+C}\epsilon_{-D}),
\n
&&\delta_{\epsilon}\lambda_+^A=\frac{1}{2}F_{ab}\gamma^{ab}\epsilon_+^A
+2D_aX^{AB}\gamma^a\epsilon_{-B}+X^{AB}\gamma^a\nabla_a\epsilon_{-B}
+2i[X^{AC},X_{CB}]\epsilon_+^B, \n
&&\delta_{\epsilon}\lambda_{-A}=\frac{1}{2}F_{ab}\gamma^{ab}\epsilon_{-A}
+2D_aX_{AB}\gamma^a\epsilon_+^B+X_{AB}\gamma^a\nabla_a\epsilon_+^B
+2i[X_{AC},X^{CB}]\epsilon_{-B}.
\label{superconformaltransformationSU(4)}
\eeqa

In the remaining of this section, 
we make a comment on the equivalence 
between ${\cal N}=4$ SYM on $R^4$ at conformal
point and ${\cal N}=4$ SYM on $R\times S^3$. 
We first see the relationship between $R^4$ and $R\times S^3$.
If one starts with the metric of $R^4$,
\begin{eqnarray}
ds^2=dr^2+r^2d\Omega_3^2,
\end{eqnarray}
makes a change of variable, $\ln r=\tau$, and
defines a new metric through a Weyl transformation,
$ds^2=e^{2\tau}{ds'}^2$, one obtains the metric of euclidean $R\times S^3$,
\begin{eqnarray}
{ds'}^2=d\tau^2+d\Omega_3^2.
\end{eqnarray}
The analytical continuation, $\tau=it$, yields the metric of $R\times S^3$.
This indicates how these two theories are related.
There is one to one correspondence between operators on 
$R^4$ and states on $R\times S^3$ as common in conformal fields theories.
Namely, one can move an operator at arbitrary point on $R^4$ to
the origin by a conformal transformation, and map it 
to an state on $R\times S^3$ because $r\rightarrow 0$ corresponds to 
$t\rightarrow -\infty$.
One can also see from $\ln r=\tau$ that the dilatation operator on $R^4$
corresponds to hamiltonian on $R\times S^3$. That is, the scaling dimension
$\Delta$ on $R^4$ corresponds to the energy $E$ on $R\times S^3$. 
More precisely, there is the Casimir energy, $E_0$, on $S^3$.
Thus $\Delta=E-E_0$. The value of $E_0$ is for instance, 
calculated through the Weyl anomaly near $R^4$ and  equal to $\frac{3}{16}N^2$
\cite{Cappelli:1988vw}.
In this paper, for simplicity, we redefine the hamiltonian
by $H\rightarrow H-E_0$ and make energy of the vacuum vanishing, so that
$\Delta=E$ holds.
Note that this equivalence holds only
at conformal point on $R^4$ and breaks for instance
in a situation where the Higgs field has a non-vanishing vev on $R^4$.

\section{Harmonic expansion on $S^3$}
\setcounter{equation}{0}
In this section, we develop the harmonic expansion on $S^3$. In section 3.1,
we consider generic spherical harmonics on $S^3$ and obtain a formula
for the integral of the product of three spherical harmonics. In section 3.2, 
we restrict ourselves to scalar, spinor and vector harmonics and describe
some useful properties. We define vertex coefficients by the integrals of 
the products of these harmonics. In section 3.3, we find the vector and
spinor harmonics that correspond to the conformal Killing vectors and spinors,
which appeared in section 2.

\subsection{Spherical harmonics on $S^3$}
First, we construct the spherical harmonics on $S^3$, following the
strategy in \cite{Salam:1981xd}, where the harmonic functions on the coset space $G/H$ are
discussed. In this case, $S^3=SO(4)/SO(3)$, namely 
$G=SO(4)=SU(2)\times \tilde{SU}(2)$
and $H=SO(3)$. The subgroup $H=SO(3)$ is naturally identified with the local 
`Lorentz' group $SO(3)$ on $S^3$. We denote the generators of the $SU(2)$
in $G$ by $J_i$ and those of the $\tilde{SU(2)}$ in $G$
by $\tilde{J}_i$ ,where $i=1,2,3$. Then, the generators of $H$ are represented
by $L_i=J_i+\tilde{J}_i$. 

The irreducible representations of $G$ are labeled
by two spins, $J$ and $\tilde{J}$, which specify the irreducible 
representations of the $SU(2)$ and the $\tilde{SU}(2)$, respectively.
We denote the basis of the $(J,\tilde{J})$ representation by 
$|Jm\rangle |\tilde{J}\tilde{m}\rangle$. The basis of 
the spin $L$ representation of $H$ is constructed in terms of 
$|Jm\rangle |\tilde{J}\tilde{m}\rangle$:
\beqa
|Ln;J\tilde{J}\rangle\rangle
=\sum_{m\tilde{m}}C^{Ln}_{Jm \; \tilde{J}\tilde{m}}
|Jm\rangle |\tilde{J}\tilde{m}\rangle,
\eeqa
where $C^{Ln}_{Jm \; \tilde{J}\tilde{m}}$ is the Clebsch-Gordan coefficient
of $SU(2)$ and 
the triangular inequality, 
\beqa
|J-\tilde{J}|\leq L \leq J+\tilde{J},
\label{triangularinequality}
\eeqa
must be satisfied. 

A definite form of the representative element of $G/H$ is given by
\beqa
\Upsilon(\Omega)=e^{-i\psi L_1}e^{-i\varphi L_3}e^{-i\theta K_1},
\label{representative}
\eeqa
where $K_i=J_i-\tilde{J}_i$ and $\Omega=(\theta,\varphi,\psi)$ is the
polar coordinates of $S^3$. Note, however, 
that the explicit form of $\Upsilon(\Omega)$ 
is barely needed in the following arguments.

The spin $L$ spherical harmonics on $S^3$ is given by
\beqa
{\cal Y}^{Ln}_{Jm,\tilde{J}\tilde{m}}(\Omega)
=N^L_{J\tilde{J}}
\langle\langle Ln;J\tilde{J}|\Upsilon^{-1}(\Omega)
|Jm\rangle |\tilde{J}\tilde{m}\rangle,
\label{spinLsphericalharmonics}
\eeqa
where $N^L_{J\tilde{J}}$ is the normalization factor. It is fixed as
\beqa
N^L_{J\tilde{J}}=\sqrt{\frac{(2J+1)(2\tilde{J}+1)}{2L+1}}.
\eeqa
such that
the spherical harmonics (\ref{spinLsphericalharmonics}) satisfies the
orthonormal condition:
\beqa
\int d\Omega \:\sum_n({\cal Y}^{Ln}_{Jm,\tilde{J}\tilde{m}})^* 
\:{\cal Y}^{Ln}_{J'm',\tilde{J}'\tilde{m}'}
=\delta_{JJ'}\delta_{\tilde{J}\tilde{J}'}\delta_{mm'}
\delta_{\tilde{m}\tilde{m}'}.
\label{orthonormality}
\eeqa
Here the measure is normalized as $\int d\Omega \: 1=1$
and can be identified with the Haar measure of $G$ since the integrand is 
invariant under the action of $H$. Then, one can easily verify 
(\ref{orthonormality})
by using the orthogonality of the representation matrices of $G$ under the
Haar measure and a relation
\beqa
\sum_{\alpha\beta}C^{c\gamma}_{a\alpha\;b\beta}C^{c'\gamma'}_{a\alpha\;b\beta}
=\delta_{cc'}\delta_{\gamma\gamma'}.
\eeqa
The equations (\ref{representative}) and (\ref{spinLsphericalharmonics})
give the complex conjugate of ${\cal Y}^{Ln}_{Jm,\tilde{J}\tilde{m}}$:
\beqa
({\cal Y}^{Ln}_{Jm,\tilde{J}\tilde{m}})^*
=(-1)^{-J+\tilde{J}-L+m-\tilde{m}+n} \:
  {\cal Y}^{L\:-n}_{J\:-m,\tilde{J}\:-\tilde{m}}.
\label{complexconjugate}
\eeqa

The covariant derivative is understood as an algebraic manipulation:
\beqa
\nabla_i\:{\cal Y}^{Ln}_{Jm,\tilde{J}\tilde{m}}(\Omega)
=N^L_{J\tilde{J}}
 \langle\langle Ln;J\tilde{J}|(-iK_i)\Upsilon^{-1}(\Omega)
 |Jm\rangle |\tilde{J}\tilde{m}\rangle.
\label{covariantderivative}
\eeqa
Using this relation, it is easy to obtain the eigenvalue of the laplacian
for the spin $L$ spherical harmonics:
\beqa
\nabla^2{\cal Y}^{Ln}_{Jm,\tilde{J}\tilde{m}}(\Omega)
=-(2J(J+1)+2\tilde{J}(\tilde{J}+1)-L(L+1))
\:{\cal Y}^{Ln}_{Jm,\tilde{J}\tilde{m}}(\Omega).
\label{laplacian}
\eeqa

We need the integral of the product of three spherical harmonics in
rewriting the interaction terms in terms of modes.
By making composition of the angular momentum repeatedly and using 
the orthogonality of the representation matrices of $G$ and a formula
for the $9-j$ symbol (\ref{9-j}),
we obtain a compact formula
\beqa
&&\int d\Omega \: \sum_{n_1n_2n_3}
                  ({\cal Y}^{L_1n_1}_{J_1m_1,\tilde{J}_1\tilde{m}_1})^* \: 
                  {\cal Y}^{L_2n_2}_{J_2m_2,\tilde{J}_2\tilde{m}_2}
                  \:{\cal Y}^{L_3n_3}_{J_3m_3,\tilde{J}_3\tilde{m}_3} \: 
                  C^{L_1n_1}_{L_2n_2 \; L_3n_3} \nonumber\\
&&=\sqrt{(2L_1+1)(2J_2+1)(2\tilde{J}_2+1)(2J_3+1)(2\tilde{J}_3+1)}\:
   \left\{   \begin{array}{ccc}
             J_1 & \tilde{J}_1 & L_1 \\
             J_2 & \tilde{J}_2 & L_2 \\
             J_3 & \tilde{J}_3 & L_3 
             \end{array}     \right\} \:
             C^{J_1m_1}_{J_2m_2 \; J_3m_3}
             C^{\tilde{J}_1\tilde{m}_1}_{\tilde{J}_2\tilde{m}_2 \; 
                \tilde{J}_3\tilde{m}_3} .  \n
\label{integralofthreeharmonics}
\eeqa
Note that the integrand on the left-hand side is again 
invariant under the action of $H$. The equation 
(\ref{integralofthreeharmonics}) is one of new results in this paper, which
can be applied to any field theory on $S^3$.

\subsection{Scalars, vectors and spinors on $S^3$}
In this subsection, as an application of the results in the previous
subsection, we consider scalars, vectors and spinors on $S^3$.

The scalar corresponds to $L=0$. From the triangular inequality
(\ref{triangularinequality}), we see that
$(J,\tilde{J})=(J,J)$. We introduce a notation for the scalar:
\beqa
Y_{JM}\equiv {\cal Y}^{L=0,n=0}_{Jm,J\tilde{m}},
\eeqa
where $M$ stands for $(m,\tilde{m})$. The vector corresponds to $L=1$.
Then, the triangular inequality implies that
$(J,\tilde{J})$ takes $(J+1,J)$ or $(J,J+1)$ or $(J,J)$.
We assign $\rho=1$, $\rho=-1$ and $\rho=0$ to these three cases, respectively.
We make a change of basis from the basis $|1n;J\tilde{J}\rangle\rangle$ to
the vector basis:
\beqa
|1;J\tilde{J}\rangle\rangle&=&\frac{1}{\sqrt{2}}
(-|1,1;J\tilde{J}\rangle\rangle+|1,-1;J\tilde{J}\rangle\rangle) \n
|2;J\tilde{J}\rangle\rangle&=&\frac{i}{\sqrt{2}}
(|1,1;J\tilde{J}\rangle\rangle+|1,-1;J\tilde{J}\rangle\rangle) \n
|3;J\tilde{J}\rangle\rangle&=&|1,0;J\tilde{J}\rangle\rangle.
\eeqa
Accordingly, the vector harmonics on $S^3$ are defined by
\beqa
{\cal Y}^{i}_{Jm,\tilde{J}\tilde{m}}
=N^1_{J\tilde{J}}\langle\langle i;J\tilde{J}|\Upsilon^{-1}(\Omega)
|Jm\rangle |\tilde{J}\tilde{m}\rangle   \;\;\;\;\;(i=1,2,3),
\eeqa
which are just a unitary transform of ${\cal Y}^{1n}_{Jm,J\tilde{m}}$.
We introduce a notation for the vector:
\beqa
Y^{\rho=1}_{JMi}&=&i{\cal Y}^{i}_{J+1\:m,J\tilde{m}}, \n
Y^{\rho=-1}_{JMi}&=&-i{\cal Y}^{i}_{Jm,J+1\:\tilde{m}}, \n
Y^{\rho=0}_{JMi}&=&{\cal Y}^{i}_{Jm,J\tilde{m}}.
\eeqa
Here the factors $\pm i$ on the right-hand side are just a convention.
Note that $Y^{0}_{J=0\:M=(0,0)i}=0$.
The spinor corresponds to $L=\frac{1}{2}$.  The triangular inequality
implies that $(J,\tilde{J})$ takes $(J+\frac{1}{2},J)$ or $(J,J+\frac{1}{2})$.
We assign $\kappa=1$ to the former and $\kappa=-1$ to the latter.
We introduce a notation for the spinor:
\beqa
Y^{\kappa=1}_{JM\alpha}
&=&{\cal Y}^{L=\frac{1}{2},\alpha}_{J+\frac{1}{2}\:m,J\tilde{m}}, \n
Y^{\kappa=-1}_{JM\alpha}
&=&{\cal Y}^{L=\frac{1}{2},\alpha}_{Jm,J+\frac{1}{2}\:\tilde{m}}, 
\eeqa
where $\alpha$ takes $\frac{1}{2}$ and $-\frac{1}{2}$.

The orthnormality condition (\ref{orthonormality}) is translated to
the scalar, the vector and the spinor as
\beqa
&&\int d\Omega \:(Y_{J_1M_1})^* Y_{J_2M_2}=\delta_{J_1J_2}\delta_{M_1M_2}, \n
&&\int d\Omega \:(Y^{\rho_1}_{J_1M_1i})^* Y^{\rho_2}_{J_2M_2i}
=\delta_{\rho_1\rho_2}\delta_{J_1J_2}\delta_{M_1M_2}, \n
&&\int d\Omega \:(Y^{\kappa_1}_{J_1M_1\alpha})^* Y^{\kappa_2}_{J_2M_2\alpha}
=\delta_{\kappa_1\kappa_2}\delta_{J_1J_2}\delta_{M_1M_2},
\label{orthnormality2}
\eeqa
while their complex conjugates are read off from (\ref{complexconjugate})
as
\beqa
&&(Y_{JM})^*=(-1)^{m-\tilde{m}}Y_{J-M}, \n
&&(Y^{\rho}_{JMi})^*=(-1)^{m-\tilde{m}+1}Y^{\rho}_{J-Mi}, \n
&& (Y^{\kappa}_{JM\alpha})^*=(-1)^{m-\tilde{m}+\kappa\alpha+1}
                             Y^{\kappa}_{J-M-\alpha}.
\label{complexconjugate2}
\eeqa 
By using (\ref{covariantderivative}), it is easy to show that the following
identities hold:
\beqa
&&\nabla_i\:Y^{\pm 1}_{JMi}=0, \n
&&\epsilon_{ijk}\:\nabla_j\:Y^{\rho}_{JMk}=-2\rho (J+1)\:Y^{\rho}_{JMi}, \n
&&\nabla_i\:Y_{JM}=-2i\sqrt{J(J+1)}\:Y^0_{JMi}.
\label{identities}
\eeqa
The eigenvalues of the laplacian can be read off from (\ref{laplacian}):
\beqa
&&\nabla^2\:Y_{JM}=-4J(J+1)\:Y_{JM}, \n
&&\nabla^2\:Y^{\pm 1}_{JMi}=-(4J(J+2)+2)\:Y^{\pm 1}_{JMi}, \n
&&\nabla^2\:Y^0_{JMi}=-(4J(J+1)-2)\:Y^0_{JMi}, \n
&&\nabla^2\:Y^{\kappa}_{JM\alpha}=-(2J(2J+3)+\frac{3}{4})\:
                                  Y^{\kappa}_{JM\alpha}.
\label{laplacian2}
\eeqa
Using (\ref{covariantderivative}) yields an identity
\beqa
\sigma^i_{\alpha\beta}\:\nabla_i\:Y^{\kappa}_{JM\beta}
=-i\kappa (2J+\frac{3}{2})\: Y^{\kappa}_{JM\alpha}.
\label{Diracoperator}
\eeqa

In what follows, we define various integrals of the product of
three scalar or spinor or vector harmonics, 
which we will call vertex coefficients.
The vertex coefficients are needed to make a mode expansion 
for the interaction part. Their expression are obtained by using the formula
(\ref{integralofthreeharmonics}). We give these expressions in appendix B.
The expressions for the vertex coefficients consisting only of scalars and
vectors are already given in \cite{Cutkosky,Hamada}, 
where the 9-j symbols are, however, not used. 
\beqa
&&{\cal C}^{J_1M_1}_{J_2M_2\;J_3M_3}
\equiv\int d\Omega \:(Y_{J_1M_1})^*Y_{J_2M_2}Y_{J_3M_3}. \n
&&{\cal C}_{J_1M_1\;J_2M_2\;J_3M_3}
\equiv\int d\Omega \:Y_{J_1M_1}Y_{J_2M_2}Y_{J_3M_3}. \n
&&{\cal D}^{JM}_{J_1M_1\rho_1\;J_2M_2\rho_2}
\equiv \int d\Omega\: (Y_{JM})^*Y^{\rho_1}_{J_1M_1i}Y^{\rho_2}_{J_2M_2i}. \n
&&{\cal D}_{JM\;J_1M_1\rho_1\;J_2M_2\rho_2}
\equiv \int d\Omega\: Y_{JM} Y^{\rho_1}_{J_1M_1i}Y^{\rho_2}_{J_2M_2i}. \n
&&{\cal E}_{J_1M_1\rho_1\;J_2M_2\rho_2\;J_3M_3\rho_3}
\equiv\int d\Omega \: \epsilon_{ijk}\: 
Y^{\rho_1}_{J_1M_1i} Y^{\rho_2}_{J_2M_2j} Y^{\rho_3}_{J_3M_3k}. \n
&&{\cal F}^{J_1M_1\kappa_1}_{J_2M_2\kappa_2\;JM}
\equiv \int d\Omega \:
           (Y^{\kappa_1}_{J_1M_1\alpha})^*Y^{\kappa_2}_{J_2M_2\alpha}Y_{JM}. \n
&&{\cal G}^{J_1M_1\kappa_1}_{J_2M_2\kappa_2\;JM\rho}
\equiv \int d\Omega \: 
         (Y^{\kappa_1}_{J_1M_1\alpha})^*\sigma^i_{\alpha\beta}
          Y^{\kappa_2}_{J_2M_2\beta}Y^{\rho}_{JMi}. 
\label{vertexcoefficients}
\eeqa

\subsection{Conformal Killing vectors and spinors}
The vector spherical harmonics that correspond to the conformal Killing
vectors were already found in \cite{Hamada}. 
The number of the independent conformal
Killing vectors is 15, which is equal to the number of the generators
of $SO(2,4)$. The conformal group $SO(2,4)$ contains $R\times SO(4)$ as
a subgroup, where $R$ corresponds to the time translation and $SO(4)$
corresponds to the isometry of $S^3$. The conformal Killing vectors 
corresponding to the generators of this subgroup is also the Killing
vectors, namely these vectors satisfy the Killing vector equation
$\nabla_a\xi_b+\nabla_b\xi_a=0$. The number of the generators of the subgroup
is $1+6=7$ so that the number of the independent Killing vectors is $1+6=7$.
It is easy to check using (\ref{covariantderivative}) 
that the 4-vectors $(1,\vec{0})$, $(0, Y^+_{0Mi})$ and
$(0,Y^-_{0Mi})$ satisfy the Killing vector equation. The first one
corresponds to the time translation, while
the second and third ones correspond to the isometry of $S^3$ and include
6 independent real vectors due to the condition (\ref{complexconjugate2}).
It is also easily verified that the remaining 8 conformal Killing vectors 
are given by $(e^{it}Y_{\frac{1}{2}M},\sqrt{3}e^{it}Y_{\frac{1}{2}Mi}^0)$.

Next, let us find the spinor spherical harmonics that correspond to the 
conformal Killing spinors \cite{Kim:2003rz}. 
If we set $\sigma_0=1_2$, it is easy to verify
that the following equation holds:
\beqa
\sum_{\beta}(\nabla_a)_{\alpha\beta}(e^{\mp \frac{i}{2}t}Y^{\pm}_{0M\beta})
=\mp\frac{i}{2}\sum_{\beta}
(\sigma_a)_{\alpha\beta}e^{\mp \frac{i}{2}t}Y^{\pm}_{0M\beta}.
\label{CKSharmonics}
\eeqa
In the next section, we will see that the conformal Killing spinors are indeed
expanded by $e^{\mp \frac{i}{2}t}Y^{\pm}_{0M\alpha}$, which include 2 
independent complex spinors for each sign.

\section{Harmonic expansion of ${\cal N}=4$ SYM on $R\times S^3$}
\setcounter{equation}{0}
In this section, we apply the results in 3 to
${\cal N}=4$ SYM on $R\times S^3$. In section 4.1, 
we make a harmonic expansion of 
${\cal N}=4$ SYM on $R\times S^3$ and rewrite the theory in terms of infinitely
many KK modes. In other words, we obtain a matrix quantum mechanics with
infinitely many matrices. In section 4.2, we quantize the free part of the
theory and obtain the KK tower.

\subsection{Harmonic expansion of ${\cal N}=4$ SYM on $R\times S^3$}
First, we fix the forms of 4-dimensional gamma matrices:
\beqa
\gamma^a=\left(\begin{array}{cc} 
               0                & i\sigma^a \\
               i\bar{\sigma}^a  & 0
               \end{array}\right),
\eeqa
where
$\sigma^0=-1_2$ and $\sigma^i\;\;\;(i=1,2,3)$ are the Pauli matrices.
$\bar{\sigma}^0=\sigma^0$ and $\bar{\sigma}^i=-\sigma^i$.
In this convention,
\beqa
\gamma_5=\left(\begin{array}{cc} 
               1_2 & 0 \\
               0   & -1_2
               \end{array}\right), \;\;\;\;\;
C_4=\left(\begin{array}{cc} 
          -\sigma^2 & 0 \\
          0         & \sigma^2
          \end{array}\right).
\eeqa
We introduce a two-component spinor:
\beqa
\lambda_+^A=\left(\begin{array}{c} \psi^A \\ 0 \end{array}\right). 
\label{4to2}
\eeqa
Using the two-component spinor, we can rewrite the action 
(\ref{actionSU(4)symmetric}) as follows:
\beqa
&&S=\frac{1}{g^2}\int dtd\Omega \: \mbox{Tr}\left(
-\frac{1}{4}F_{ab}F^{ab}-\frac{1}{2}D_aX_{AB}D^aX^{AB}
-\frac{1}{2}X_{AB}X^{AB}+i\psi_A^{\dagger}D_0\psi^A
+i\psi_A^{\dagger}\sigma^iD_i\psi^A \right.\n
&&\hspace{3.7cm}\left.
+\psi_A^{\dagger}\sigma^2[X^{AB},(\psi_B^{\dagger})^T]
-\psi^{AT}\sigma^2[X_{AB},\psi^B]
+\frac{1}{4}[X_{AB},X_{CD}][X^{AB},X^{CD}]\right), \n
\label{actiontwo-component}
\eeqa
where $g^2\equiv \frac{g_{YM}^2}{2\pi^2}$ since the area of unit $S^3$ is 
$2\pi^2$. $A_0$ and $X^{AB}$ are scalars on $S^3$, $A_i$ is a vector on $S^3$
and $\psi^A$ is a spinor on $S^3$. $\nabla_0=\partial_t$ and $\nabla_i$ is
the covariant derivative on $S^3$.

To quantize the system, we need a gauge-fixing. We take the Coulomb gauge,
\beqa
\nabla_i A_i=0,
\label{Coulombgauge}
\eeqa
for convenience. The residual gauge symmetry which is realized by
a gauge parameter that depends only on time is fixed by\footnote{In the
theory on $S^1\times S^3$, the zero mode of the lefthand side of 
(\ref{residualgaugesymmetryfixing}), 
which is given by its integral on $S^1$, becomes dynamical and plays an
important role \cite{Aharony,Aharony:2005bq}.}
\beqa
\int d\Omega \: A_0=0.
\label{residualgaugesymmetryfixing}
\eeqa
The gauge-fixing and Faddeev-Popov terms for the above gauge-fixing
are given by
\beqa
S_{GF+FP}=\int dtd\Omega\: \mbox{Tr}(-i\bar{c}\nabla_iD_ic).
\eeqa
It should be understood that the condition (\ref{Coulombgauge}) is always
imposed by the delta function in the path-integral.
The free part of the gauge-fixed action, $I=S+S_{GF+FP}$, is
\beqa
&&I_0=\int dtd\Omega \: \mbox{Tr} \left(
-\frac{1}{2}A_0\nabla^2A_0+\frac{1}{2}\partial_0A_i\partial_0A_i
+\frac{1}{2}A_i\nabla^2A_i-A_iA_i \right. \n
&&\hspace{3.2cm}+\frac{1}{2}\partial_0X_{AB}\partial_0X^{AB}
+\frac{1}{2}X_{AB}\nabla^2X^{AB}-\frac{1}{2}X_{AB}X^{AB}  \n
&&\hspace{3.2cm}\left.+i\psi_A^{\dagger}\partial_0\psi^A
+i\psi_A^{\dagger}\sigma^i\nabla_i\psi^A
-i\bar{c}\nabla^2c \right),
\label{gauge-fixedactionfree}
\eeqa
while the interaction part of the gauge-fixed action is
\beqa
&&I_{int}=\int dtd\Omega \: \mbox{Tr} \left(
-ig\partial_0A_i[A_0,A_i]+ig\nabla_iA_0[A_0,A_i]
+\frac{ig}{2}(\nabla_iA_j-\nabla_jA_i)[A_i,A_j] \right.\n
&&\hspace{3.5cm}-\frac{g^2}{2}[A_0,A_i]^2+\frac{g^2}{4}[A_i,A_j]^2
-ig\partial_0X_{AB}[A_0,X^{AB}]+ig\nabla_iX_{AB}[A_i,X^{AB}] \n
&&\hspace{3.5cm}-\frac{g^2}{2}[A_0,X_{AB}][A_0,X^{AB}]
+\frac{g^2}{2}[A_i,X_{AB}][A_i,X^{AB}]+g\psi_A^{\dagger}[A_0,\psi^A] \n
&&\hspace{3.5cm}+g\psi_A^{\dagger}\sigma^i[A_i,\psi^A]
+g\psi_A^{\dagger}\sigma^2[X^{AB},(\psi_B^{\dagger})^T]
-g\psi^{AT}\sigma^2[X_{AB},\psi^B] \n
&&\hspace{3.5cm}\left.+\frac{g^2}{4}[X_{AB},X_{CD}][X^{AB},X^{CD}]
+g\nabla_i\bar{c}[A_i,c] \right).
\label{gauge-fixedactioninteraction}
\eeqa
In (\ref{gauge-fixedactionfree}) and (\ref{gauge-fixedactioninteraction}),
we have rescaled the fields by $1/g$.

We make the mode expansion for the fields as
\beqa
&&A_0(t,\Omega)=\sum_{(JM)\neq(00)}B_{JM}(t)Y_{JM}(\Omega),\;\;\;\;\;
A_i(t,\Omega)=\sum_{\rho=\pm 1}\sum_{JM}A_{JM\rho}(t)Y_{JMi}^{\rho}(\Omega), \n
&&X_{AB}(t,\Omega)=\sum_{JM}X_{AB}^{JM}(t)Y_{JM}(\Omega),\;\;\;\;\;
X^{AB}(t,\Omega)=\sum_{JM}X^{AB}_{JM}(t)Y_{JM}(\Omega), \n
&&\psi^A_{\alpha}(t,\Omega)
=\sum_{\kappa=\pm 1}\sum_{JM}\psi^A_{JM\kappa}(t)Y_{JM\alpha}^{\kappa}(\Omega), \n
&&c(t,\Omega)=\sum_{(JM)\neq(00)}c_{JM}(t)Y_{JM}(\Omega),\;\;\;\;\;
\bar{c}(t,\Omega)=\sum_{(JM)\neq(00)}\bar{c}_{JM}(t)Y_{JM}(\Omega)
\label{KKmodeexpansion}
\eeqa
The condition $(JM)\neq (00)$ for the summation in $A_0$,
$c$ and $\bar{c}$ comes from the gauge-fixing condition
(\ref{residualgaugesymmetryfixing}).
Each mode is $N\times N$ matrix. Due to (\ref{complexconjugate2}), 
$A_0^{\dagger}=A_0$, $A_i^{\dagger}=A_i$
and $X_{AB}^{\dagger}=X^{AB}$ imply
\beqa
&&(B_{JM})^{\dagger}=(-1)^{m-\tilde{m}}B_{J\:-M},\;\;\;
(A_{JM\rho})^{\dagger}=(-1)^{m-\tilde{m}+1}A_{J\:-M\rho},\n
&&(X_{AB}^{JM})^{\dagger}=(-1)^{m-\tilde{m}}X^{AB}_{J\:-M}.
\eeqa
Note that $\rho$ takes only $\pm 1$ in (\ref{KKmodeexpansion}) because of 
the gauge-fixing condition (\ref{Coulombgauge}) and the first identity in
(\ref{identities}).

In order to express (\ref{gauge-fixedactionfree}) and 
(\ref{gauge-fixedactioninteraction}) in terms of the modes in
(\ref{KKmodeexpansion}), we use 
(\ref{orthnormality2})$\sim$(\ref{vertexcoefficients}). For the four-point
interaction terms, we also use product expansions such as
\beqa
Y_{J_1M_1}(\Omega)Y_{J_2M_2}(\Omega)
=\sum_{J_1M_1J_2M_2}{\cal C}^{J_3M_3}_{J_1M_1\;J_2M_2}Y_{J_3M_3}(\Omega).
\eeqa
The result is
\beqa
&&I=I_0+I_{int},\;\;\;\;\; I_0=\int dt\:L_0,\;\;\;\;\;
I_{int}=\int dt\:(L_{int}^{(1)}+L_{int}^{(2)}), \\
&&L_0=\mbox{Tr}\left[
\sum_{(JM)\neq(00)}(-1)^{m-\tilde{m}}2J(J+1)B_{J-M}B_{JM} \right.\n
&&\hspace{1.8cm}+\sum_{\rho=\pm 1}\sum_{JM}(-1)^{m-\tilde{m}+1}\frac{1}{2}
(\dot{A}_{J\:-M\rho}\dot{A}_{JM\rho}-{\omega_J^A}^2A_{J\:-M\rho}A_{JM\rho}) \n
&&\hspace{1.8cm}+\sum_{JM}(-1)^{m-\tilde{m}}\frac{1}{2}
(\dot{X}_{AB}^{J\:-M}\dot{X}_{JM}^{AB}-{\omega_J^X}^2X_{AB}^{J\:-M}X_{JM}^{AB})
\n
&&\hspace{1.8cm}+\sum_{\kappa=\pm 1}\sum_{JM}
(i\psi_{JM\kappa A}^{\dagger}\dot{\psi}_{JM\kappa}^A
+\kappa\omega_J^{\psi}\psi_{JM\kappa A}^{\dagger}\psi^A_{JM\kappa}) \n
&&\hspace{1.8cm}\left.+\sum_{(JM)\neq(00)}(-1)^{m-\tilde{m}}4iJ(J+1)
\bar{c}_{J\:-M}c_{JM} \right], 
\label{freepart} \\
&&L_{int}^{(1)}
=\mbox{Tr}\left[
-ig\rho_1(J_1+1){\cal E}_{J_1M_1\rho_1\;J_2M_2\rho_2\;J_3M_3\rho_3}
A_{J_1M_1\rho_1}[A_{J_2M_2\rho_2},A_{J_3M_3\rho_3}] \right.\n
&&\qquad\qquad\;
+\frac{g^2}{4}{\cal D}^{JM}_{J_1M_1\rho_1\;J_3M_3\rho_3}
{\cal D}_{JM\;J_2M_2\rho_2\;J_4M_4\rho_4}
[A_{J_1M_1\rho_1},A_{J_2M_2\rho_2}][A_{J_3M_3\rho_3},A_{J_4M_4\rho_4}]) \n
&&\qquad\qquad\; 
+2g\sqrt{J_1(J_1+1)}{\cal D}_{J_2M_2\;J_1M_10\;JM\rho}
X_{AB}^{J_1M_1}[A_{JM\rho},X^{AB}_{J_2M_2}] \n
&&\qquad\qquad\;
+\frac{g^2}{2}{\cal C}^{JM}_{J_2M_2\;J_4M_4}
{\cal D}_{JM\;J_1M_1\rho_1\;J_3M_3\rho_3}
[A_{J_1M_1\rho_1},X_{AB}^{J_2M_2}][A_{J_3M_3\rho_3},X^{AB}_{J_4M_4}]) \n
&&\qquad\qquad\;
+g{\cal G}^{J_1M_1\kappa_1}_{J_2M_2\kappa_2\;JM\rho}
\psi_{J_1M_1\kappa_1A}^{\dagger}[A_{JM\rho},\psi_{J_2M_2\kappa_2}^A] \n
&&\qquad\qquad\;
-ig(-1)^{m_2-\tilde{m}_2+\frac{\kappa_2}{2}}
{\cal F}^{J_1M_1\kappa_1}_{J_2-M_2\kappa_2\;JM}
\psi_{J_1M_1\kappa_1A}^{\dagger}[X^{AB}_{JM},\psi_{J_2M_2\kappa_2B}^{\dagger}] 
\n
&&\qquad\qquad\;
+ig(-1)^{-m_1+\tilde{m}_1+\frac{\kappa_1}{2}}
{\cal F}^{J_1-M_1\kappa_1}_{J_2M_2\kappa_2\;JM} 
\psi_{J_1M_1\kappa_1}^A [X_{AB}^{JM},\psi_{J_2M_2\kappa_2}^B]) \n
&&\qquad\qquad\;\left. 
+\frac{g^2}{4}{\cal C}^{JM}_{J_1M_1\;J_2M_2}
{\cal C}_{JM\;J_3M_3\;J_4M_4}
[X_{AB}^{J_1M_1},X_{CD}^{J_2M_2}][X_{J_3M_3}^{AB},X_{J_4M_4}^{CD}]) \right], 
\label{interactionpart1} \\
&&L_{int}^{(2)}=\mbox{Tr}\left[ 
-ig{\cal D}_{JM\;J_1M_1\rho_1\;J_2M_2\rho_2}
\dot{A}_{J_1M_1\rho_1}[B_{JM},A_{J_2M_2\rho_2}] \right. \n
&&\qquad\qquad\;
+2g\sqrt{J_1(J_1+1)}{\cal D}_{J_2M_2\;J_1M_10\;JM\rho}
B_{J_1M_1}[B_{J_2M_2},A_{JM\rho}]  \n
&&\qquad\qquad\; 
-\frac{g^2}{2}{\cal C}^{JM}_{J_1M_1\;J_3M_3}
{\cal D}_{JM\;J_2M_2\rho_2\;J_4M_4\rho_4}
[B_{J_1M_1},A_{J_2M_2\rho_2}][B_{J_3M_3},A_{J_4M_4\rho_4}] \n
&&\qquad\qquad\; 
-ig{\cal C}_{JM\;J_1M_1\;J_2M_2}
\dot{X}_{AB}^{J_1M_1}[B_{JM},X^{AB}_{J_2M_2}])  \n
&&\qquad\qquad\; 
-\frac{g^2}{2}{\cal C}^{JM}_{J_1M_1\;J_2M_2}
{\cal C}_{JM\;J_2M_3\;J_4M_4}
[B_{J_1M_1},X_{AB}^{J_2M_2}][B_{J_3M_3},X^{AB}_{J_4M_4}]) \n
&&\qquad\qquad\; 
+g{\cal F}^{J_1M_1\kappa_1}_{J_2M_2\kappa_2\;JM}
\psi_{J_1M_1\kappa_1A}^{\dagger}[B_{JM},\psi_{J_2M_2\kappa_2}^A]) \n
&&\qquad\qquad\;\left.
-2ig\sqrt{J_1(J_1+1)}{\cal D}_{J_2M_2\;J_1M_10\;JM\rho}
\bar{c}_{J_1M_1}[A_{JM\rho},c_{J_2M_2}] \right],
\label{interactionpart2}
\eeqa
where
\beqa
&&\omega_J^A=2J+2, \n
&&\omega_J^X=2J+1, \n
&&\omega_J^{\psi}=2J+\frac{3}{2}.
\eeqa
We have classified the interaction terms into two categories. $L_{int}^{(1)}$
consists of the terms that do not contain $B$ or $c$ or $\bar{c}$ while
$L_{int}^{(2)}$ consists of the terms that contain $B$ or $c$ or $\bar{c}$.
In each term in $L_{int}^{(1)}$ and $L_{int}^{(2)}$, 
the summation over indices that appear twice or more than twice is assumed.
Of course, `$J$' in $B$, $c$ and $\bar{c}$ cannot take zero.
Note that the way to express the four-point interaction using the vertex
coefficients is not unique. The expressions for 
$L_{int}^{(1)}$ and $L_{int}^{(2)}$, (\ref{interactionpart1}) 
and (\ref{interactionpart2}), are one of new results in this paper.

\subsection{Quantization of free part and the Kaluza-Klein tower}
The free theory in which $g=0$ is easy to quantize.
In the free theory, one can set
$B_{JM}=0$ and $c_{JM}=\bar{c}_{JM}=0$. $A_{JM\rho}$, $X_{JM}^{AB}$ and
$\psi_{JM\kappa}^A$ behave as free particles. 
We can construct the hamiltonian of the free theory from $L_0$ as
\beqa
&&H_0=\mbox{Tr}\left[
\sum_{JM\rho}(-1)^{m-\tilde{m}+1}\frac{1}{2}
(P_{J\:-M\rho}P_{JM\rho}+{\omega_J^A}^2A_{J\:-M\rho}A_{JM\rho}) 
\right.\n
&&\hspace{1cm}\left.+\sum_{JM}(-1)^{m-\tilde{m}}\frac{1}{2}
(P_{AB}^{J\:-M}P_{JM}^{AB}+{\omega_J^X}^2X_{AB}^{J\:-M}X_{JM}^{AB})
-\sum_{JM\kappa}
\kappa\omega_J^{\psi}\psi_{JM\kappa A}^{\dagger}\psi_{JM\kappa}^A \right], \n
\label{freehamiltonian}
\eeqa
where $P_{JM\rho}$ and $P_{AB}^{JM}$ are the canonical conjugate momenta of
$A_{JM\rho}$ and $X^{AB}_{JM}$, respectively, while the canonical conjugate of
$\psi^A_{JM\kappa}$ is $i\psi_{JM\kappa A}^{\dagger}$. 
The (anti-)commutation relations are
\beqa
&&[(A_{JM\rho})_{kl},(P_{J'M'\rho'})_{k'l'}]
=i\delta_{J_1J_2}\delta_{M_1M_2}\delta_{\rho_1\rho_2}\delta_{kl'}\delta_{lk'}, 
\n
&&[(X^{AB}_{JM})_{kl},(P_{A'B'}^{J'M'})_{k'l'}]
=i\frac{1}{2}(\delta^A_{A'}\delta^B_{B'}-\delta^A_{B'}\delta^B_{A'})
\delta_{JJ'}\delta_{MM'}\delta_{kl'}\delta_{lk'}, \n
&&\{(\psi^A_{JM\kappa})_{kl},(\psi_{J'M'\kappa' A'}^{\dagger})_{k'l'}\}
=\delta^A_{A'}\delta_{JJ'}\delta_{MM'}\delta_{\kappa\kappa'}
\delta_{kl'}\delta_{lk'}.
\eeqa
$A_{JM\rho}$, $X_{JM}^{AB}$ and
$\psi_{JM\kappa}^A$ and their
canonical conjugates are expanded in terms of the creation and annihilation
operators as
\beqa
&&A_{JM\rho}=\frac{1}{\sqrt{2\omega_J^A}}(a_{JM\rho}e^{-i\omega_J^At}
+(-1)^{m-\tilde{m}+1}a_{J-M\rho}^{\dagger}e^{i\omega_J^At}), \n
&&P_{JM\rho}=-i\sqrt{\frac{\omega_J^A}{2}}
((-1)^{m-\tilde{m}+1}a_{J-M\rho}e^{-i\omega_J^At}
-a^{\dagger}_{JM\rho}e^{i\omega_J^At}), \n
&&X^{AB}_{JM}=\frac{1}{\sqrt{2\omega_J^X}}(\alpha^{AB}_{JM}e^{-i\omega_J^Xt}
+(-1)^{m-\tilde{m}}\alpha^{AB\dagger}_{J-M}e^{i\omega_J^Xt}), \n
&&P^{AB}_{JM}=-i\sqrt{\frac{\omega_J^X}{2}}
((-1)^{m-\tilde{m}}\alpha^{AB}_{J-M}e^{-i\omega_J^Xt}
-\alpha^{AB\dagger}_{JM}e^{i\omega_J^Xt}), \n
&&\psi^A_{JM+}=d_{J-M}^{A\dagger}e^{i\omega_J^{\psi}}\;, \;\;\;\;\; 
\psi^A_{JM-}=b_{JM}^Ae^{-i\omega_J^{\psi}} \;.
\label{modeexpansion}
\eeqa
The (anti-)commutation relations for the creation and annihilation operators
are
\beqa
&&[(a_{JM\rho})_{kl},(a_{J'M'\rho'}^{\dagger})_{k'l'}]
=\delta_{JJ'}\delta_{MM'}\delta_{\rho\rho'}\delta_{kl'}\delta_{lk'}, \;\;\;\;
[(\alpha_{JM}^{AB})_{kl},(\alpha_{J'M'}^{A'B'\dagger})_{k'l'}]
=\frac{1}{2}\epsilon^{ABA'B'}\delta_{JJ'}\delta_{MM'}\delta_{kl'}\delta_{lk'}, 
\n
&&\{(b_{JM}^A)_{kl},(b_{J'M'A'}^{\dagger})_{k'l'}\}
=\delta^A_{A'}\delta_{JJ'}\delta_{MM'}\delta_{kl'}\delta_{lk'},\;\;\;\;
\{(d_{JMA})_{kl},(d_{J'M'}^{A'\dagger})_{k'l'}\}
=\delta_A^{A'}\delta_{JJ'}\delta_{MM'}\delta_{kl'}\delta_{lk'}.
\eeqa
The free hamiltonian is rewritten in terms of the creation and annihilation
operators:
\beqa
H_0=:\mbox{Tr}\left[
\sum_{JM\rho}\omega_J^Aa_{JM\rho}^{\dagger}a_{JM\rho}
+\sum_{JM}\omega_J^X\alpha_{JM}^{AB\dagger}\alpha_{AB}^{JM}
+\sum_{JM}\omega_J^{\psi}(b_{JMA}^{\dagger}b_{JM}^A+d_{JM}^{A\dagger}d_{JMA})
\right]:.
\label{freehamiltonianmode}
\eeqa
In section 6.2, we will make a comment 
on the constant which we discarded when we obtained
the above normal-ordered expression.

\begin{figure}
	\begin{center}
		\includegraphics[width=13cm]{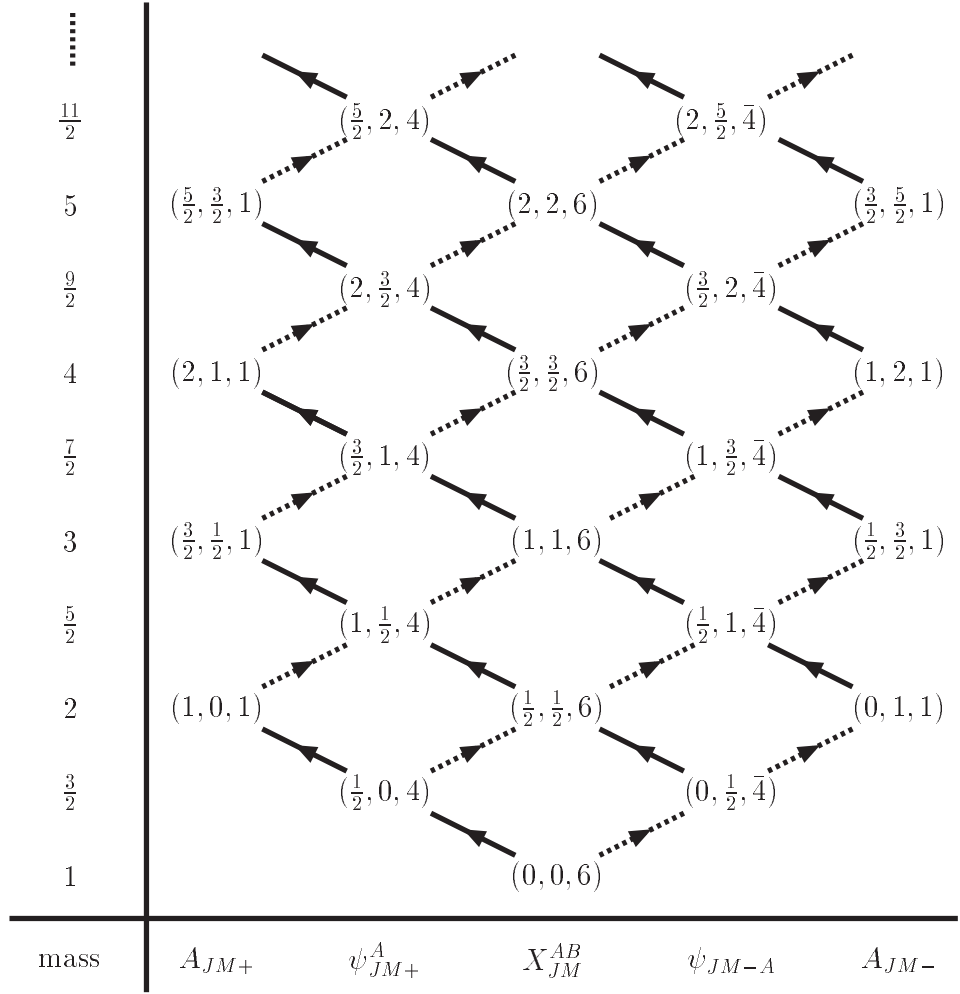}
	\end{center}
\caption{The KK tower of $\mathcal{N}=4$ super Yang-Mills 
on $R \times S^3$. The first number, the second number and the third
number in the parentheses represent $J$, $\tilde{J}$ and the dimension of
the representation of $SU(4)$, respectively. 
The solid and dotted arrows represent
the superconformal transformation in the free theory
for the creation operator caused 
by $\eta_{m+}$ and $\eta_{m-}^*$, respectively.}
\label{KKtower}
\end{figure}

As in \cite{Deger:1998nm,Kim:2003rz}, 
the mass spectrum of the free theory in which $g=0$ can be
read off from (\ref{freehamiltonian}).   
These forms the infinitely high
KK tower. As stated in introduction,
there exists a mass gap and the mass spectrum is discrete. The mass spectrum
is summarized in Fig.\ref{KKtower}. Note that there is no mass 
multiplicity between the bosons and the fermions unlike 
the supersymmetric theories in flat space.

In the case of the free theory, given an operator on $R^4$, 
one can easily construct the corresponding state on
$R\times S^3$ in terms of the creation operators.
For instance, the state that corresponds to 
\beqa
\mbox{Tr}(X^{A_1B_1}X^{A_2B_2}\cdots X^{A_lB_l})
\label{SO(6)spinchain}
\eeqa
on $R^4$ is
\beqa
\frac{2^{\frac{l}{2}}}{N^{\frac{l}{2}}}
\mbox{Tr}(\alpha_{00}^{A_1B_1\dagger}\alpha_{00}^{A_2B_2\dagger}\cdots
\alpha_{00}^{A_lB_l\dagger})|0\rangle,
\label{SO(6)spinchainstate}
\eeqa
where $|0\rangle$ is the Fock vacuum and the vacuum of the free theory.
Note that this state is normalized in the large $N$ limit. 
In general, the operators that contain derivatives correspond to the states
constructed by the higher modes of the creation operators.
It was shown \cite{Minahan:2002ve} that the l-loop dilatation operator for
a set of the operators (\ref{SO(6)spinchain}) with fixed $l$ 
is regarded as the hamiltonian of
the integrable $SO(6)$ spin chain. In this sence, the operators
(\ref{SO(6)spinchain}) are regarded as the integrable $SO(6)$ spin chain.
In section 6, we will obtain this dilatation operator by calculating
the energy corrections of the states (\ref{SO(6)spinchainstate}).

For later convenience, we rewrite the 
superconformal transformation (\ref{superconformaltransformationSU(4)})
for the free theory in terms of the modes. We introduce the two-component
spinor $\eta^A$ for the conformal Killing spinor:
\beqa
&&\epsilon_+^A=\left( \begin{array}{c}
                    \eta^A \\
                    0       
                    \end{array} \right), \n
&&\nabla_a\epsilon_+^A=\pm\frac{i}{2}\gamma_a\gamma^0\epsilon_+^A
\;\;\leftrightarrow \;\;
\nabla_a\eta^A=\pm \frac{i}{2}\sigma_a\eta^A.
\label{conformalKillingspinor2-component}
\eeqa
Using the two-components spinors, we rewrite 
(\ref{superconformaltransformationSU(4)}) with $g=0$ as
\beqa
&&\delta_{\eta}A_i=i(-\psi_A^{\dagger}\sigma_i\eta^A
+\eta_A^{\dagger}\sigma_i\psi^A), \n
&&\delta_{\eta}X^{AB}=i(-\eta^{AT}\sigma^2\psi^B+\eta^{BT}\sigma^2\psi^A
-\epsilon^{ABCD}\psi_C^{\dagger}\sigma^2(\eta_D^{\dagger})^T),\n
&&\delta_{\eta}\psi^A=-F_{0i}\sigma_i\eta^A
+\frac{i}{2}F_{ij}\epsilon_{ijk}\sigma_k\eta^A
-2\partial_0X^{AB}\sigma^2(\eta_B^{\dagger})^T
+2\nabla_iX^{AB}\sigma_i\sigma^2(\eta_B^{\dagger})^T
-2iX^{AB}\sigma^2(\eta_B^{\dagger})^T. \n
\label{susy2-component}
\eeqa
As anticipated in section 3,
(\ref{CKSharmonics}) and  (\ref{conformalKillingspinor2-component})
show that $\eta^A$ is expanded in terms of 
$e^{\mp\frac{i}{2}t}Y_{0M\alpha}^{\pm}$:
\beqa
\eta^A_{\alpha}
=\sum_{m=\pm \frac{1}{2}}\eta^A_{m+}e^{-\frac{i}{2}t}Y_{0M\alpha}^+
+\sum_{m=\pm \frac{1}{2}}\eta^A_{m-}e^{\frac{i}{2}t}Y_{0M\alpha}^-.
\label{etamodes}
\eeqa
The superconformal transformation for the KK modes are read off by 
substituting
(\ref{KKmodeexpansion}) and (\ref{etamodes}) into (\ref{susy2-component}).
In Fig.\ref{KKtower}, the solid and dotted arrows represent
the superconformal transformation for the creation operator caused 
by $\eta_{m+}$ and $\eta_{m-}^*$, respectively. 
In particular, the transformation
of the lowest creation operators caused by $\eta_{m+}$ is
\beqa
&&\delta_{\eta_+} \alpha_{00}^{AB\dagger}
=i\sqrt{2}\sum_{m=\pm\frac{1}{2}}
(-1)^m(\eta^A_{m+}d_{0M}^{B\dagger}-\eta^B_{m+}d_{0M}^{A\dagger}), \n
&&\delta_{\eta_+} d_{0M}^{A\dagger}
=2\sqrt{2}\sum_{m_1=\pm\frac{1}{2},m_2=0,\pm 1}(-1)^{m+\frac{1}{2}}
C^{1m_2}_{\frac{1}{2}m_1\;\frac{1}{2}m}\eta^A_{m_1+}a_{0M_2+}^{\dagger}, \n
&&\delta_{\eta_+} a_{0M\rho}^{\dagger}=0.
\label{superconformaltransformationmode}
\eeqa
We will use these equations in section 6.

\section{Consistent truncations}
\setcounter{equation}{0}
In this section we describe the consistent truncations 
of $\mathcal N=4$ SYM on $R \times S^3$ to the theories with 16 supercharges, 
in terms of the mode expansion performed in the previous section.
This description helps us to extract various results for the theories 
with 16 supercharges from ones for $\mathcal N=4$ SYM on $R\times S^3$, 
such as the 1-loop hamiltonian for the $SO(6)$ sector (section 6) and 
the 1-loop effective action around a BPS solution (section 7).
In section~\ref{sec:cons-trunc-theor}, we make the consistent 
truncations of $\mathcal N=4$ SYM on $R \times S^3$ to the theories 
with 16 supercharges in terms of the KK modes.
In section~\ref{sec:comp-with}, we compare the mass spectrum 
of $\mathcal N=4$ SYM on $R\times S^2$ with that of the theory
obtained by quotienting the original theory by $U(1)$.
We clarify how 
quotienting by $U(1)$ yields $\mathcal N=4$ SYM on $R\times S^2$.
In section~\ref{sec:non-trivial-vacua}, we examine
the vacua of $\mathcal N=4$ SYM on $R\times S^2$ 
in terms of the KK modes.

\subsection{Consistent truncations to theories with 16 supercharges}
\label{sec:cons-trunc-theor}
The original SYM on $R\times S^3$ has the superconformal 
$SU(2,2|4)$, whose bosonic subgroup is $SO(2,4)\times SO(6)$.
$SO(2,4)$ has a subgroup $SO(4)$ that is the isometry of 
the $S^3$ on which the theory defined.
In section 2, we decomposed the $SO(4)$ as 
$SU(2)\times \tilde{SU}(2)$ and developed the harmonic expansion.
We consider a subgroup of $\tilde{SU}(2)$.
We project out all fields of $\mathcal N=4$ SYM on $R \times S^3$ 
which are not invariant under the subgroup of $\tilde{SU}(2)$ and 
consider the same interactions for the remaining fields as the ones 
in $\mathcal N=4$ SYM on $R\times S^3$.
Taking full $\tilde{SU}(2)$, U(1), and $Z_k$ as the subgroup 
of $\tilde{SU}(2)$ leads to the plane wave matrix model, 
$\mathcal N=4$ SYM on $R\times S^2$ and $\mathcal N=4$ SYM on $S^3/Z_k$, 
respectively \cite{Lin:2005nh}.

Let us describe the above truncations in terms of the KK modes.
The plane wave matrix model is obtained by keeping only the modes that 
are singlet with respect to $\tilde{SU}(2)$, namely $(0,0,6)$ 
as ($X_{00}^{AB}$), $(\frac{1}{2},0,4)$ as ($\psi_{0M+}^A$) and 
$(1,0,1)$ as ($A_{0M+}$) in the KK tower\cite{Kim:2003rz}.
The $\mathcal N=4$ SYM on $R\times S^3/Z_k$ is obtained by keeping 
only the modes with $\tilde{m}=\pm\frac{k}{2}q$, 
where $q \in \bm{Z}_{\ge 0}$.\footnote{The set ``$\bm{Z}_{\ge 0}$'' consists 
of zero and positive integers.}
For later convenience, we examine the multiplicity of the remaining modes
for fixed $\tilde{J}$.
When $k$ is even, the remaining modes after the truncation 
have the following quantum numbers of $\tilde{SU}(2)$:
\begin{eqnarray}
 \tilde J=\frac{n}{2}+\frac{v}{2},
\end{eqnarray}
where $n \in \bm{Z}_{\ge 0} $ and $v=0, 2, \cdots, k-2$, and
\begin{eqnarray}
\tilde{m}=0,\pm\frac{k}{2},\cdots,\pm\frac{k}{2}n
\end{eqnarray}
for each $v$.
Then the multiplicity of the remaining modes for fixed $n$ and $v$ is $2n+1$.
Note that all the modes with $\tilde J$ a half odd integer
should be projected out, 
because such modes cannot have $\tilde m=\frac{k}{2} \bm{Z}_{\ge 0}$.

In the odd $k$ case the discussion is similar to the above one.
The quantum number $\tilde{J}$ for the remaining modes in this case
takes the following values:
\begin{eqnarray}
\tilde J=\frac{n}{2}+\frac{v}{2},
\end{eqnarray}
where $n \in \bm{Z}_{\ge 0}$ and $v=$ $0$, $1$, $\cdots$, $k-1$.
Note that the range of $v$ for odd $k$ is different from that for even $k$.
The values of $\tilde{m}$ and  the multiplicity for
fixed $n$ and $v$ 
are summarized in Table~\ref{tab:k=odd}.
\begin{table}[h]
\renewcommand{\arraystretch}{1.5}
\begin{center}
\begin{tabular}{c|c||c|c}
$n$  & $v$  & $\tilde{m}$                                                            & multiplicity  \\ \hline
even & even & $0$,  $\pm\frac{2k}{2}$ ,$\cdots$, $\pm\frac{nk}{2}$                   & $n+1$ \\ \hline
even & odd  & $\pm\frac{k}{2}$, $\pm \frac{3k}{2}$, $\cdots$, $\pm \frac{k}{2}(n-1)$ & $n$ \\ \hline
odd  & even & $\pm \frac{k}{2}$, $\pm \frac{3k}{2}$, $\cdots$, $\pm \frac{k}{2}n$    & $n+1$ \\ \hline
odd  & odd  & $0$, $\pm \frac{2}{2} k$, $\cdots$, $\pm \frac{k}{2}(n-1)$             & $n$
\end{tabular}
\end{center}
\caption{The remaining modes for $\mathcal N=4$ SYM on $R\times S^3/Z_k$ for odd $k$.}
\label{tab:k=odd}
\end{table}

The $\mathcal N=4$ SYM on $R\times S^2$ is obtained 
by keeping only the modes with $\tilde{m}=0$.
We will discuss this truncation in the next subsection in detail.

We close this subsection by showing the consistency of the above truncations in terms of the KK modes.
Let us first consider the cases of $\mathcal N=4$ SYM on $R\times S^3/Z_k$ and on $R\times S^2$.
The conservation of $\tilde{m}$ implies that 
each term in the action of the original theory
includes no KK mode or more than one KK mode that are projected out in
the truncations. This fact ensures that 
the equation of motion in the original theory
for a KK mode projected out in the truncations becomes trivial after
the truncations. Hence, every classical solution of the truncated theories
can be lifted up to a classical solution of the original theory.

In a similar way, one can show that the 16 supercharges for
the supersymmetry transformations caused by $\eta_{m+}$ and 
$\eta_{m+}^*$ are preserved in the truncations. These parameters
have $\tilde{m}=0$. 
The conservation of $\tilde{m}$ again implies that after the truncations
the transformations of the KK modes that are projected out in the truncations
become trivial and those of the remaining modes
are still nontrivial. This means that the truncated theories have the 16 
supercharges corresponding to $\eta_{m+}$ and $\eta_{m+}^*$.

In the case of the plane wave matrix model one must also use the conservation of $\tilde J$ to show the consistency of the truncation.
Indeed the consistency of the truncation was checked explicitly in \cite{Kim:2003rz}.

\subsection{Comparison with $\mathcal N=4$ SYM 
on $R\times S^2$}\label{sec:comp-with}
In this subsection, we compare the remaining KK modes in the $U(1)$ truncation with the KK modes of $\mathcal N=4$ SYM on $R\times S^2$.
Due to the mixing terms in $\mathcal N=4$ SYM on $R\times S^2$ 
this comparison is not trivial.

We begin by recalling the action of $\mathcal N=4$ SYM on 
$R\times S^2$ \cite{Maldacena:2002rb}\footnote{The coefficient 
of the fermion mass term in (\ref{sym2}) is different from the one 
in \cite{Maldacena:2002rb}. This originates from the difference 
of the coordinate systems.}
\begin{eqnarray}
\label{sym2}
S_2 &=& \frac{1}{g^{\prime 2}} \int dt \frac{d\Omega'}{\mu^2} {\rm Tr} \bigg\{
-\frac{1}{4} F_{a'b'}F^{a'b'} 
-\frac{1}{2}(D_{a'} X_m)^2 -\frac{\mu^2}{8}X_m^2  
- \frac{1}{2}(D_{a'}\Phi)^2 -\frac{\mu^2}{2}\Phi^2
\n
&&-\frac{i}{2} \bar{\lambda} \Gamma^{a'} D_{a'} \lambda +\frac{i \mu}{8}\bar \lambda \Gamma^{12\Phi}\lambda 
-\frac{1}{2}\bar \lambda \Gamma^m\left[X_m,\lambda \right] 
+\frac{1}{2} \bar \lambda \Gamma^\Phi\left[\Phi,\lambda\right] \n
&&+\frac{1}{4}\left[X_m,X_n\right]^2
+\frac{1}{2}[\Phi,X_m]^2
-\mu\Phi F_{12}
\bigg\}, 
\end{eqnarray}
where $a'=0,1,2$, and $m=1,\cdots, 6$ and 
$(\Gamma^{a'}, \Gamma^{\Phi}, \Gamma^m)$ are ten dimensional gamma matrices.
The radius of $S^2$ is $\mu^{-1}$ and
the effective Yang-Mills coupling $g^{\prime 2}$ is defined 
by $ g^{\prime 2} = {g_{YM2}^2}/{4 \pi} $, 
since the area of $S^2$ is $4\pi$ times square of the radius.
We set $\mu=2$ since this value is obtained by the ${U}(1)$ 
truncating of $\mathcal N=4$ SYM on unit $S^3$.
The volume integration over $S^2$ is normalized as
\begin{eqnarray}
\int_{S^2} d\Omega' = \int_{S^2} \frac{d\Omega_2}{4 \pi \mu^{-2}}=1.
\end{eqnarray}
Note that the last term in (\ref{sym2}) mixes $\Phi$ with $A_{a'}$.

For later convenience we write down the mode expansion 
for the fields on $S^2$ here.
The details for the harmonics on $S^2$ are left to appendix C.
The mode expansions for the scalars, the vectors and the spinors on $S^2$ 
are given by
\footnote{The set $\bm{Z}_{> 0}$  consist of only ``positive'' integers, although the set $\bm{Z}_{\ge 0}$ consists of zero and positive integers.} 
\begin{eqnarray}
\label{eq:3}
X_{AB}(t, \Omega') &=& \sum_{J \in \bm{Z}_{\ge 0}} \sum_{m=-J}^J X_{AB}^{Jm}(t) {Y}_{Jm}(\Omega'),
\quad 
\Phi(t, \Omega') = \sum_{J \in \bm{Z}_{\ge 0}} \sum_{m=-J}^J \Phi_{Jm}(t) {Y}_{Jm}(\Omega'), 
\\
\label{eq:4}
A_{i}(t,\Omega') &=& \sum_{J \in \bm{Z}_{>0}} 
\sum_{m=-J}^J  \Big[ A_{Jm}^t(t) {Y}_{Jmi}^t(\Omega') +A_{Jm}^l(t) Y_{Jmi}^{l}(\Omega') \Big] 
\quad (\mbox{for } i=1,2),
\\
\label{eq:5}
\psi_{\alpha}^A (t,\Omega') &=& \sum_{J \in \frac{1}{2}+\bm{Z}_{\ge 0}} \sum_{m=-J}^J \psi_{Jm}^{A \alpha}(t) {Y}_{Jm\alpha}(\Omega') \quad (\mbox{for } \alpha=\pm\frac{1}{2}),
\end{eqnarray}
where the spinor $\psi_\alpha^A$ is a two component one on $S^2$.
Here $A_{Jm}^t$ and $A_{Jm}^l$ are the transverse and the longitudinal modes for the gauge fields.
In the Coulomb gauge, the longitudinal modes ($A_{Jm}^l$) in (\ref{eq:4}) vanish because $\nabla_{i} A_{i}=0$ and $\nabla_{i} Y_{Jmi}^t=0$.
Note that the range of $J$ is different form one for $S^3$, that is, $J$ takes
zero and positive integers for the scalar, positive integers for the vector
and positive half odd integers for the spinor. 
The hermicity of the fields 
implies together with (\ref{calYjm}) the following relations:
\begin{eqnarray}
\label{eq:6}
\left(X_{AB}^{Jm}\right)^\dagger &=& (-1)^m X^{AB}_{J-m}, 
\qquad
\left(\Phi_{Jm}\right)^\dagger = (-1)^m \Phi_{J-m}, 
\\
\label{eq:7}
\left(A_{Jm}^t \right)^\dagger &=& (-)^{-m} A_{J-m}^t,
\qquad
\left(A_{Jm}^l \right)^\dagger = (-1)^{-m} A_{J-m}^l.
\end{eqnarray}

\bigskip
\par
Let us first consider the spectrum of the $SO(6)$ scalar modes.
In this case the comparison of the spectrum is straightforward.
The mass term for the $SO(6)$ scalars in the $SU(4)$ notation 
is read off from (\ref{sym2}) as
\footnote{For a moment, we omit the common factor $1/(\mu g')^2$ 
for convenience since it is irrelevant here.}
\begin{eqnarray}
S_X &=& \int dt d\Omega' {\rm Tr} \left\{ \frac{1}{2}X_m \nabla^2 X_m -\frac{\mu^2}{8} X_m^2 \right\}
\nonumber\\
&=&
\int dt \sum_{J \in \bm{Z}_{\ge 0}} \sum_{m=-J}^J \left\{-\frac{1}{2} \left[ \mu \big(J+\frac{1}{2}\big)\right]^2 {\rm Tr} \left\{ (X_{Jm}^{AB})^\dagger X_{Jm}^{AB}
\right\}\right\},
\label{eq:9}
\end{eqnarray}
where in the second line we made the mode expansion by using (\ref{eq:3}) 
and used the formulae (\ref{calYjm}) and (\ref{calYjmS}).
It is clear that this equation is the same as the third line in (\ref{freepart})with the modes with integer $J$ and $\tilde m=0$ kept.
Note that all the scalar modes with half odd integer $J$ in (\ref{freepart}) 
should be projected out in this truncation because these modes 
cannot have  $\tilde m=0$.
The mass for the scalars on $S^2$ are immediately read off 
as $\mu(J+\frac{1}{2})$.
The multiplicity for fixed $J$ is given by
\begin{eqnarray*}
\sum_{m=-J}^J 1 =2J+1. 
\end{eqnarray*}
The result is summarized in Table~\ref{XS2}.
\begin{table}[h]
\begin{center}
\renewcommand{\arraystretch}{1.5}
\begin{tabular}{c|c||c}
mass        & multiplicity & $X_{JM}^{AB}$ \\ \hline
$\mu(J+\frac{1}{2})$  & $2J+1$     & $(J,J,6)$
\end{tabular}
\caption{The $SO(6)$ scalar mass spectrum of $\mathcal N=4$ SYM on $R \times S^2$ :
The range of $J$ is $J \in \bm{Z}_{\ge 0}$.
Note that $\mu=2$.
The column of $X_{JM}^{AB}$ shows the corresponding $\mathcal N=4$ scalar modes on $S^3$ with the same mass.}
\label{XS2}
\end{center}
\end{table}

We next consider the gauge field $A_{i}$ and the  scalar $\Phi$ together.
As mentioned before this comparison is not straightforward 
due to the mixing between $A_{i}$ and $\Phi$.
We obtain their mass terms using 
the mode expansions (\ref{eq:3}) and (\ref{eq:4}) as follows:
\begin{eqnarray}
S_{A\Phi} &=& \int dt {d\Omega'} {\rm Tr} \left[ \frac{1}{2} A_{i} \nabla^2 A_{i} -\frac{\mu^2}{2} A_{i} A_{i}
+\frac{1}{2}\Phi \nabla^2 \Phi  -\frac{\mu^2}{2} \Phi^2 -\mu \Phi F_{12}\right] 
\\ &=&
 \int dt {\rm Tr} \left\{
\frac{\mu^2}{2} \sum_{J \in \bm{Z}_{\ge 0}} \sum_{m=-J}^J \left[A_{JM}^{t\dagger}, \Phi^{\dagger}_{Jm} \right]
\left[\begin{array}{@{\,}cc@{\,}}
 -J(J+1) & \sqrt{J(J+1)}\\
 \sqrt{J(J+1)} & -J(J+1)-1     \end{array}
\right]
\left[\begin{array}{@{\,}c@{\,}}
 A_{Jm}^t\\
 \Phi_{Jm}      \end{array}
\right] \right\}. \nonumber
\end{eqnarray}
Here we took the Coulomb gauge, 
so that there is no longitudinal mode $A_{Jm}^l$ in this expression.
A unitary matrix that diagonalizes the above mass matrix is given by
\begin{eqnarray}
 U =\frac{1}{\sqrt{2J+1}}
\left[\begin{array}{@{\,}cc@{\,}}
 \sqrt{J+1} & -\sqrt{J} \\
 +\sqrt{J}  & \sqrt{J+1} \end{array}
\right].
\end{eqnarray}
By redefining the modes for $A_{Jm}$ and $\Phi_{Jm}$ as
\begin{eqnarray}
 iA_{(J-1)m+} &\equiv& \sqrt{\frac{1+J}{1+2J}}A_{Jm}^t+\sqrt{\frac{J}{1+2J}} \Phi_{Jm}, \qquad (\mbox{for } J \ge 1)
\\
iA_{Jm-} &\equiv& -\sqrt{\frac{J}{1+2J}} A_{Jm}^t + \sqrt{\frac{1+J}{1+2J}} \Phi_{Jm}, \qquad (\mbox{for } J \ge 0)
\end{eqnarray}
we find
\begin{eqnarray}
\label{AS2}
S_{A\Phi}= \!\int \! dt\mbox{Tr}\left\{\!-\frac{1}{2} \sum_{J \in \bm{Z}_{\ge 0}} \sum_{m=-J-1}^{J+1}  \mu^2(J+1)^2 A_{Jm+}^{\dagger} A_{Jm+} -\frac{1}{2} \sum_{J \in \bm{Z}_{\ge 0}} \sum_{m=-J}^J \mu^2 (J+1)^2 A_{Jm-}^{\dagger} A_{Jm-}
\!\right\}. 
\n
\end{eqnarray}
It is clear  that this expression is the same as the second line 
in (\ref{freepart}) with the modes with $\tilde m=0$ kept.
Note that all the vector modes with half odd integer $J$ in (\ref{freepart}) 
should be projected out in this truncation 
because these modes cannot have  $\tilde m=0$.
The result are summarized in Table~\ref{tab:A}.
\begin{table}[h]
\begin{center}
\renewcommand{\arraystretch}{1.5}
\begin{tabular}{c|c||c}
mass        & multiplicity & $A_{JM\pm}$ \\ \hline
$\mu(J+1)$  & $2J+1$     & $(J,J+1,1)$ \\ \hline
$\mu (J+1)$ & $2J+3$     & $(J+1,J,1)$
\end{tabular}
\caption{The gauge boson and $\Phi$ mass spectrum of $\mathcal N=4$ SYM on $R \times S^2$ :
The range of $J$ is $J \in \bm{Z}_{\ge 0}$.
Note that $\mu=2$.
The column of $A_{JM\pm}$ shows  the corresponding gauge field modes on $S^3$ with the same mass.}
\label{tab:A}
\end{center}
\end{table}

Finally, in a similar way, we examine the mass spectrum of the fermions.
The fermion mass term in (\ref{sym2}) is
\begin{eqnarray}
 S_{\lambda} &=& \int dt {d\Omega'} {\rm Tr} \left[-\frac{i}{2}\bar \lambda \Gamma^{i} \nabla_{i} \lambda +\frac{i \mu}{8} \bar \lambda \Gamma^{12\Phi} \lambda \right]
=
 \int dt d\Omega' {\rm Tr} \left[ i \psi_A^\dagger \sigma^i \nabla_i \psi^A + \frac{\mu}{4} \psi_A^\dagger \psi^A \right]
\n
&=& {\rm Tr} \int dt \sum_{J \in \frac{1}{2}+\bm{Z}_{\ge 0}}\sum_{m=-J}^J \mu \left[ \psi_{JmA}^{1/2 \dagger} \,\, \psi_{JmA}^{-1/2 \dagger} \right]
\left[
\begin{array}{@{\,}cc@{\,}}
 \frac{1}{4} & J+\frac{1}{2}\\
J+\frac{1}{2} & \frac{1}{4}
\end{array}
\right]
\left[
\begin{array}{@{\,}c@{\,}}
 \psi_{Jm}^{\frac{1}{2} A}\\
 \psi_{Jm}^{-\frac{1}{2} A}
\end{array}
\right],
\label{FS2}
\end{eqnarray}
In the first line we decomposed the sixteen component spinor $\lambda$ 
into the two component one $\psi_\alpha$ using (\ref{10to4}) and (\ref{4to2}).
In the second line we made the mode expansion by using (\ref{eq:5}).
Then a unitary matrix that diagonalize the fermion mass matrix 
in  (\ref{FS2}) is given by
\begin{eqnarray}
 V = \frac{1}{\sqrt 2}
\left[
\begin{array}{@{\,}cc@{\,}}
 1 & 1\\
 -1 & 1
\end{array}
\right].
\end{eqnarray}
After redefining the modes as
\begin{eqnarray}
 \psi_{(J-\frac{1}{2})(m,0)+}^A \equiv \frac{1}{\sqrt 2}\left[\psi_{Jm}^{\frac{1}{2} A}- \psi_{Jm}^{-\frac{1}{2} A} \right],
\qquad
 \psi_{J(m,0)-}^A \equiv \frac{1}{\sqrt 2}\left[\psi_{Jm}^{\frac{1}{2} A}+ \psi_{Jm}^{-\frac{1}{2} A} \right],
\end{eqnarray}
one finds
\begin{eqnarray}
\label{DFS2}
S_{\lambda} &=&  \int dt {\rm Tr} \left\{\sum_{J \in \bm{Z}_{\ge 0}} \sum_{m=-J-\frac{1}{2}}^{J+\frac{1}{2}} -\mu\left[J+\frac{3}{4} \right] \psi_{J(m,0)+A}^{\dagger} \psi_{J(m,0)+}^A \right.\n
&&\left.+ \sum_{J \in \frac{1}{2}+\bm{Z}_{\ge 0}} \sum_{m=-J}^J  \mu \left[J+\frac{3}{4}\right]\psi_{J(m,0)-A}^{\dagger} \psi_{J(m,0)-}^A\right\}.
\end{eqnarray}
It is clear that this expression  is the same as the forth line in  
(\ref{freepart}) with the modes $\tilde m=0$ kept.
The multiplicity for the modes with $J$ is $2J+1$.
Notice that all the fermion mode $(J+\frac{1}{2},J,4)$ 
with half odd integer $J$ in (\ref{freepart}) should be projected out 
because these modes cannot have $\tilde m=0$.
For the same reason all the fermion mode $(J,J+\frac{1}{2},4)$ 
with integer $J$ in (\ref{freepart}) should be projected out.
The result for the fermion is summarized in Table~\ref{tab:f}. 
\begin{table}[h]
\begin{center}
\begin{tabular}{c|c|c||c}
 $J$ &mass                 & multiplicity & $\psi_{JM\pm}$ \\ \hline
$J \in \bm{Z}_{\ge 0}$ & $\mu(J+\frac{3}{4})$ & $2J+2$     & $(J+\frac{1}{2},J,4)$ \\ \hline
$J \in \frac{1}{2}+\bm{Z}_{\ge 0}$ & $\mu(J+\frac{3}{4})$ & $2J+1$     & $(J,J+\frac{1}{2}, \bar 4)$
\end{tabular}
\caption{The fermion mass spectrum of $\mathcal N=4$ SYM on $R\times S^2$ :The column of $\psi_{JM\pm}$ shows the corresponding fermion modes of $\mathcal N=4$ SYM on $R\times S^3$ with the same mass.
Note that $\mu=2$.}
\label{tab:f}
\end{center}
\end{table}

\subsection{Non-trivial vacua of $\mathcal N=4$ SYM on $R\times S^2$}\label{sec:non-trivial-vacua}
It is discussed in \cite{Lin:2005nh} that $\mathcal N=4$ super Yang-Mills 
on $R\times S^2$ has 
many non-trivial vacua.
Then it is valuable to describe these non-trivial vacua in terms of 
the modes to investigate the dynamics of this theory there, 
although we will study this theory in the trivial vacuum in this paper.
%
%
\par
Let us start with writing down the potential terms in (\ref{sym2}) 
that we focus on :
\begin{eqnarray}
S_{pot} =  \frac{1}{ g^{\prime  2} \mu^2} \int dt d\Omega'
{\rm Tr} \left\{ - \frac{1}{2}\Big(F_{12}+\mu \Phi \Big)^2 - \frac{1}{2} \Big(\nabla_i \Phi - i \left[A_i, \Phi\right] \Big)^2   \right\}.
\end{eqnarray}
Because the potential consist of the sum of the two complete square terms, one immediately reads off the conditions for the zero-energy vacua:
\begin{eqnarray}
\label{eq:10}
&&F_{12} + \mu \Phi =0,
\\
\label{eq:11}
&&\nabla_i \Phi -i \left[A_i, \Phi \right] = 0 \qquad (i=1,2).
\end{eqnarray}
These equations are rewritten  in terms of the KK modes (\ref{eq:3}) 
and (\ref{eq:4}) as
\begin{eqnarray}
\label{F12}
&&-\mu \sqrt{J(J+1)} A_{Jm}^t + \mu \Phi_{Jm}+\frac{n_{J_1}^0 n_{J_2}^0}{4 n_J^0} \big\{ 1- (-1)^{J_1+J_2-J} \big\} C_{J_1 1 \, J_2 -1}^{J0} C_{J_1 m_1 J_2 m_2}^{Jm} \left[A_{J_1 m_1}^t, A_{J_2 m_2}^t\right]
=0, \n
&&\\
&& \n
\label{lng}
&&\mu J(J+1) \Phi_{Jm} -\frac{n_{J_1}^0 n_{J_2}^0}{2 n_J^0} \sqrt{J_2 (J_2+1)} \{ 1- (-1)^{J_1+J_2-J} \big\} C_{J_1 1 \, J_2 -1}^{J0} C_{J_1 m_1 J_2 m_2}^{Jm} \left[A_{J_1 m_1}^t, \Phi_{J_2 m_2}\right]
=0, \n
&&\\
\label{trs}
&&n_{J_1}^0 n_{J_2}^0  \{ 1 + (-1)^{J_1+J_2-J} \big\} C_{J_1 1 \, J_2 0}^{J 1} C_{J_1 m_1 J_2 m_2}^{Jm} \left[A_{J_1 m_1}^t, \Phi_{J_2 m_2}\right]=0,
\end{eqnarray}
with no summation over $J$ and $m$.
Here we took the Coulomb gauge $\nabla_i A_i=0$, so that there is no longitudinal mode $A_{Jm}^l$ in the above expressions.
The equations (\ref{lng}) and (\ref{trs}) correspond to
the longitudinal and transverse components of (\ref{eq:11}), respectively.

Unfortunately, it is difficult
to find general solutions for (\ref{F12})-~(\ref{trs}).
Then we would like to solve them with some assumptions.
Let us first make an ansatz that the non-vanishing modes are 
only $A_{1m}$ and $\Phi_{1m}$ and that they are related as
\begin{equation}
\Phi_{1m}= \alpha A_{1m}^t.
\end{equation}
Then it is easily verified using the relation 
$C^{Jm}_{1m_11m_2}=(-1)^{1+1-J}C^{Jm}_{1m_21m_1}$ that
the equation (\ref{trs}) is trivially satisfied.
When we set $\alpha=\frac{1}{\sqrt 2}$, the equations (\ref{F12}) and (\ref{lng}) are reduced to three non-trivial ones:
\begin{eqnarray}
 \left[ A_{10}, A_{1\pm1}\right]= \mp \sqrt{\frac{2}{3}} \mu A_{1\pm 1},
\qquad
\left[A_{11}, A_{1-1}\right] = \sqrt{\frac{2}{3}} \mu A_{10}.
\end{eqnarray}
This is nothing but the $SU(2)$ algebra.
Then the non-trivial solution is
\begin{equation}
\label{eq:2}
 A_{1-1}=\frac{\mu}{\sqrt 3} L_+, \quad 
A_{11}=-\frac{\mu}{\sqrt 3} L_-,
\quad
A_{10}=\sqrt{\frac{2}{3}} \mu L_3,
\quad
\Phi_{1m} = \frac{1}{\sqrt 2} A_{1m},
\end{equation}
where $L_i$'s are the $SU(2)$ generators.
It is easily checked that this solution are consistent with the hermicity conditions for the KK modes (\ref{eq:6}) and (\ref{eq:7}), of course, as it should be.
When we consider the $\mathcal N=4$ $U(N)$ SYM on $R\times S^2$, our solution is expressed by an irreducible or reducible $SU(2)$ representation of dimension $N$.
Then the number of the vacua that our solution (\ref{eq:2}) can represent  
is equal to the partitions of $N$, that is, $P(N)$.
This number coincides with the number of vacua of the plane wave matrix model
\cite{Lin:2005nh}.
Note that our solution corresponds to a part of the solutions discussed in 
\cite{Maldacena:2002rb,Lin:2005nh}, where the total number of the vacua of this theory and the tunneling amplitude between them are discussed.

\section{1-loop calculations and the $SO(6)$ spin chains}
\setcounter{equation}{0}
In this section, we examine the 1-loop corrections. 
We consider those in the original theory in sections 6.1$\sim$6.3, and
those in the truncated theories in section 6.4. In section 6.1, we illustrate
the calculation of the 1-loop diagrams with the 1-loop self-energy of $X_{AB}$.
In section 6.2, we introduce cut-offs for loop angular momenta as a 
regularization scheme and calculate the divergent parts of the self-energies
of all the fields and some interaction vertices. We see that the coefficients
of the logarithmic divergences are consistent with the vanishing of the
beta function and the Ward identity. In section 6.3, we determine some 1-loop
counter terms by examining the energy corrections of the BPS states. We
examine the 1-loop energy corrections of the states that correspond
to the operators on $R^4$ which are
regarded as the integrable $SO(6)$ spin chain. We show that 
the energy corrections are actually given by 
the hamiltonian of the spin chain.
In section 6.4, we determine some couter terms in the truncated theories
by examining the 1-loop energy corrections of the BPS states.
We find that the states viewed as the integrable
$SO(6)$ spin chain in the original theory are also viewed
as the same spin chain in the truncated theories.

\subsection{Calculation of 1-loop diagrams}
In the calculation of the 1-loop Feynman diagrams, 
we need the propagators, which are read off from (\ref{freepart}) as
\beqa
&&\langle X_{AB}^{JM}(q)_{kl}X_{A'B'}^{J'M'}(-q)_{k'l'}\rangle
=\frac{1}{2}\varepsilon_{ABA'B'}(-1)^{m-\tilde{m}}\delta_{JJ'}
\delta_{M\:-M'}\delta_{kl'}\delta_{lk'}
\frac{i}{q^2-{\omega_J^X}^2}, \label{Xpropagator}\\
&&\langle B_{JM}(q)_{kl}B_{J'M'}(-q)_{k'l'}\rangle
=(-1)^{m-\tilde{m}}\delta_{JJ'}\delta_{M\:-M'}\delta_{kl'}\delta_{lk'}
\frac{i}{4J(J+1)}, \label{Bpropagator}\\
&&\langle A_{JM\rho}(q)_{kl}A_{J'M'\rho'}(-q)_{k'l'}\rangle
=(-1)^{m-\tilde{m}+1}\delta_{JJ'}
\delta_{M\:-M'}\delta_{\rho\rho'}\delta_{kl'}\delta_{lk'}
\frac{i}{q^2-{\omega_J^A}^2}, \label{Apropagator}\\
&&\langle \psi_{JM\kappa}^A(q)_{kl}\psi_{J'M'\kappa'A'}^{\dagger}(q)_{k'l'}
\rangle
=\delta_{JJ'}\delta_{MM'}\delta^A_{\;A'}\delta_{\kappa\kappa'}
\frac{i(q-\kappa\omega_J^{\psi})}{q^2-{\omega_J^{\psi}}^2}, 
\label{psipropagator}\\
&&\langle c_{JM}(q)_{kl}\bar{c}_{J'M'}(-q)_{k'l'}\rangle
=(-1)^{m-\tilde{m}}\delta_{JJ'}\delta_{M\:-M'}\delta_{kl'}\delta_{lk'}
\frac{1}{4J(J+1)},
\label{cpropagator}
\eeqa
where $q$ is conjugate to $t$.

\begin{figure}[htbp]
\begin{center}
 \begin{minipage}{0.2\hsize}
  \begin{center}
\includegraphics[width=30mm]{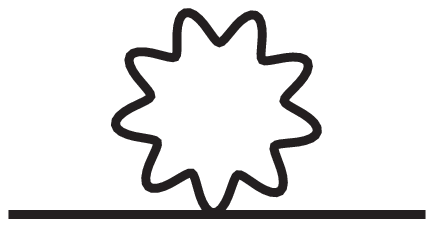}
  (X-a)
\end{center}
 \end{minipage}
 \begin{minipage}{0.2\hsize}
  \begin{center}
  {\includegraphics[width=30mm]{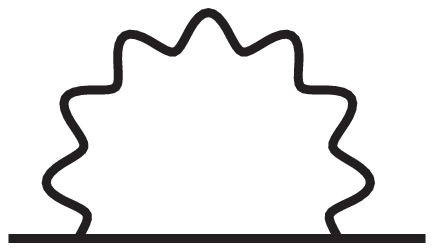}}
  (X-b)
\end{center}
 \end{minipage}
 \begin{minipage}{0.2\hsize}
  \begin{center}
  {\includegraphics[width=30mm]{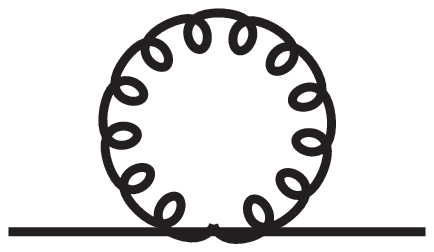}}
  (X-c)
  \end{center}
\end{minipage}
 \begin{minipage}{0.2\hsize}
  \begin{center}
   \includegraphics[width=30mm]{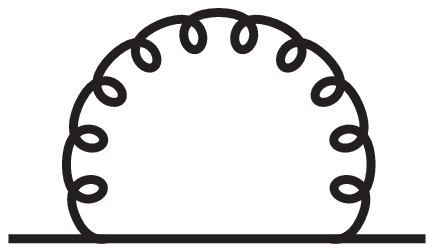}
   (X-d)
  \end{center}
 \end{minipage}
\\
 \begin{minipage}{0.2\hsize}
  \begin{center}
   {\includegraphics[width=30mm]{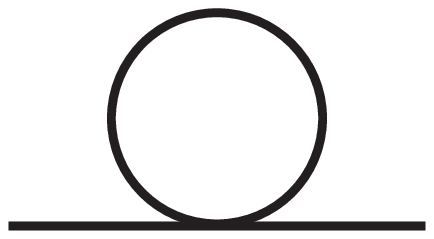}}
   (X-e)
\end{center}
 \end{minipage}
 \begin{minipage}{0.2\hsize}
  \begin{center}
   {\includegraphics[width=30mm]{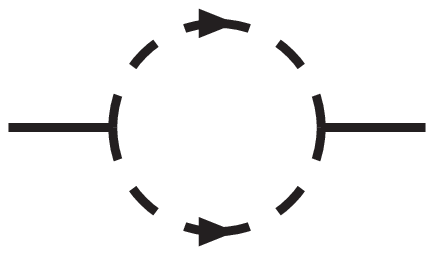}}
   (X-f)
  \end{center}
 \end{minipage}
\caption{Diagrams for the one-loop self-energy of $X_{AB}$. The curly line
represents the propagator of $A_i$. The wavy line represents the propagator
of $A_0$.
The solid line represents the propagator of $X_{AB}$. The dashed line
represents the propagator of $\psi^A$.}
\label{diagrams for self-energy of X}
\end{center}
\end{figure}

Here we consider the 1-loop self-energy of $X_{AB}$, 
which is $(-i)$ times the 1-loop contribution to the 1PI part of the truncated
2-point function
$\langle X_{AB}^{JM}(q)_{kl}X_{A'B'}^{J'M'}(-q)_{k'l'}\rangle$.
We will consider the self-energy of the other fields 
and the 1-loop corrections to
some interaction vertices in the next subsection.
The six diagrams for the self-energy of $X_{AB}$ are shown 
in Fig. 2.
We illustrate our method by calculating
one of the diagrams, $(X-f)$. 
By using the vertices in (\ref{interactionpart1}) and 
the propagator (\ref{psipropagator}), 
we obtain an expression for this diagram.
\beqa
&&4ig^2N\delta_{kl'}\delta_{lk'}\frac{1}{2}\varepsilon_{ABA'B'}
\sum_{J_1M_1J_2M_2\kappa_1\kappa_2} \n
&&\times\int \frac{dp}{2\pi}\left(
\frac{i(p-\kappa_1\omega_{J_1}^{\psi})}{p^2-{\omega_{J_1}^{\psi}}^2}
\frac{i(-p+q-\kappa_2\omega_{J_2}^{\psi})}{(-p+q)^2-{\omega_{J_2}^{\psi}}^2}
{\cal F}^{J_1\:-M_1\kappa_1}_{J_2M_2\kappa_2\;J-M}
{\cal F}^{J_2M_2\kappa_2}_{J_1\:-M_1\kappa_1\;J'-M'} \right.\n
&&\qquad\qquad\;\;\;\left.
+\frac{i(p-\kappa_1\omega_{J_1}^{\psi})}{p^2-{\omega_{J_1}^{\psi}}^2}
\frac{i(-p-q-\kappa_2\omega_{J_2}^{\psi})}{(p+q)^2-{\omega_{J_2}^{\psi}}^2}
{\cal F}^{J_1\:-M_1\kappa_1}_{J_2M_2\kappa_2\;J'-M'}
{\cal F}^{J_2M_2\kappa_2}_{J_1\:-M_1\kappa_1\;J-M} \right) \n
&&=-8g^2N\delta_{kl'}\delta_{lk'}\frac{1}{2}\varepsilon_{ABA'B'}
\sum_{J_1M_1J_2M_2\kappa_1}
{\cal F}^{J_1M_1\kappa_1}_{J_2M_2\kappa_1\;J-M}
{\cal F}^{J_2M_2\kappa_1}_{J_1M_1\kappa_1\;J'-M'} 
\frac{\omega_{J_1}^{\psi}+\omega_{J_2}^{\psi}}
{q^2-(\omega_{J_1}^{\psi}+\omega_{J_2}^{\psi})^2}.\n
\eeqa
Here we plug in the expression for ${\cal F}$ in (\ref{vertexfunctions}),
take summations over $M_1$ and $M_2$ using the formulae
(\ref{C-Gand3-j}) and (\ref{3-jsummation}). We also take
a summation over $\kappa_1$ and plug in
the expression for the $9-j$ symbol available in \cite{vmk}.
We eventually obtain
\beqa
&&-16g^2N\delta_{kl'}\delta_{lk'}\frac{1}{2}\varepsilon_{ABA'B'}
(-1)^{m-\tilde{m}}\delta_{JJ'}\delta_{M-M'} \n
&&\;\;\times\sum_{J_1J_2}\frac{(2J_1+2J_2+3)(J_1+J_2+J+2)(J_1+J_2-J+1)}
{(q^2-(2J_1+2J_2+3)^3)(2J+1)},
\label{expressionforX-f}
\eeqa
where $J_1$ and $J_2$ take non-negative half-integers 
$(0,\frac{1}{2},1,\frac{3}{2},\cdots,)$,
and summations over $J_1$ and $J_2$ are taken such that
they satisfy $|J_1-J_2|\leq J \leq J_1+J_2$.
Because the summations give rise to divergence, we must introduce
a regularization. In the next subsection, we give a method for regularization
and calculate the divergent parts of the 1-loop diagrams.

In the following, we list unregularized expressions for all the diagrams 
in Fig. 2. The 1-loop self-energy of $X_{AB}$ takes the form
\beqa
g^2N\delta_{kl'}\delta_{lk'}\frac{1}{2}\epsilon_{ABA'B'}(-1)^{m-\tilde{m}}
\delta_{JJ'}\delta_{M-M'}\Pi^X_J(q).
\eeqa
We write down the contributions of each diagram
to $\Pi^X_J(q)$.
\beqa
&&(X-a)=\sum_{J_1\neq0,J_2M_1M_2}
\frac{i(-1)^{m_1-\tilde{m}_1+m_2-\tilde{m}_2} 
\delta(0)}{2J_1(J_1+1)} 
{\cal C}_{J\:-M\;J_1M_1\;J_2\:-M_2}{\cal C}_{J'\:-M'\;J_1\:-M_1\;J_2M_2}, \n
&&(X-b)=-\sum_{J_1\neq0,J_2M_1M_2}
\frac{i(-1)^{m_1-\tilde{m}_1+m_2-\tilde{m}_2}\delta(0)}{2J_1(J_1+1)} 
{\cal C}_{J\:-M\;J_1M_1\;J_2\:-M_2}{\cal C}_{J'\:-M'\;J_1\:-M_1\;J_2M_2} \n
&&\qquad\qquad\;\;
- \frac{1}{4}\sum_{J_1\neq 0,J_2}\frac{(2J_1+1)\left[q^2+(2J_2+1)^2\right]}
{J_1(J_1+1)(2J+1)} \left\{J, J_1, J_2\right\}, \n
&&(X-c)=-2\sum_{J_1} \frac{(2J_1+1)(2J_1+3)}{2J_1+2}, \n
&&(X-d)=-4\sum_{J_1 J_2} \n
&&\qquad\qquad\;\;
\frac{(2J_1+2J_2+3)(J+J_1+J_2+2)(J_1+J_2-J+1)(J-J_1+J_2+1)(J+J_1-J_2)}
{ (2J+1) (J_2+1)^2 \big[q^2-(2J_1+2J_2+3)^2 \big] }  
\nonumber
\\ &&\qquad\qquad\;\;
\times \{ J,J_1,J_2 \} \{ J,J_1,J_2+1 \},\n
&&(X-e)=-5\sum_{J_1 J_2} \frac{2J_2+1}{2J+1}  
\left\{J,J_1,J_2\right\}, \n
&&(X-f)=-16\sum_{J_1 J_2} 
\frac{(2J_1+2J_2+3)(J_1+J_2+J+2)(J_1+J_2-J+1)}
{(2J+1)\big[q^2-(2J_1+2J_2+3)^2 \big]}  \{J,J_1,J_2\}
\label{expressionsforselfenergyofscalars} 
\eeqa
where $\{J,J_1,J_2\}$ represents the constraint $|J_1-J_2|\leq J \leq J_1+J_2$.
Note that the terms proportional to $\delta (0)$ cancel in
$(X-a)$ and $(X-b)$ \cite{Aharony:2005bq}. 
We will later need the 1-loop on-shell self-energy for 
the lowest mode of $X_{AB}$, which is obtained by plugging in 
$q=1$ and $J=0$ into (\ref{expressionsforselfenergyofscalars}).
\beqa
&&(X-a)+(X-b)=-\frac{1}{4}
\sum_{J_1\neq 0}\frac{(2J_1+1)(1+(2J_1+1)^2)}{J_1(J_1+1)}, \n
&&(X-c)=-2\sum_{J_1}\frac{(2J_1+1)(2J_1+3)}{2J_1+2}, \n
&&(X-d)=0, \n
&&(X-e)=-5\sum_{J_1}(2J_1+1), \n
&&(X-f)=4\sum_{J_1}(4J_1+3).
\label{on-shellself-energy}
\eeqa

\subsection{1-loop divergences and the Ward identity}
All the expressions in (\ref{expressionsforselfenergyofscalars}) are divergent
and must be regularized.
As a regularization method, we introduce a cut-off for the loop angular
momentum. Again, as an example, we explicitly regularize $(X-f)$. 
We introduce the cut-off $\Lambda_f$ for $J_1$. (Of course, we could
introduce it for $J_2$.) The suffix `$f$' indicates that
the cut-off is the one for the loop of $\psi^A_{JM\kappa}$. 
Fig. 3 shows the region of 
the regularized summations over $J_1$ and $J_2$. 
\begin{figure}[htbp]
\begin{center}
   {\includegraphics[width=70mm]{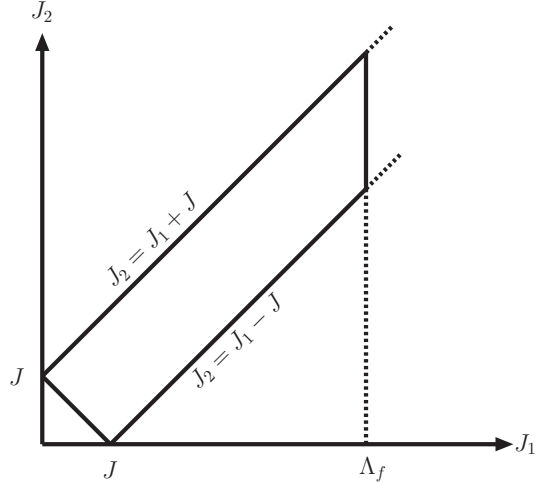}}
   \caption{Region of the regularized summations over $J_1$ and $J_2$}
   \label{cut-off}
\end{center}
\end{figure}
We define new variables
$P=J_1+J_2$ and $Q=J_2-J_1$, which take integers for integer $J$ and
half odd integers for half odd integer $J$.
Then, we obtain the regularized expression for $(X-f)$.
\beqa
-16
\left(\sum_{P=J}^{2\Lambda_f-J}\sum_{Q=-J}^{J}
            +\left.\sum_{r=-J+1}^J\sum_{Q=r}^J\right|_{P=2\Lambda_f+2r}\right)
\frac{(2P+3)(P+J+2)(P-J+1)}{(q^2-(2P+3)^2)(2J+1)}.
\eeqa
It is difficult to calculate this analytically, however the divergent part is
easily evaluated as
\beqa
&&8\sum_{P=J}^{2\Lambda_f-J}(P+\frac{3}{2})
+16\sum_{r=-J+1}^{J}\sum_{Q=r}^{J}\frac{\Lambda_f}{2J+1}
+2(q^2-(2J+1)^2)\sum_{P=J}^{\Lambda_f-J}\frac{1}{P} \n
&&=16\Lambda_f^2+32\Lambda_f+2(q^2-(2J+1)^2)\ln(2\Lambda_f).
\eeqa

We list the divergent parts of the expressions in 
(\ref{expressionsforselfenergyofscalars}). 
\beqa
&&(X-a)+(X-b)
=-2\Lambda_s^2-3\Lambda_s
+\left[ -q^2 - \frac{4}{3}J(J+1) -1\right]  \log(2 \Lambda_s), \n
&&(X-c)=-4 \Lambda_v^2 -10 \Lambda_v + 2\log(2\Lambda_v),  \n
&&(X-d)=\frac{16}{3}J(J+1) \log(2 \Lambda_v),  \n
&&(X-e)= -10 \Lambda_s^2 -15 \Lambda_s, \n
&&(X-f)=16 \Lambda_f^2 + 32 \Lambda_f 
+2 \left[ q^2 - (2J+1)^2 \right] \log (2 \Lambda_f ),
\label{divergentpartsofselfenergyofscalars}
\eeqa
where $\Lambda_v$ and $\Lambda_s$ represent the cut-off for the loop of
$A_{JM\rho}$ and the cut-off for the loop of $X_{AB}^{JM}$ or $B_{JM}$,
respectively.
It is natural that $\Lambda_s$, $\Lambda_v$ and $\Lambda_f$ are the same
order quantities, so that 
we can set $\log(2\Lambda_s)=\log(2\Lambda_v)=\log(2\Lambda_f)=\log(2\Lambda)$ 
in the divergent parts. In appendix D, we list the divergent parts
of the 1-loop self-energies of the other fields and those of 
the 1-loop corrections to some interaction vertices. 

It should be remarked that all the 1-loop divergences here and in appendix D 
are local ones, namely they can be canceled by the local counter terms.
This property is crucial in renormalizing the theory.
In order to keep this property, one must introduce the cut-off
for the angular momentum of a certain internal propagator in each diagram.
For instance, one is not allowed to introduce the cut-offs for the angular 
momenta of several internal propagators or divide a contribution of a diagram 
into several parts and introduce the cut-off for the angular momentum of
a different internal propagator in each part. Indeed, in the above example,
we have introduced the cut-off $\Lambda_f$ only for $J_1$.
Of course, 
the finite part as well as the divergent part in a 1-loop diagram 
generally depends
on for which angular momentum the cut-off is introduced. As discussed in the
following, however, this ambiguity does not matter. 
Our regularization method 
breaks the gauge symmetry and the superconformal symmetry though it preserves
the $R\times SO(4)$ symmetry. As in \cite{'tHooft}, these symmetries would be
recovered by introducing the counter terms that breaks the gauge invariance
or the superconformal invariance and making the fine-tuning for 
the coefficients of these counter terms including the finite renormalization. 
Our gauge fixing also respects only $R\times SO(4)$ symmetry. 
We have to consider, therefore, 
all the terms whose dimension is less than or equal to four and which
are invariant under $R\times SO(4)$, as the counter terms.
The counter terms quadratic in $A_i$, $A_0$, $c$, $X_{AB}$ and
$\psi^A$ take the following forms.
\beqa
A_i &:& \alpha_A\mbox{Tr}\left(\frac{1}{2}(\partial_0 A_i)^2
+\frac{1}{2}A_i\nabla^2A_i-A_iA_i\right)
+\frac{\beta_A}{2}\mbox{Tr}(A_i\nabla^2A_i+2A_iA_i) \n
&&-\gamma_A\mbox{Tr}(A_iA_i), 
\label{Aicounterterm} \\
A_0 &:& -\alpha_B\mbox{Tr}\left(\frac{1}{2}A_0\nabla^2A_0\right)
+\frac{\gamma_B}{2}\mbox{Tr}(A_0)^2, 
\label{A0counterterm} \\
c &:&\alpha_c\mbox{Tr}(-i\bar{c}\nabla^2c)
+\gamma_C\mbox{Tr}(\bar{c}c), 
\label{ghostscounterterm} \\
 X_{AB} &:& \alpha_X\mbox{Tr}\left(
\frac{1}{2}\partial_0 X_{AB}\partial_0 X^{AB}
+\frac{1}{2}X_{AB}\nabla^2X^{AB}-\frac{1}{2}X_{AB}X^{AB}\right) \n
&&+\frac{\beta_X}{2}\mbox{Tr}(X_{AB}\nabla^2X^{AB}) 
-\frac{\gamma_X}{2}\mbox{Tr}(X_{AB}X^{AB}), 
\label{Xcounterterm} \\
\psi^A  &:&
\alpha_{\psi}\mbox{Tr}(i\psi^{\dagger}_A\partial_0\psi^A
+i\psi^{\dagger}_A\sigma^i\nabla_i\psi^A)
+\beta_{\psi}\mbox{Tr}(i\psi^{\dagger}_A\sigma^i\nabla_i\psi^A).
\label{psicounterterm} 
\eeqa
The first term in each line is absorbed by the wave function renormalization
of the corresponding field.

Let us see that our results of the 1-loop calculation are consistent with
the vanishing of the beta function, which is characteristic of conformal
field theories. We immediately see that the quadratic and linear divergences
in (\ref{divergentpartsofselfenergyofscalars}) are absorbed in $\gamma_X$.
The sum of the logarithmic divergences in 
(\ref{divergentpartsofselfenergyofscalars}) is
$(q^2-{\omega_J^X}^2)\log(2\Lambda)$.
This shows that the cut-off dependent part of $\alpha_X$ is
\beqa
\alpha_X \sim -\log(2\Lambda)g^2N.
\label{alphaXdivergence}
\eeqa
Eqs.(\ref{divergentpartsofselfenergyofAi}), 
(\ref{divergentpartsofselfenergyofA0}), 
(\ref{divergentpartsofselfenergyofghosts}) 
and (\ref{divergentpartsofselfenergyoffermions}) in appendix D
show the divergent parts of the diagrams for the 1-loop self-energies
of $A_i$, $A_0$, $c$ and $\psi^A$, respectively.
The quadratic and linear divergences in
(\ref{divergentpartsofselfenergyofAi}) and  
(\ref{divergentpartsofselfenergyofA0}) are absorbed 
in $\gamma_A$ and $\gamma_B$, respectively,
while the self-energies of $c$ and $\psi^A$ contain only the 
logarithmic divergences.
The sum of the logarithmic divergences in
(\ref{divergentpartsofselfenergyofAi}) is 
$\frac{4}{3}(q^2-{\omega_J^A}^2)\log(2\Lambda)$.
The sum of those in (\ref{divergentpartsofselfenergyofA0}) vanishes.
The sum of those in (\ref{divergentpartsofselfenergyofghosts}) is
$-\frac{8i}{3}J(J+1)\log(2\Lambda)$. The sum of 
those in (\ref{divergentpartsofselfenergyoffermions}) is
$2(q+\kappa\omega_J^{\psi})\log(2\Lambda)$.
All of these logarithmic divergences are absorbed by the wave function 
renormalization.
We can determine the cut-off dependent parts of 
$\alpha_A$, $\alpha_B$, $\alpha_c$ and $\alpha_{\psi}$ as follows:
\beqa
&&\alpha_A \sim -\frac{4}{3}\log(2\Lambda)g^2N, 
\label{alphaAdivergence} \\
&&\alpha_B \sim 0,
\label{alphaBdivergence} \\
&&\alpha_c \sim \frac{2}{3}\log(2\Lambda)g^2N,  
\label{alphacdivergence} \\
&&\alpha_{\psi} \sim -2\log(2\Lambda)g^2N.
\label{alphapsidivergence}
\eeqa
As seen in (\ref{divergentpartsofghostghostgaugeinteraction}),
the diagrams for the 1-loop correction to the ghost-ghost-gauge
interaction term are not divergent. The counter term proportional to
$\mbox{Tr}(\nabla_i\bar{c}[A_i,c])$ does not depend on the cut-off.
This means together with
(\ref{alphaAdivergence}) and (\ref{alphacdivergence}) that
the bare coupling constant can coincide with the renormalized one,
namely the beta function vanishes.
Similarly, the divergent parts of the diagrams for the 1-loop correction
to the Yukawa interaction term are listed
in (\ref{divergentpartsofghostghostgaugeinteraction}) and contain
only the logarithmic divergences. The sum of those divergences is 
$\frac{5}{2}\log(2\Lambda)$. The cut-off dependent part of the coefficient
of the counter term proportional to $\mbox{Tr}(\psi^{\dagger}_A\sigma^2
[X^{AB},(\psi^{\dagger}_B)^T])$ is $-\frac{5}{2}\log(2\Lambda)g^3N$.
This again means together with (\ref{alphaXdivergence})
and (\ref{alphapsidivergence}) that the beta function vanishes.

In general, the coefficients of the logarithmic divergences do not depend on
the details of regularization, so that they respect the symmetries.
This is consistent with the fact that
we were able to check the vanishing of the beta function through
the logarithmic divergences in our 1-loop calculation.
Because our gauge choice only keeps the $R\times S^3$,
it is difficult to examine the Ward identities for the superconformal symmetry.
Here we content ourselves to see that the coefficients of the 1-loop logarithmic
divergences satisfy the Ward identity for the gauge symmetry. 
As in \cite{Aharony:2005bq}, we consider the Ward identity in the 
flat limit that relates
the 1-loop self-energy $\tilde{\Pi}_{ab}$ of
the gauge field with the coefficient $\Phi^a$ of the $K_ac$ term in
the 1-loop effective action, 
where $K_a$ is the source added for
the operator $[Q_{BRST},c]$.\footnote{Here
the longitudinal components of the gauge fields are included in
the definition of $\tilde{\Pi}_{ab}$.} It takes the form
\beqa
\partial^a\tilde{\Pi}_{ab}+(\partial^2\eta_{ab}-\partial_a\partial_b)\Phi^a=0.
\label{Wardidentity}
\eeqa
As discussed above, the logarithmic divergent parts of $\tilde{\Pi}_{ab}$
and $\Phi_a$ should satisfy this identity.
As explained in \cite{Aharony:2005bq}, the logarithmic divergent parts of
$\tilde{\Pi}_{ab}$ take the forms
\beqa
&&\tilde{\Pi}_{ij}^{div}=C((p_0^2-p_kp_k)\delta_{ij}+p_ip_j)
g^2N\log(2\Lambda), \n
&&\tilde{\Pi}_{0i}^{div}=Dp_ip_0g^2N\log(2\Lambda), \n
&&\tilde{\Pi}_{00}^{div}=(-C+2D)p_ip_ig^2N\log(2\Lambda),
\eeqa
where $C$ and $D$ are certain numerical constants.
The logarithmic divergent parts of $\Phi_a$ are determined by the Ward
identity (\ref{Wardidentity}) as
\beqa
\Phi_0^{div}=0, \;\;\;\;\; \Phi_i^{div}=(-C+D)p_ig^2N\log(2\Lambda).
\eeqa
We saw above that $C=\frac{4}{3}$ and $-C+2D=0$, namely $D=\frac{2}{3}$.
In our case, $\Phi_0$ obviously vanishes and $\Phi_i$ is determined by
calculating the diagram in Fig. 4.
Its divergent part is 
\begin{eqnarray}
\int \!\! dt d\Omega \, {\rm Tr}  \left( K_i \nabla_i c \right) 
\times \left[ -\frac{2}{3} g^2 N  \log(2 \Lambda)\right].
\end{eqnarray}
This means $-C+D=-\frac{2}{3}$, which is indeed consistent with $C=\frac{4}{3}$
and $D=\frac{2}{3}$. We can also read off $C$ and $D$ for the pure Yang Mills
sector by considering only $(A-a)\sim (A-f)$ in Fig. 7 and
$(B-a)\sim (B-c)$ in Fig. 8. The result is $C=-\frac{1}{2}$ and
$D=-\frac{7}{6}$ for the pure Yang Mills sector, which gives
$-C+D=-\frac{2}{3}$ again. This is consistent because $\Phi_a$ for the 
pure Yang Mills sector is the same as that for ${\cal N}=4$ SYM.
This consistency in pure Yang Mills is actually shown in \cite{Aharony:2005bq}.

\begin{figure}[htbp]
\begin{center}
   {\includegraphics[width=50mm]{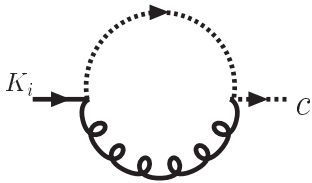}}
   \caption{Diagram determining $\Phi_i$. The curly line represents 
the propagator of $A_i$. The dotted line represents the propagator of the
ghost.}
   \label{fig:A1}
\end{center}
\end{figure}

We close this subsection with an interesting observation.
The quadratic and linear divergences appear
in (\ref{divergentpartsofselfenergyofscalars}), 
(\ref{divergentpartsofselfenergyofAi}) and 
(\ref{divergentpartsofselfenergyofA0}).
If we set 
\beqa
\Lambda_v=\Lambda_s-\frac{1}{2}, \;\;\;\;\; 
\Lambda_f=\Lambda_s-\frac{1}{4}, 
\label{shift}
\eeqa
those quadratic and linear divergences
cancel and only the logarithmic divergences are left.
Furthermore, these constant shifts of the cut-offs enable us to reproduce
the Casimir energy in the free theory as follows. 
When we rewrote the naive expression to
the normal ordered one in (\ref{freehamiltonianmode}), we discarded the
constant
\beqa
N^2\left(2\sum_{J}(2J+1)(2J+3)\frac{1}{2}\omega_J^A 
+6\sum_{J}(2J+1)^2\frac{1}{2}\omega_J^X
-8\sum_{J}(2J+1)(2J+2)\frac{1}{2}\omega_J^{\psi}\right),
\label{Casimirenergy}
\eeqa
where the first, second and third terms are the contributions of
the gauge fields, the scalars and the fermions, respectively.
Each term in (\ref{Casimirenergy}) is quartic divergent in the angular momentum
and must be regularized. If we set
the upper end in the summation over $J$
in the first term at $\Lambda_v$, in the second term at $\Lambda_s$ and 
in the third term at $\Lambda_f$ and 
assume the above constant shifts of the cut-offs (\ref{shift}), we
remarkably obtain the finite value, $\frac{3}{16}N^2$,
which is independent of $\Lambda_s$. This is equal to the Casimir energy and
is reasonably obtained as the zero point energy.
The constant shifts of the cut-offs correspond
to a complete specification of the regularization scheme.
The physical meaning of these shifts is unclear
at present and its understanding is an open problem. Here we only point out
that these shifts are obtained by requiring that the average of
$J$ and $\tilde{J}$ of the internal propagator agree for all the fields.
That we are left only with the logarithmic divergences after the shifts of 
the cut-offs does not mean that
we need no counter terms that break the gauge invariance.
We need in general the finite couter terms that break the gauge invariance
even in this situation.

\subsection{Determination of counter terms and the $SO(6)$ spin chain}
In this subsection, we obtain the 1-loop dilatation operator
for the operators (\ref{SO(6)spinchain}) in ${\cal N}=4$ SYM on $R^4$  
by calculating the order $g^2N$
corrections to the energy of the states (\ref{SO(6)spinchainstate}) in
${\cal N}=4$ SYM on $R\times S^3$.
One can also consider the states (\ref{SO(6)spinchainstate}) in the 
truncated theories. We show in the next subsection that the order $g^2N$
energy corrections of these states agree with that in the original theory,
namely these states in the truncated theories are also regarded as the 
same integrable $SO(6)$ spin chain.

For the above purpose, we need the $\Pi^X_{J=0}(1)$,
which is the coefficient of the on-shell self-energy for the lowest mode.
The determination of this value is equivalent to fixing $\gamma_X$ 
in (\ref{Xcounterterm}), because the first and second terms in 
(\ref{Xcounterterm}) vanishes for $J=0$ and $q=1$.
We determine this value by considering the BPS state.
In addition, we similarly determine 
$\Pi^A_{J=0}(2)$ and $\Pi^{\psi}_{J=0}(-\frac{3}{2})$.
The determination of the former is equivalent to fixing 
$\gamma_A$ in (\ref{Aicounterterm}), 
while that of $\Pi^{\psi}_{J=0}(-\frac{3}{2})$
is equivalent to fixing 
$\beta_{\psi}$ in (\ref{psicounterterm}).

We consider the half-BPS state in the free theory, which corresponds to
a special case with $l=2$ in (\ref{SO(6)spinchainstate}):
\beqa
\frac{2}{N}\mbox{Tr}(\alpha_{00}^{34\dagger}\alpha_{00}^{34\dagger})|0\rangle,
\label{chiralprimary}
\eeqa
This state is mapped to the chiral primary operator $\mbox{Tr}((X^{34})^2)$
on $R^4$.
The energy of this state is 2.
We also focus on the states that correspond to the descendant operators 
generated by the superconformal transformation caused by $\eta_{m+}$.
Their forms are determined by (\ref{superconformaltransformationmode}) as
\beqa
&&\frac{\sqrt{2}}{N}\mbox{Tr}
(d_{0M}^{3\dagger}\alpha_{00}^{34\dagger})|0\rangle, \;\;\;\;\;
\frac{\sqrt{2}}{N}\mbox{Tr}
(d_{0M}^{4\dagger}\alpha_{00}^{34\dagger})|0\rangle, 
\label{E=5/2} \\
&&\frac{1}{\sqrt{3}N}\mbox{Tr}
(d^{3\dagger}_{0(\pm\frac{1}{2}0)}d^{4\dagger}_{0(\pm\frac{1}{2}0)}
+2a_{0(\pm 10)+}^{\dagger}\alpha_{00}^{34\dagger})|0\rangle, 
\label{E=3J=1m=pm1} \\
&&\frac{1}{\sqrt{6}N}\mbox{Tr}
(d^{3\dagger}_{0(\frac{1}{2}0)}d^{4\dagger}_{0(-\frac{1}{2}0)}
+d^{3\dagger}_{0(-\frac{1}{2}0)}d^{4\dagger}_{0(\frac{1}{2}0)}
-2\sqrt{2}a_{0(00)+}^{\dagger}\alpha_{00}^{34\dagger})|0\rangle, 
\label{E=3J=1m=0} \\
&&\frac{1}{N}\mbox{Tr}
(d^{3\dagger}_{0(\pm\frac{1}{2}0)}d^{3\dagger}_{0(\mp\frac{1}{2}0)})|0\rangle
,\;\;\;\;\;
\frac{1}{N}\mbox{Tr}
(d^{4\dagger}_{0(\pm\frac{1}{2}0)}d^{4\dagger}_{0(\mp\frac{1}{2}0)})|0\rangle,
\label{E=3J=0(1)}\\
&&\frac{1}{\sqrt{2}N}\mbox{Tr}
(d^{3\dagger}_{0(\frac{1}{2}0)}d^{4\dagger}_{0(-\frac{1}{2}0)}
-d^{3\dagger}_{0(-\frac{1}{2}0)}d^{4\dagger}_{0(\frac{1}{2}0)})|0\rangle.
\label{E=3J=0(2)}
\eeqa
The energy of (\ref{E=5/2}) is $\frac{5}{2}$. The energy of
(\ref{E=3J=1m=pm1}), (\ref{E=3J=1m=0}), (\ref{E=3J=0(1)}) and 
(\ref{E=3J=0(2)}) is 3.
All the above states are half-BPS, and their energy must not receive any
correction when the interactions are turned on.
The BPS state (\ref{E=3J=1m=pm1}) may mix with the non-BPS state whose 
energy is 3,
\beqa
\sqrt{\frac{3}{2}}\frac{1}{N}\mbox{Tr}
(d^{3\dagger}_{0(\pm\frac{1}{2}0)}d^{4\dagger}_{0(\pm\frac{1}{2}0)}
-a_{0(\pm 10)+}^{\dagger}\alpha_{00}^{34\dagger})|0\rangle, 
\label{E=3J=1m=pm1nonBPS}
\eeqa
while the BPS state (\ref{E=3J=1m=0}) may mix with the non-BPS state
whose energy is 3,
\beqa
\frac{1}{\sqrt{3}N}\mbox{Tr}
(d^{3\dagger}_{0(\frac{1}{2}0)}d^{4\dagger}_{0(-\frac{1}{2}0)}
+d^{3\dagger}_{0(-\frac{1}{2}0)}d^{4\dagger}_{0(\frac{1}{2}0)}
+\sqrt{2}a_{0(00)+}^{\dagger}\alpha_{00}^{34\dagger})|0\rangle.
\label{E=3J=1m=0nonBPS}
\eeqa
On the other hand, the BPS states (\ref{chiralprimary}), 
(\ref{E=5/2}), (\ref{E=3J=0(1)}) and (\ref{E=3J=0(2)}) 
cannot mix with the other states.

We need to develop the hamiltonian formalism for the interacting theory
to calculate the corrections to the energy.
The canonical conjugate momenta obtained from (\ref{freepart}),
(\ref{interactionpart1}) and (\ref{interactionpart2}) have the corrections
proportional to $g$, compared with those in the free energy, as follows.
\beqa
&&P_{JM\rho}=\frac{\delta I}{\delta \dot{A}_{JM\rho}}
=(-1)^{m-\tilde{m}+1}\dot{A}_{J-M\rho}-ig{\cal D}_{J_1M_1\;JM\rho\;J_2M_2\rho_2}[B_{J_1M_1},A_{J_2M_2\rho_2}], \n
&&P_{AB}^{JM}=\frac{\delta I}{\delta \dot{X}^{AB}_{JM}}
=(-1)^{m-\tilde{m}}\dot{X}_{AB}^{J-M}-ig{\cal C}_{J_1M_1\;JM\;J_2M_2}
[B_{J_1M_1},X^{AB}_{J_2M_2}], \n
&&P_{JM\kappa A}=\delta I/\delta \dot{\psi}^A_{JM\kappa}
=i\psi^{\dagger}_{JM\kappa A}.
\label{canonicalconjugatemomenta}
\eeqa
We solve the equations of motion for the auxiliary fields $B_{JM}$ and $c_{JM}$
iteratively with respect to $g$ and obtain
\beqa
&&\hat{B}_{JM}=\frac{g}{4J(J+1)}\mbox{Tr}\left[ 
(i(-1)^{m_2-\tilde{m}_2+1}{\cal D}^{JM}_{J_1M_1\rho_1\;J_2-M_2\rho_2}
[A_{J_1M_1\rho_1},P_{J_2M_2\rho_2}]  \right.\n
&&\qquad\quad\;\left. +i(-1)^{m_2-\tilde{m}_2}{\cal C}^{JM}_{J_1M_1\;J_2-M_2}
[X^{AB}_{J_1M_1},P_{AB}^{J_2M_2}]
+(-1)^{m-\tilde{m}}{\cal F}^{J_2M_2\kappa_2}_{J_1M_1\kappa_1\;J-M}
\{\psi^A_{J_1M_1\kappa_1},\psi^{\dagger}_{J_2M_2\kappa_2A}\}) \right] \n
&&\qquad\quad\;+{\cal O}(g^2), \n
&&\hat{c}_{JM}=0.
\label{Bandc}
\eeqa
By substituting (\ref{canonicalconjugatemomenta}) and (\ref{Bandc}) into 
the hamiltonian, 
\beqa
H=\sum_{JM\rho}P_{JM\rho}\dot{A}_{JM\rho}
+\sum_{JM}P_{AB}^{JM}\dot{X}^{AB}_{JM}
+\sum_{JM\kappa}P_{JM\kappa A}\dot{\psi}^A_{JM\kappa}-L,
\eeqa
we obtain
\beqa
&&H=H_0+H_{int}, \n
&&H_{int}=-L_{int}^{(1)}
+\sum_{J\neq 0,M}\frac{(-1)^{m-\tilde{m}}}{2}4J(J+1)
\hat{B}_{J-M}\hat{B}_{JM}+{\cal O}(g^3),
\label{hamiltonian}
\eeqa
where $H_0$ takes the same form as that in the free theory, and
$L_{int}^{(1)}$ is given in (\ref{interactionpart1}).

In order to obtain the order $g^2N$ corrections to the energy, we calculate
for the degenerate states, $|S_n\rangle$, 
the matrix elements
\beqa
\Delta E^{g^2N}_{mn}
=\langle S_m|H_{int,4}
+H_{int,3}\frac{1-\sum_n|S_n\rangle\langle S_n|}{E_0-H_0}H_{int,3}
+H_{2}^{1-loop}|S_n\rangle
\equiv \langle S_m|H^{g^2N}_{eff}|S_n\rangle,
\label{energycorrection}
\eeqa
where $E_0$ is the unperturbed energy, and $H_{int,3}$ and $H_{int,4}$
is the 3-point and 4-point interaction terms in $H_{int}$, respectively,
while $H_2^{1-loop}$ comes from the 1-loop counter terms quadratic in the
fields and is proportional to $g^2N$.

We first calculate $H_{eff}^{g^2N}$ for the states (\ref{SO(6)spinchainstate}).
It is easy to see that the matrix elements among
the states (\ref{SO(6)spinchainstate}) with fixed $l$
are closed in the $g^2N$ corrections. 
As an example, let us see the contribution
of the 4-point interaction in (\ref{hamiltonian}), 
\beqa
H_{int}^X&=&-\frac{g^2}{4}\int d\Omega 
\mbox{Tr}([X_{AB},X_{CD}][X^{AB},X^{CD}])
\n
&=&-\frac{g^2}{2}
{\cal C}^{j_3}_{j_1j_2}{\cal C}_{j_3j_4j_5}
(\delta^{AB}_{EF}\delta^{CD}_{GH}-\delta^{AB}_{GH}\delta^{CD}_{EF})
\mbox{Tr}(X_{AB}^{j_1}X_{CD}^{j_2}X^{EF}_{j_4}X^{GH}_{j_5}),
\label{HintX}
\eeqa
where we have introduced the abbreviated notations. 
$j$ represents a pair of $(J,M)$. 
$-j$ represents $(J,-M)$, and $j=0$ represents
to $(J=0,M=0)$ in the following.
We substitute 
\beqa
X^{AB}_j=\frac{1}{\sqrt{2\omega_J^X}}
(\alpha^{AB}_j+(-1)^{m-\tilde{m}}\alpha^{AB\dagger}_{-j})
\eeqa
into (\ref{HintX}). We take the Wick contractions to obtain the normal ordered
form. After the contractions, we are forced to set $j=0$ for
the creation and annihilation operators that are left in the normal ordering,
because we consider the matrix elements among (\ref{SO(6)spinchainstate}).
The result is
\beqa
&&H_{int}^X \n
&&=-\frac{g^2}{8}
:\mbox{Tr}(
2[\alpha_0^{AB\dagger},\alpha_0^{CD\dagger}][\alpha^0_{AB},\alpha^0_{CD}]
-[\alpha_0^{AB\dagger},\alpha^0_{AB}][\alpha^{CD\dagger}_0,\alpha^0_{CD}]
+[\alpha_0^{AB\dagger},\alpha^0_{CD}][\alpha^{0\dagger}_{AB},\alpha^{CD}_0]):
\n
&&\;\;\;+\frac{5g^2N}{2}\sum_{j_2j_3}
\frac{(-1)^{m_2-\tilde{m}_2}}{\omega_{J_2}^X}
{\cal C}^{j_3}_{0j_2}{\cal C}_{j_3-j_20}
:\mbox{Tr}(\alpha^{AB\dagger}_0\alpha^0_{AB}): \n
&&\;\;\;+\frac{15g^2N^3}{4}\sum_{j_1j_2j_3}
\frac{(-1)^{m_1-\tilde{m}_1+m_2-\tilde{m}_2}}{\omega_{J_1}^X\omega_{J_2}^X}
{\cal C}^{j_3}_{j_1j_2}{\cal C}_{j_3-j_2-j_1},
\label{X 4-point}
\eeqa
where $\alpha^0_{AB}\equiv \alpha^{(JM)=(00)}_{AB}$, and 
we have used ${\cal C}^{j_3}_{00}=1$ in the first term in the righthand
side. 
We further evaluate the second term using 
$\sum_{M_2M_3}(-1)^{m_2-\tilde{m}_2}{\cal C}^{j_3}_{0j_2}{\cal C}_{j_3-j_20}
=(2J_2+1)^2\delta_{J_2J_3}$ and obtain
\beqa
\frac{5g^2N}{2}\sum_{J_2}(2J_2+1)
:\mbox{Tr}(\alpha^{AB\dagger}_0\alpha^0_{AB}):,
\label{numberoperator}
\eeqa
We see from (\ref{on-shellself-energy}) that the coefficient of the number
operator in  (\ref{numberoperator}) is nothing but
$-\frac{g^2N}{2\omega_{J=0}^X}=-\frac{g^2N}{2}$ times 
the contribution of $(X-e)$ to
$\Pi_{J=0}^X(1)$. Indeed, the contribution of
the other 4-point interactions and the 3-point interactions to this coefficient
correspond to the contribution of the other diagrams
in (\ref{on-shellself-energy}). Note that the contribution of $(X-a)+(X-b)$
comes from the second term of $H_{int}$ in (\ref{hamiltonian}).
Moreover, the contribution of $H_2^{1-loop}$ to
this coefficient is $\frac{\gamma_X}{2}$.
The third term in (\ref{X 4-point})
is a constant that contributes equally to any 
$\langle S_m| H_{eff}^{g^2N} |S_n\rangle$. The sum of such constants that
all the interactions yield must be zero due to the supersymmetry. We ignore
these constants hereafter. 
As in \cite{Kim:2003rz}, we rewrite 
$\mbox{Tr}([\alpha_0^{AB\dagger},\alpha^0_{AB}]
[\alpha^{CD\dagger}_0,\alpha^0_{CD}])$ in the first term as
\beqa
:\mbox{Tr}([\alpha_0^{AB\dagger},\alpha^0_{AB}]T^a):
:\mbox{Tr}(T^a[\alpha^{CD\dagger}_0,\alpha^0_{CD}]):
-2N:\mbox{Tr}(\alpha^{AB\dagger}_0\alpha^0_{AB}):,
\eeqa
where $T^a$ is the generators of $U(N)$. As shown in \cite{Kim:2003rz}, 
the first term annihilates the states (\ref{SO(6)spinchainstate}).
We eventually obtain for the states (\ref{SO(6)spinchain})
\beqa
H_{eff}^{g^2N}
&=&
\left( -\frac{g^2N}{2}\Pi_{J=0}^X(1)+\frac{1}{2}\gamma_X-\frac{g^2N}{4}\right)
:\mbox{Tr}(\alpha^{AB\dagger}_0\alpha^0_{AB}): \n
&&-\frac{g^2}{8}
:\mbox{Tr}(
2[\alpha_0^{AB\dagger},\alpha_0^{CD\dagger}][\alpha^0_{AB},\alpha^0_{CD}]
+[\alpha_0^{AB\dagger},\alpha^0_{CD}][\alpha^{0\dagger}_{AB},\alpha^{CD}_0]):.
\label{HeffforSO(6)spinchain}
\eeqa
The expectation value of $H_{eff}^{g^2N}$ 
with respect to the state (\ref{chiralprimary}) must vanish, because
it is BPS and does not mix with other states.
The second term in (\ref{HeffforSO(6)spinchain}) annihilates
the state (\ref{chiralprimary}). Thus the coefficient of the number operator
in the first term must vanish. Namely, $\gamma_X$ is determined as
\beqa
\gamma_X=g^2N\left(\Pi_{J=0}^X(1)+\frac{1}{2}\right),
\label{gammaX}
\eeqa
which in general depends on the cut-off and includes the finite
renormalization.

The dilatation operator for the operators (\ref{SO(6)spinchain})
on $R^4$ \cite{Minahan:2002ve,Beisert:2003tq}
is 
\beqa
D_2=-\frac{g_{YM}^2}{32\pi^2}
:\mbox{Tr}\left(2[X^{AB},X^{CD}][\frac{d}{dX^{AB}},\frac{d}{dX^{CD}}]
+[X^{AB},\frac{d}{dX^{CD}}][X_{AB},\frac{d}{dX_{CD}}]\right):.
\label{dilatationoperator}
\eeqa
Recalling $g^2=\frac{g_{YM}^2}{4\pi^2}$ and comparing the remaining
second term in (\ref{HeffforSO(6)spinchain}) and (\ref{dilatationoperator}),
we find that the matrix elements of the order $g^2N$ corrections to
the energy of the states (\ref{SO(6)spinchainstate}) completely agree with
those of the 1-loop dilatation operator for the operators 
(\ref{SO(6)spinchain}), as expected.

Let us determine other counter terms.
For the states (\ref{E=5/2}),
\beqa
H_{eff}^{g^2N}
&=&\frac{g^2N}{4}
:\mbox{Tr}(\alpha^{AB\dagger}_0\alpha^0_{AB}):
+\left(g^2N\Pi_{J=0}^{\psi}(-\frac{3}{2})+\frac{3}{2}\beta_{\psi}\right)
:\mbox{Tr}(d^{A\dagger}_{m}d_{mA}): \n
&&+2g^2:\mbox{Tr}(d^{C\dagger}_m\alpha^{AB\dagger}_0d_{mA}\alpha_{BC}^0):,
\label{HeffforE=5/2}
\eeqa
where $d_{mA}\equiv d_{0(m,0)A}$ and $m$ takes $\pm\frac{1}{2}$.
The states (\ref{E=5/2}) do not mix with the other states, either.
The expectation value of $H_{eff}^{g^2N}$ with respect to the states must
vanish. It is evaluated as
\beqa
\frac{g^2N}{4}
+\left(g^2N\Pi_{J=0}^{\psi}(-\frac{3}{2})+\frac{3}{2}\beta_{\psi}\right)
-g^2N=0,
\eeqa
from which we obtain 
\beqa
\beta_{\psi}
=-\frac{2g^2N}{3}\Pi_{J=0}^{\psi}(-\frac{3}{2})+\frac{g^2N}{2}.
\label{betapsi}
\eeqa
For the states (\ref{E=3J=1m=0})$\sim$(\ref{E=3J=1m=0nonBPS}),
\beqa
H_{eff}^{g^2N}
&=&\frac{g^2N}{4}
:\mbox{Tr}(\alpha^{AB\dagger}_0\alpha^0_{AB}):
+\frac{3g^2N}{4}:\mbox{Tr}(d^{A\dagger}_{m}d_{mA}): \n
&&+\left(-\frac{g^2N}{2\omega_{J=0}^A}\Pi_{J=0}^A(2) 
+\frac{\gamma_A}{\omega_{J=0}^A}\right)
:\mbox{Tr}(a^{\dagger}_ma_m): \n
&&-\frac{g^2}{4}:\mbox{Tr}(a_m^{\dagger}\alpha^{AB\dagger}_0a_m\alpha_{AB}^0):
+\sqrt{6}g^2(-1)^{m_1+\frac{1}{2}}
C^{\frac{1}{2}-m_1}_{\frac{1}{2}m_2\;1m_3}
:\mbox{Tr}(d^{B\dagger}_{m_2}d^{A\dagger}_{m_1}\alpha_{AB}^0a_{m_3}):
+(c.c.) \n
&&+\frac{g^2}{4}:\mbox{Tr}(
d^{A\dagger}_{-m}d^{B\dagger}_md_{-mA}d_{mB}
-d^{A\dagger}_md^{B\dagger}_{-m}d_{-mA}d_{mB}
+d^{A\dagger}_md^{B\dagger}_md_{mB}d_{mA}
+d^{A\dagger}_md^{B\dagger}_{-m}d_{mB}d_{-mA}):, \n
\label{HeffforE=3}
\eeqa
where $a_m=a_{0(1m)+}$ and $m$ takes $0,\:\pm1$.
The matrix elements of $H_{eff}^{g^2N}$ among
(\ref{E=3J=1m=pm1}) and (\ref{E=3J=1m=pm1nonBPS}) form the $2\times 2$ matrix
\beqa
\left(\begin{array}{cc}
\frac{2}{3}(\chi-g^2N) & \frac{\sqrt{2}}{3}(\chi-g^2N) \\
\frac{\sqrt{2}}{3}(\chi-g^2N) & \frac{1}{3}(\chi+8g^2N)
\end{array}\right),
\eeqa
where 
\beqa
\chi=-\frac{g^2N}{4}\Pi_{J=0}^A(2)+\frac{\gamma_A}{2}.
\eeqa
Those among (\ref{E=3J=1m=0}) and (\ref{E=3J=1m=0nonBPS}) also form
the same $2\times 2$ matrix. In order for the BPS energy not to receive
any correction, one of the eigenvalues of this matrix must vanish.
This is true if and only if $\chi=g^2N$,
namely, we obtain 
\beqa
\gamma_A=g^2N\left(\frac{1}{2}\Pi_{J=0}^A(2)+2\right).
\label{gammaA}
\eeqa
In this case, the other eigenvalue is $3g^2N$, and 
(\ref{E=3J=1m=pm1}) and (\ref{E=3J=1m=0}) are the eigenvector for the
zero eigenvalue, while (\ref{E=3J=1m=pm1nonBPS}) and (\ref{E=3J=1m=0nonBPS})
are the eigenvector for the other eigenvalue.
There is no correction to the BPS energy, and there is no mixing between
the BPS and non-BPS states. 
It is also easy to see
that the matrix elements among the BPS states (\ref{E=3J=0(1)}) and
(\ref{E=3J=0(2)}), which
have no mixing with the other states, vanish.


\subsection{1-loop analysis of the truncated theories}
So far we have been examining the 1-loop corrections in the original theory.
It is easy to generalize the analysis in sections 6.1$\sim$6.3 to
the 1-loop perturbation theory 
around the trivial vacua of the truncated theories. 
Consider the expression for a certain diagram in the original theory.
By keeping only the KK modes to be remained in each truncated theory,
in the external and internal propagators, one obtains the expression
for the corresponding diagram in the truncated theory.
The plane wave matrix model is at least perturbatively a finite theory,
where no regularization is needed in the perturbative expansion,
while ${\cal N}=4$ SYM on $R\times S^2$ and ${\cal N}=4$ SYM 
on $R\times S^3/Z_k$ give rise to divergences and must be regularized.
In the perturbative expansion of the latters, as a regularization scheme,
introducing the cut-offs
for the loop angular momenta should be useful as in the original theory,
although we have not explicitly calculated the divergent parts of
the diagrams in those theories
which are regularized in such a way.
At any rate, we can proceed the following arguments
assuming ${\cal N}=4$ SYM on $R\times S^2$ and ${\cal N}=4$ SYM 
on $R\times S^3/Z_k$ are appropriately regularized in terms of a certain 
regularization scheme.

One can also develop the hamiltonian formalism for the truncated theories.
In particular, considering the states in 
(\ref{SO(6)spinchainstate}) and 
(\ref{chiralprimary})$\sim$(\ref{E=3J=1m=0nonBPS}) makes sence, because
$X^{AB}_{00}$, $\psi_{0M+}$ and $A_{0M+}$ are remained in all the truncated
theories although the correspondence with the operators on $R^4$ no longer
exist. Furthermore, the truncated theories possess 16 supercharges, and
the states (\ref{chiralprimary})$\sim$(\ref{E=3J=0(2)}) are also half-BPS,
namely preserve 8 supercharges. 
Their mass spectrum must not receive any
quantum correction. The mixing of these states with other states
is the same as the original theory. 
The analysis of the $g^2N$ correction to the energy of the states
(\ref{SO(6)spinchainstate}) and 
(\ref{chiralprimary})$\sim$(\ref{E=3J=1m=0nonBPS}) runs parallel to
the one in the original theory, which is given below (\ref{energycorrection}).
It is easy to see that (\ref{HeffforSO(6)spinchain}), (\ref{HeffforE=5/2})
and (\ref{HeffforE=3}) hold for the truncated theories, and $\gamma_X$,
$\beta_{\psi}$ and $\gamma_A$ are determined as (\ref{gammaX}), 
(\ref{betapsi}) and (\ref{gammaA}), respectively, in such a way that
the supersymmetry is realized.
Of course, the values of $\Pi_{J=0}^X(1)$, $\Pi_{J=0}^{\psi}(-\frac{3}{2})$
and $\Pi_{J=0}^A(2)$ depend on which theory is considered.
In particular, in the plane wave matrix model, $\gamma_X$,
$\beta_{\psi}$ and $\gamma_A$ are all zero, namely
\beqa
\Pi_{J=0}^X(1)=-\frac{1}{2}, \;\;\;\;\;
\Pi_{J=0}^{\psi}(-\frac{3}{2})=\frac{3}{4}, \;\;\;\;\;
\Pi_{J=0}^A(2)=-4
\label{on-shell self-energy in plane wave matrix model}
\eeqa
must hold. Indeed, from (\ref{on-shellself-energy}), 
we can calculate the contribution of each diagram to
$\Pi_{J=0}^X$ as 
\beqa
(X-c)=-\frac{3}{2},\;\;\;\;\; (X-e)=-5,\;\;\;\;\; (X-f)=6,
\eeqa
The total of these values amounts to $-\frac{1}{2}$.
Note that the diagrams $(X-a)$, $(X-b)$, $(X-d)$ and $(X-g)$ do not exist 
in this theory.
Similarly, we obtained 
$\Pi_{J=0}^{\psi}(-\frac{3}{2})$ and $\Pi_{J=0}^A(2)$ in
(\ref{on-shell self-energy in plane wave matrix model}) 
by calculating the diagrams in the plane wave matrix model.

The above arguments lead us to a following interesting conclusion.
In the truncated theories, the matrix elements
of the $g^2N$ corrections to the energy of the states
(\ref{SO(6)spinchain}) are mapped to the hamiltonian of the same integrable
$SO(6)$ spin chain that appear in the original theory.
Indeed, the authors of \cite{Kim:2002if} verified 
this fact in the plane wave matrix
model by direct calculation.
In \cite{Kim:2002if}, the matrix elements of (\ref{HeffforE=5/2}) 
in the plane wave
matrix model are also obtained by direct calculation, and are consistent with
the above arguments.

As a side remark, we checked that as in the original theory
by making shifts of the cut-offs in (\ref{shift}) one can obtain
the finite zero point energy in the truncated theories with $g=0$.
Its value is zero for ${\cal N}=4$ SYM on $R\times S^2$ and 
$\frac{3}{16k}N^2$ for ${\cal N}=4$ SYM on $R\times S^3/Z_k$.
These two values are consistent, since in the $k\rightarrow \infty$
limit ${\cal N}=4$ SYM on $R\times S^3/Z_k$ is reduced to
${\cal N}=4$ SYM on $R\times S^2$ \cite{Lin:2005nh}.

\section{Time-dependent BPS solution}
\setcounter{equation}{0}
In this section, we examine a classical time-dependent BPS solution and
the 1-loop effective action around it in the original and truncated
theories. In section 7.1, we construct the time-dependent BPS solution of
the original and truncated theories.
In section 7.2, we calculate the 1-loop effective action around it
in the original theory, and in section 7.3 that in the truncated theories.

\subsection{Classical time-dependent BPS solution}
\begin{figure}[tbp]
\begin{center}
\begin{minipage}[b]{0.49\textwidth}
   \begin{center}
  \includegraphics[width=6cm]{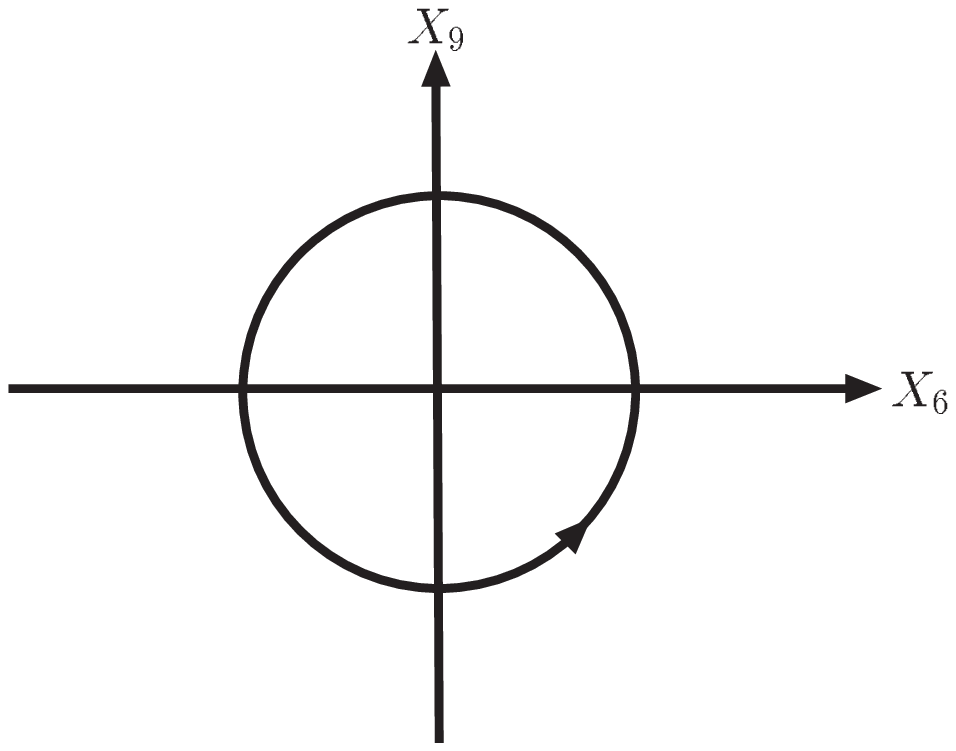}\\
   \caption{BPS solution}
   \label{fig:bps}
   \end{center}
\end{minipage}
\begin{minipage}[b]{0.49\textwidth}
   \begin{center}
   \includegraphics[width=6cm]{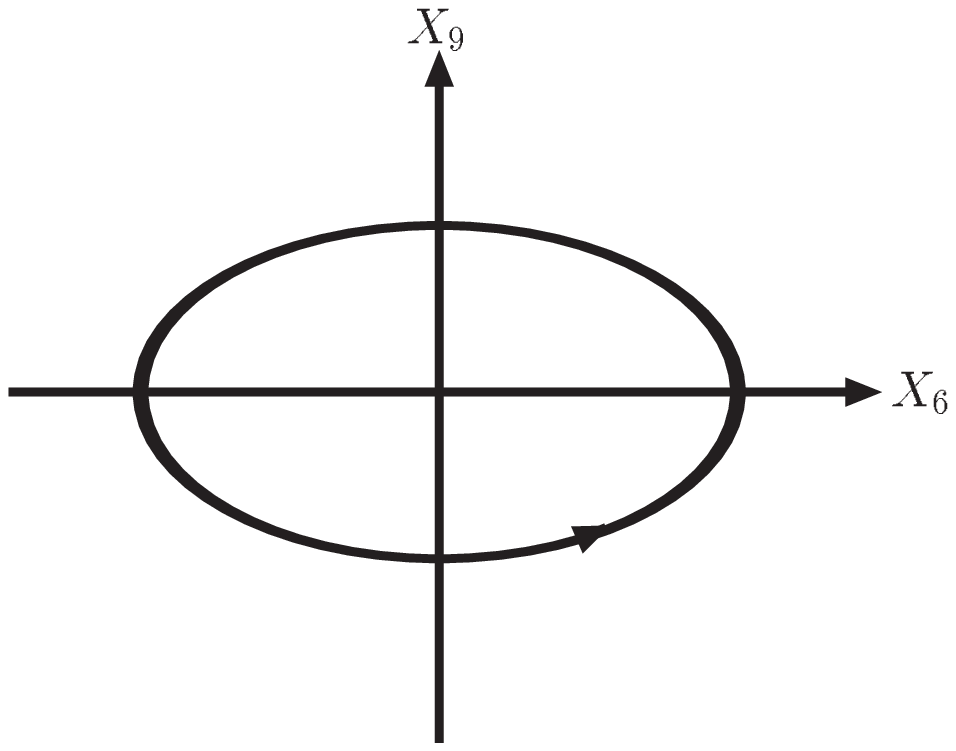}\\
   \caption{Non-BPS solution}
   \label{fig:nonbps}
   \end{center}
\end{minipage}   
\end{center}
\end{figure}

We consider a configuration in which all the KK modes and matrix
components except the $(1,1)$ component of 
$X_{34}^{00}$ vanish. Namely,
\beqa
X_{34}^{00}=X^{12}_{00}=(X^{34})^{\dagger} 
=(X_{12}^{00})^{\dagger}=
\left(\begin{array}{cccc}
\frac{1}{2}\rho(t)\:e^{i\eta(t)} & 0 & \cdots & 0 \\ 
0 & 0 & \cdots & 0 \\
\vdots & \vdots & & \vdots \\
0 & 0 & \cdots & 0 
\end{array}\right).
\label{bpssol}
\eeqa
It is easy to see that this assumption is a consistent truncation in
the original and truncated theories.
Under this assumption, the classical action becomes
\beqa
S_{c}=\int dt \: \frac{1}{2}(\dot{\rho}^2+\rho^2\dot{\eta}^2-\rho^2).
\eeqa
The canonical momenta are read off as
\beqa
&&p_{\rho}=\frac{\delta S_{c}}{\delta \dot{\rho}}=\dot{\rho}, \n
&&l=\frac{\delta S_{c}}{\delta \dot{\eta}}=\rho^2\dot{\eta}.
\eeqa
The angular momentum in the $(6,9)$ plane, $l$, is conserved and corresponds
to the R charge (Recall $X_{34}=(X_6+iX_9)/2$).
The energy possesses the BPS bound:
\beqa
E=\frac{1}{2}{p_{\rho}}^2+\frac{l^2}{2\rho^2}+\frac{1}{2}\rho^2
\geq |l|
\label{lowerboundforE}
\eeqa
When $p_{\rho}=0$ and $l^2=\rho^4$, the BPS bound is saturated. 
In this case, $\rho=\sqrt{|l|}=\mbox{const.}$ and $\eta=\pm t+\mbox{const.}$.
We can set $\rho=\sqrt{l}$ and $\eta=t$ without loss of generality.
That is, we consider the solution\footnote{This solution on $R\times S^3$
is formally mapped to
a vacuum with a nontrivial Higgs vev,
$(X_{34})_{11}=\frac{1}{2}\sqrt{l}$, on $R^4$. However,
in this situation the correspondence between the two theories breaks down,
so that it seem rather nontrivial to examine the quantum correction around 
this solution.}
\beqa
(X_{34}^{00})_{11}=\frac{1}{2}\sqrt{l}\:e^{it}.
\eeqa
For this solution, non-vanishing elements in 
(\ref{superconformaltransformationSU(4)}) are 
\beqa
&&\delta_{\epsilon}(\lambda_+^A)_{11}
=2(\partial_0(X^{AB})_{11}\mp i(X^{AB})_{11})\gamma^0\epsilon_{-B}, \n
&&\delta_{\epsilon}(\lambda_{-A})_{11}
=2(\partial_0(X_{AB})_{11}\pm i(X_{AB})_{11})\gamma^0\epsilon_+^B .
\eeqa
The requirement $\delta_{\epsilon}\lambda_+^A=0$ and
$\delta_{\epsilon}\lambda_{-A}=0$ leads to
$\epsilon_{-3}=\epsilon_+^3=\epsilon_{-4}=\epsilon_+^4=0$ for the upper sign
and $\epsilon_{-1}=\epsilon_+^1=\epsilon_{-2}=\epsilon_+^2=0$ 
for the lower sign.
The solution is, therefore, a half BPS solution.
It preserves 16 supercharges for the original theory and 8 supercharges
for the truncated theories. The BPS solution corresponds to a circular
motion in the $(6,9)$ plane (see Fig. \ref{fig:bps})
while generic non-BPS solutions correspond to elliptical motions 
(see Fig. \ref{fig:nonbps}).
The BPS solution is the classical counterpart of the lowest Landau level in
the Landau problem. 
The BPS solution is interpreted as the AdS giant graviton in the original
theory \cite{Hashimoto}, 
and corresponds to a particular one of the spherical membrane
solutions in the plane wave
matrix model, which were studied in \cite{Berenstein:2002jq}.

\subsection{1-loop effective action around the solution in the original SYM}
We calculate the 1-loop effective action around  the BPS solution
in the original ${\cal N}=4$ SYM, 
which was obtained in the previous subsection.
Following the background field method, we make a 
substitution
\beqa
&&(X_{34})_{kl} \rightarrow \frac{1}{2}\sqrt{l}e^{it}\delta_{k1}\delta_{l1}
                            +(X_{34})_{kl}, \n
&&(X_{12})_{kl} \rightarrow \frac{1}{2}\sqrt{l}e^{-it}\delta_{k1}\delta_{l1}
                            +(X_{12})_{kl}
\label{backgroundfieldmethod}
\eeqa  
in the gauge-fixed action $I$ and keep 
the second-order in all fields.\footnote{In this subsection, we 
rescale all the fields back by $g$.}
Then we immediately see that
$I_{int}$ are only written by the $(1,k)$ and $(k,1)$ components, 
where $k \neq 1$, and as far as the other components are concerned, 
$I$ takes the same form as the free theory. We can therefore 
forget the contribution
of the other components. Moreover, the fields with different $k$'s are 
decoupled and $I$ takes the same form for each $k$. We can calculate
the effective action for a fixed $k$ and multiply the result by $N-1$
to obtain the final answer. (In the 't Hooft limit, the factor $N-1$ can be
replaced with $N$.)
We omit the suffices for the matrix components
and absorb explicit time dependence into the fields:
\beqa
&&(X_{34})_{1k} \rightarrow \frac{1}{\sqrt{2}}e^{it}Z_1 ,\;\;\;
  (X_{12})_{1k} \rightarrow \frac{1}{\sqrt{2}}e^{-it}Z_2^{\ast}, \n
&&(X_{24})_{1k} \rightarrow \frac{1}{\sqrt{2}}Y_1 ,\;\;\;
  (X_{31})_{1k} \rightarrow \frac{1}{\sqrt{2}}Y_2^{\ast} ,\;\;\;
  (X_{14})_{1k} \rightarrow \frac{1}{\sqrt{2}}Y_3 ,\;\;\;
  (X_{23})_{1k} \rightarrow \frac{1}{\sqrt{2}}Y_4^{\ast}, \n
&&(A_0)_{1k} \rightarrow A_0, \;\;\; (A_i)_{1k} \rightarrow A_i, \n
&&(\psi^3)_{1k} \rightarrow e^{-\frac{i}{2}t}\varphi_1,\;\;\;
  (\psi_3^{\dagger T})_{1k}^{\ast} \rightarrow e^{-\frac{i}{2}t}\varphi_2, 
   \;\;\;
  (\psi^4)_{1k} \rightarrow e^{-\frac{i}{2}t}\varphi_3,\;\;\;
  (\psi_4^{\dagger T})_{1k}^{\ast} \rightarrow e^{-\frac{i}{2}t}\varphi_4, \n
&&(\psi^1)_{1k} \rightarrow e^{\frac{i}{2}t}\phi_5,\;\;\;
  (\psi_1^{\dagger T})_{1k}^{\ast} \rightarrow e^{\frac{i}{2}t}\varphi_6, 
   \;\;\;
  (\psi^2)_{1k} \rightarrow e^{\frac{i}{2}t}\varphi_7,\;\;\;
  (\psi_2^{\dagger T})_{1k}^{\ast} \rightarrow e^{\frac{i}{2}t}\varphi_8 .
\eeqa
The resultant quadratic action is
\beqa
&&I=\frac{1}{g^2}\int dtd\Omega \left[
\sum_{r=1,2}Z_r^{\ast}(-\partial_0^2-2i\partial_0+\nabla^2-\frac{l}{2})Z_r
+\frac{l}{2}(Z_1^{\ast}Z_2^{\ast}+Z_1Z_2) \right.\n
&&\qquad
+\sum_{r=1}^4 Y_r^{\ast}(-\partial_0^2+\nabla^2-1-l)Y_r
+A_0^{\ast}(-\nabla^2+l)A_0 \n
&&\qquad 
+\sqrt{2l}(A_0(Z_1^{\ast}-Z_2)+A_0^{\ast}(Z_1-Z_2^{\ast}))
+i\sqrt{\frac{l}{2}}(A_0(\partial_0Z_1^{\ast}+\partial_0Z_2)
                      -A_0^{\ast}(\partial_0Z_1+\partial_0Z_2^{\ast})) \n
&&\qquad 
+A_i^{\ast}(-\partial_0^2+\nabla^2-2-l)A_i 
+\sum_{s=1}^8\varphi_s^{\dagger}(i\partial_0+i\sigma^i\nabla_i)
                      \varphi_s
+\frac{1}{2}\sum_{s=1}^4\varphi_s^{\dagger}\varphi_s
-\frac{1}{2}\sum_{s=5}^8\varphi_s^{\dagger}\varphi_s \n
&&\qquad
+\sqrt{l}(
\varphi_4^{\dagger}\sigma^2\varphi_1^{\dagger T}+\varphi_4^T\sigma^2\varphi_1
-\varphi_2^{\dagger}\sigma^2\varphi_3^{\dagger T}-\varphi_2^T\sigma^2\varphi_3
\n
&&\qquad\quad\;\;\;\left.
+\varphi_8^{\dagger}\sigma^2\varphi_5^{\dagger T}+\varphi_8^T\sigma^2\varphi_5
-\varphi_6^{\dagger}\sigma^2\varphi_7^{\dagger T}-\varphi_6^T\sigma^2\varphi_7)
\right]. 
\label{quadraticaction}
\eeqa
Note that the ghosts do not contribute to this calculation of
the  1-loop effective action because of the Coulomb gauge.
We must also take into account the contribution 
of the 1-loop counter terms consisting only of $X^{AB}$.
We substitute the background in (\ref{backgroundfieldmethod}) into them. 
As far as the counter terms quadratic in $X^{AB}$ (\ref{Xcounterterm})
are concerned,
there is the contribution only from 
$-\frac{\gamma_X}{2}\mbox{Tr}(X_{AB}X^{AB})$, which results in
$-\int dt\frac{\gamma_X}{2}l$, where $\gamma_X$ is given in (\ref{gammaX}).
We will see below that this contribution is consistently 
needed for vanishing of
the 1-loop effective action around the time-dependent BPS solution.
Among possible counter terms quartic in $X^{AB}$, the single trace ones are 
\beqa
&&\mbox{Tr}([X_{AB},X_{CD}][X^{AB},X^{CD}]), \label{commutatorX^4} \\
&&\mbox{Tr}(X_{AB}X^{AB}X_{CD}X^{CD}),
\label{X^4}
\eeqa
and the double trace ones are
\beqa
\frac{1}{N}\mbox{Tr}(X_{AB}X^{AB})\mbox{Tr}(X_{CD}X^{CD}),\;\;\;\;\;
\frac{1}{N}\mbox{Tr}(X_{AB}X_{CD})\mbox{Tr}(X^{AB}X^{CD}).
\label{doubletrace}
\eeqa
(\ref{commutatorX^4}) vanishes when the background is plugged
in, while the double trace ones (\ref{doubletrace}) do not contribute 
in this case
due to $1/N$ suppression.
We can, therefore, determine the coefficient of (\ref{X^4})
from the requirement of vanishing of the 1-loop
effective action.

We make a mode expansion for all fields in (\ref{quadraticaction}).
We first integrate over $A_0$ and obtain new terms that are quadratic
in $Z_r$ and $Z_r^{\ast}$.
After the redefinition, $(-1)^{m-\tilde{m}}Z_2^{JM}\rightarrow Z_2^{JM}$,
the action concerning $Z_r$ and $Z_r^{\ast}$ becomes
\beqa
&&\frac{1}{g^2}\int dt \left[
{Z_1^{00}}^{\ast}
(-\partial_0^2-2i\partial_0-\frac{1}{2}l)
Z_1^{00}
+{Z_2^{00}}^{\ast}
(-\partial_0^2+2i\partial_0-\frac{1}{2}l)
Z_2^{00} \right. \n
&&
\left.
+\frac{1}{2}l({Z_1^{00}}^{\ast}{Z_2^{00}}^{\ast}+Z_1^{00}Z_2^{00}) 
\right] \n
&&+\sum_{J\neq 0,M}\int dt \left[
{Z_1^{JM}}^{\ast}(-(1-K_J)\partial_0^2-2i(1-2K_J))\partial_0
-({\omega_J^X}^2-1+\frac{1}{2}l+4K_J)Z_1^{JM}
\right.\n
&&
+{Z_2^{J\:-M}}(-(1-K_J)\partial_0^2+2i(1-2K_J))\partial_0
-({\omega_J^X}^2-1+\frac{1}{2}l+4K_J){Z_2^{J\:-M}}^{\ast} \n
&&
\left.
+{Z_1^{JM}}^{\ast}(K_J\partial_0^2+\frac{1}{2}l+4K_J)
{Z_2^{J\:-M}}^{\ast} 
+Z_2^{J\:-M}(K_J\partial_0^2+\frac{1}{2}l+4K_J)Z_1^{JM} \right], 
\eeqa
where 
\beqa
K_J=\frac{l}{2}\frac{1}{4J(J+1)+l}.
\eeqa
In order to evaluate the 1-loop effective action, we use a formula
\beqa
\mbox{Tr}\ln(\partial_0^2-2ip\partial_0+m^2)=i\int dt \sqrt{p^2+m^2}.
\eeqa
It is easy to see that
the contribution of $Z_1^{00}$ and $Z_2^{00}$ to the effective action
is
\beqa
\Gamma_{eff}^{Z0}=-g^2N\int dt \sqrt{4+l},
\label{GammaZ0}
\eeqa
and the contribution of $Z_1^{JM}$ and $Z_2^{JM}$ $((JM)\neq (00))$ is 
\beqa
\Gamma_{eff}^{Z}&=&-g^2N\int dt \sum_{(JM)\neq (00)}
(\sqrt{4J^2+l}+\sqrt{(2J+2)^2+l}\:) \n
&=&-g^2N\int dt \sum_{J\neq 0}(2J+1)^2 
(\sqrt{4J^2+l}+\sqrt{(2J+2)^2+l}\:).
\label{GammaZ}
\eeqa
We can evaluate the contribution of $Y_r$, $A_i$ and the fermions 
in a similar way.
The contribution of $Y_r$ is 
\beqa
\Gamma_{eff}^{Y}=-4g^2N\int dt \sum_J (2J+1)^2 \sqrt{(2J+1)^2+l}.
\label{GammaY}
\eeqa
The contribution of $A_i$ is
\beqa
\Gamma_{eff}^{A}=-2g^2N\int dt \sum_J(2J+1)(2J+3)\sqrt{(2J+2)^2+l}.
\label{GammaA}
\eeqa
The contribution of the fermions is
\beqa
\Gamma_{eff}^{F}=4g^2N\int dt \sum_J(2J+1)(2J+2)
(\sqrt{(2J+2)^2+l}+\sqrt{(2J+1)^2+l}\:).
\label{GammaF}
\eeqa
We also have the contribution of the 1-loop counter term,
$-\frac{\gamma_X}{2}\mbox{Tr}(X_{AB}X^{AB})$,
\beqa
\Gamma_{eff}^{c.t.(1)}=-g^2N\int dt\frac{l}{2}
\left(\Pi_{J=0}^X(1)+\frac{1}{2}\right).
\label{Gammact}
\eeqa
Besides, there can be a contribution of the 1-loop counter term (\ref{X^4}),
which is quadratic in $l$ and
denoted by $\Gamma_{eff}^{c.t.(2)}$.
We denote the sum of all the contribution by $\Gamma_{eff}$:
\beqa
\Gamma_{eff}=\Gamma_{eff}^{Z0}+\Gamma_{eff}^Z+\Gamma_{eff}^Y+\Gamma_{eff}^A
+\Gamma_{eff}^F+\Gamma_{eff}^{c.t.(1)}+\Gamma_{eff}^{c.t.(2)}.
\eeqa

Let us see that the sum of (\ref{GammaZ0})$\sim$(\ref{GammaF}) vanishes.
First, comparing 
the order $l^0$ contribution in (\ref{GammaZ0})$\sim$(\ref{GammaF})
with (\ref{Casimirenergy}), we find that it is nothing but the contribution
of the $(1,k)$ and $(k,1)$ components of the fields
to the zero point energy, and we can ignore it here.
Next, the order $l^1$ contribution
is evaluated as follows (we omit the common
factor $lg^2N \int dt$):
\beqa
\Gamma_{eff}^{Z0} &\rightarrow& -\frac{1}{4}, \n
\Gamma_{eff}^{Z}  
&\rightarrow& -\frac{1}{4}\sum_{J\neq 0}\frac{(2J+1)^3}{J(J+1)}, \n
\Gamma_{eff}^Y 
&\rightarrow& -2\sum_{J}(2J+1), \n
\Gamma_{eff}^{A}
&\rightarrow& -\sum_{J}\frac{(2J+1)(2J+3)}{2J+2}, \n
\Gamma_{eff}^F 
&\rightarrow& 2\sum_{J}(4J+3).
\label{firstorderinl}
\eeqa
Comparing (\ref{firstorderinl}) with (\ref{on-shellself-energy}),
we find that the total of (\ref{firstorderinl})
is equal to
\beqa
\frac{1}{2}\Pi_{J=0}^X(1)+\frac{1}{4}.
\eeqa
This is canceled by (\ref{Gammact}). Namely, we find
\beqa
\mbox{the order $l^1$ contribution in }  \Gamma_{eff}=0.
\eeqa
Note that the righthand sides in (\ref{firstorderinl})
except the first line  
have correspondence with  those in (\ref{on-shellself-energy}).
If this correspondence also held for the first line in (\ref{firstorderinl}),
the order $l^1$ contribution in $\Gamma_{eff}^{Z0}$ would be $-\frac{1}{2}$ 
rather than $-\frac{1}{4}$ and the total of the righthand sides in
(\ref{firstorderinl}) would agree with 
$\frac{1}{2}\Pi_{J=0}^X(1)$. This agreement is naively anticipated
because the background field
method usually gives the generating function of the 1PI diagrams. 
However, this is not true in this case. Our result shows that in this case
the loop expansion and the expansion in $l$ do not commute.

Finally the order $l^2$ contribution in (\ref{GammaZ})$\sim$(\ref{GammaF}) is
logarithmically divergent, while the contribution of orders higher
than second in $l$ are finite.
At the second and higher orders, 
therefore, one can shift $J$, over which the summation
is taken.
We set $2J=n$ and shift $n$ appropriately in (\ref{GammaZ})$\sim$(\ref{GammaF})
to obtain the following expressions, where we focus only on these
orders in $l$. For the second order, the upper bounds of the summations 
are $\Lambda_s$ or $\Lambda_v$ or $\Lambda_f$ depending on the angular
momentum of which field is summed. For higher orders, they are set at infinity.
\beqa
&&\Gamma_{eff}^{Z0}+\Gamma_{eff}^{Z}
=-g^2N\int dt \left(\sum_{n=1}(n+1)^2\sqrt{n^2+l} 
                +\sum_{n=0}(n+1)^2\sqrt{(n+2)^2+l}\right), \n
&&\Gamma_{eff}^{Y}=-4g^2N\int dt 
\sum_{n=0}(n+1)^2\sqrt{(n+1)^2+l}, \n
&&\Gamma_{eff}^{A}=-g^2N\int dt \left(
\sum_{n=0}(n+1)(n+3)\sqrt{(n+2)^2+l}
+\sum_{n=1}(n-1)(n+1)\sqrt{n^2+l}\right), \n
&&\Gamma_{eff}^{F}=g^2N\int dt \left(
\sum_{n=0}(2(n+1)(n+2)\sqrt{(n+2)^2+l}
+4(n+1)^2\sqrt{(n+1)^2+l}) \right. \n
&&\qquad\qquad\qquad\;\left. 
+2\sum_{n=1}n(n+1)\sqrt{n^2+l} \right).
\label{secondandhigher}
\eeqa
A naive sum of the righthand sides in (\ref{secondandhigher}) is zero.
This means that the sum of higher orders in $l$ of the righthand sides
vanishes,
\beqa
\mbox{the $l^q$ contribution in } \Gamma_{eff}=0 \;\;\;(q\ge 3),
\eeqa
and the second order also vanishes if $\Lambda_s$, $\Lambda_v$
and $\Lambda_f$ differ only by constants. Otherwise, we are left with
certain finite
contribution of the second order in $l$, which must be canceled by
the counter term (\ref{X^4}).
Thus we can determine the coefficient of (\ref{X^4}).
In particular, in the case in which $\Lambda_s$, $\Lambda_v$
and $\Lambda_f$ differ only by constants, the coefficient is determined as 
zero.
It should be emphasized
that the value of $\gamma_X$ which is determined in section 6.3
is consistent with vanishing of the 1-loop effective action
around the time-dependent BPS solution.
We conclude that if the counter term quartic in $X^{AB}$ is appropriately
fixed,
\beqa
\Gamma_{eff}=0.
\eeqa

\subsection{1-loop effective action in the truncated theories}
As in section 6.4, it is easy to obtain the 1-loop effective action
around the time-dependent BPS solution in the truncated theories by using
the result in the original theory. 
What should be done is to keep only the modes remaining 
in the truncations in 
(\ref{GammaZ0})$\sim$(\ref{GammaF}).
Here we can make use of the multiplicities that we described in section 5.

We write down explicitly the expressions for
$\Gamma_{eff}^{Z0}$, $\Gamma_{eff}^{Z}$, $\Gamma_{eff}^Y$,
$\Gamma_{eff}^A$, $\Gamma_{eff}^F$ and $\Gamma_{eff}^{c.t(1)}$
in appendix E, where $\Gamma_{eff}^{c.t(1)}$ is again the contribution from
the counter term, $-\frac{\gamma_X}{2}\mbox{Tr}(X_{AB}X^{AB})$.
Besides, there can be the contribution from the counter term (\ref{X^4})
also in the truncated theories.
Those for the plane wave matrix model are given in 
(\ref{1-loop effective action in plane wave matrix model}).
Of course, in this case, all the expressions are finite and 
there is no contribution from the counter terms. 
Indeed the sum of the expressions in 
(\ref{1-loop effective action in plane wave matrix model}) vanishes.
In particular, the total of the first order in $l$ is again
$g^2Nl(\frac{1}{2}\Pi_{J=0}^X(1)+\frac{1}{4})$, which vanishes by itself as
seen in (\ref{on-shell self-energy in plane wave matrix model}).
The expressions for ${\cal N}=4$ SYM on $R\times S^2$,
${\cal N}=4$ SYM on $R\times S^3/Z_k$ with $k$ even and 
${\cal N}=4$ SYM on $R\times S^3/Z_k$ with $k$ odd are given in
(\ref{1-loop effective action in N=4 SYM on R times S^2}),
(\ref{1-loop effective action in N=4 SYM on R times S^3/Z_k with k even})
and (\ref{1-loop effective action in N=4 SYM on R times S^3/Z_k with k odd}),
respectively. 
As for these three cases, one can ignore the zero-th order in $l$ 
on the same ground as
the case of the original theory.
The first order in $l$ in each case vanishes if the value of 
$\gamma_X$ that was determined in section 6.4 is applied.
The requirement of vanishing of the second order in $l$ fixes
the coefficient of (\ref{X^4}). It is easy to check that a naive sum
in each of (\ref{1-loop effective action in N=4 SYM on R times S^2}),
(\ref{1-loop effective action in N=4 SYM on R times S^3/Z_k with k even})
and (\ref{1-loop effective action in N=4 SYM on R times S^3/Z_k with k odd})
vanishes (These expressions are counterparts of (\ref{secondandhigher}).
This means that the contribution of orders higher than second in $l$ in
(\ref{1-loop effective action in N=4 SYM on R times S^2}),
(\ref{1-loop effective action in N=4 SYM on R times S^3/Z_k with k even})
and (\ref{1-loop effective action in N=4 SYM on R times S^3/Z_k with k odd})
and, in addition,
when $\Lambda_s$, $\Lambda_v$ and $\Lambda_f$ differ only by
constants, no contribution from the counter term (\ref{X^4}) is needed and 
the coefficient of (\ref{X^4}) is fixed to zero.
To summarize, the contribution of the first order and orders higher than
second in $l$ in 1-loop effective action vanishes, and 
the coefficient of (\ref{X^4}) should be fixed 
in such a way that the second order in $l$ vanishes.

\section{Summary and discussion}
\setcounter{equation}{0}
In this paper we studied the dynamics of the original $\mathcal N=4$ SYM on $R\times S^3$ and the truncated theories by making a harmonic expansion of the original theory on $S^3$.
We first developed the harmonic expansion on $S^3$.
We obtained the new compact formula for the integral of the product of three harmonics (\ref{integralofthreeharmonics}).
Then we carried out the harmonic expansion of $\mathcal N=4$ SYM on $R\times S^3$ including the interaction terms.
Second, we described the consistent truncations of the original SYM to the theories with 16 supercharges.
We realized the truncations by keeping a part of the KK modes of the original theory.
In particular, we verified that quotienting by the subgroup $U(1)$ of
$\tilde{SU}(2)$ 
indeed yields $\mathcal N=4$ SYM on $R\times S^2$, by comparing the modes of $\mathcal N=4$ SYM on $R\times S^2$ and those of the orignal theory with 
the modes with $\tilde m=0$ kept ((\ref{eq:3}), (\ref{AS2}) and (\ref{DFS2})).
In addition, we explicitly constructed some of the non-trivial vacua of the $\mathcal N=4$ SYM on $R\times S^2$ in terms of the KK modes (\ref{eq:2}), which are  a part of the solutions discussed in \cite{Maldacena:2002rb,Lin:2005nh}.
Third, we calculated the 1-loop diagrams in the orignal theory by introducing the cut-offs for loop angular momenta.
We saw that this cut-off scheme gave the correct coefficients of the logarithmic divergences, which are consistent with vanishing of the beta function and the Ward identity (\ref{Wardidentity}).
We determined the counter terms in the original and the truncated theories in the trivial vacuum, by using the non-renormalization theorem of energy of the BPS states.
This told us that the 1-loop effective hamiltonians of the $SO(6)$ sector for the orignal and the truncated theories are the hamiltonian of 
the same integrable $SO(6)$ spin chain.
Finally we examine the time-dependent BPS solution (\ref{bpssol}) in the original and truncated theories, which are considered to correspond to the AdS giant graviton in the original theory.
We found that the 1-loop effective action around this solution vanishes
if the counter term quartic in $X^{AB}$ is appropriately fixed.
This implied that the BPS configuration is stable 
against the quantum corrections at the 1-loop level, as is expected.

\par
There are some directions as extension of the present work.
First, it is interesting to consider the  the non-BPS configuration 
(Fig.~\ref{fig:nonbps}) for the original and the truncated theories.
In particular, in the case of the plane wave matrix model,
a series of such investigations is done 
\cite{Shin:2003np,Shin:2004az,Shin:2005tb}.
It is also interesting to investigate the dynamics of ${\cal N}=4$ SYM
on $R\times S^2$
in the non-trivial vacua (\ref{eq:2}).
It would be also interesting to explore possibilities of another solution 
for (\ref{F12})-(\ref{trs}).
In addition it would be nice to construct the vacua for $\mathcal N=4$ SYM on $R\times S^3/Z_k$ explicitly, to study the dynamics around those non-trivial vacua
and to find the electrostatic picture for the vacua of 
the truncated theories discussed in \cite{Lin:2005nh}.
Another interesting future direction is thermodynamics of the original and the truncated theories\cite{Witten:1998zw,Sundborg,Aharony,Kawahara:2006hs,Yamada,Wadia}.
We will work in these directions and report the result in the near future.
We expect our findings in this paper to give some insight to these subjects.

\section*{Acknowledgements}
We would like to thank H. Aoki, K. Hamada, M. Hatsuda, Y. Hosotani, S. Iso,
H. Kawai, N. Kim, 
T. Miwa, J. Nishimura, H. Suzuki,
T. Yoneya and K. Yoshida for discussions.
Y.T. would like to thank APCTP for hospitality while this work
was in progress. A.T. would like to thank Kyung Hee University for hospitality
during the initial stage of this work.
The work of Y.T. is supported in part by The 21st Century COE Program
``Towards a New Basic Science; Depth and Synthesis."
The work of A.T. is supported in part by Grant-in-Aid for Scientific
Research (No.16740144) from the Ministry of Education, Culture, Sports,
Science and Technology.

\appendix

\section*{Appendices}

\section{Useful formulae for representations of $SU(2)$}
\setcounter{equation}{0}
\renewcommand{\theequation}{A.\arabic{equation}}
In this appendix, we gather some useful formulae concerning
the representation of $SU(2)$, most of which are found in \cite{vmk}.
The relationship between the Clebsch-Gordan coefficient and the $3-j$ symbol is
\beqa
\left( \begin{array}{ccc}
        J_1 & J_2 & J_3 \\
        m_1 & m_2 & m_3  
       \end{array}  \right)
=(-1)^{J_3+m_3+2J_1}\frac{1}{\sqrt{2J_3+1}}\:C^{J_3m_3}_{J_1\:-m_1\;J_2\:-m_2}.
\label{C-Gand3-j}
\eeqa
The $3-j$ symbol possesses the following symmetries
\beqa
&&\left( \begin{array}{ccc}
        J_1 & J_2 & J_3 \\
        m_1 & m_2 & m_3  
       \end{array}  \right)
=\left( \begin{array}{ccc}
        J_2 & J_3 & J_1 \\
        m_2 & m_3 & m_1 
       \end{array}  \right)
=\left( \begin{array}{ccc}
        J_3 & J_1 & J_2 \\
        m_3 & m_1 & m_2  
       \end{array}  \right)      \n
&&=(-1)^{a+b+c} 
   \left( \begin{array}{ccc}
         J_1 & J_3 & J_2 \\
         m_1 & m_3 & m_2  
        \end{array}  \right)
=(-1)^{a+b+c} 
 \left( \begin{array}{ccc}
        J_2 & J_1 & J_3 \\
        m_2 & m_1 & m_3  
       \end{array}  \right)
=(-1)^{a+b+c} 
 \left( \begin{array}{ccc}
        J_3 & J_2 & J_1 \\
        m_3 & m_2 & m_1  
       \end{array}  \right) ,     \n
&&\left( \begin{array}{ccc}
        J_1 & J_2 & J_3 \\
        m_1 & m_2 & m_3  
       \end{array}  \right)
=(-1)^{a+b+c} \left( \begin{array}{ccc}
                     J_1  & J_2  & J_3 \\
                     -m_1 & -m_2 & -m_3  
                    \end{array}  \right) .
\eeqa
In section 6 and appendix D, 
we frequently use a summation formula for the $3-j$ symbol
\beqa
\sum_{m_1m_2}
\left( \begin{array}{ccc}
        J_1 & J_2 & J_3 \\
        m_1 & m_2 & m_3  
       \end{array}  \right)
\left( \begin{array}{ccc}
        J_1 & J_2 & {J_3}' \\
        m_1 & m_2 & {m_3}'  
       \end{array}  \right)
=\frac{1}{2J_3+1}\delta_{J_3{J_3}'}\delta_{m_3{m_3}'}.
\label{3-jsummation}
\eeqa
In section 3, we use a formula for the $9-j$ symbol
\beqa
&&\left\{   \begin{array}{ccc}
             a & b & c \\
             d & e & f \\
             g & h & j 
             \end{array}     \right\}  \n
&&=[(2c+1)(2f+1)(2g+1)(2h+1)]^{-\frac{1}{2}}(2j+1)^{-1}
   \sum_{\alpha\beta\gamma\delta\epsilon\varphi\eta\mu\nu}
   C^{c\gamma}_{a\alpha \; b\beta}C^{f\varphi}_{d\delta \; e\epsilon}
   C^{j\nu}_{c\gamma \; f\varphi}C^{g\eta}_{a\alpha \; d\delta}
   C^{h\mu}_{b\beta \; e\epsilon}C^{j\nu}_{g\eta \; h\mu}. \n
\label{9-j}
\eeqa

\section{Vertex coefficients}
\setcounter{equation}{0}
\renewcommand{\theequation}{B.\arabic{equation}}
In this appendix, we give expressions for
the vertex coefficients we defined in section 3.
These expressions are obtained by using the formula 
(\ref{integralofthreeharmonics}). In the following,
$Q\equiv J+\frac{(1+\rho)\rho}{2}$, 
$\tilde{Q}\equiv J-\frac{(1-\rho)\rho}{2}$, $U\equiv J+\frac{1+\kappa}{4}$
and $\tilde{U}\equiv J+\frac{1-\kappa}{4}$. Suffices on these variables must
be understood appropriately.
\beqa
&&{\cal C}^{J_1M_1}_{J_2M_2\;J_3M_3}
=\sqrt{\frac{(2J_2+1)(2J_3+1)}{2J_1+1}}
C^{J_1m_1}_{J_2m_2\;J_3m_3}
C^{J_1\tilde{m}_1}_{J_2\tilde{m}_2\;J_3\tilde{m}_3}, \\
&&{\cal D}^{JM}_{J_1M_1\rho_1\;J_2M_2\rho_2}
=(-1)^{\frac{\rho_1+\rho_2}{2}+1}
\sqrt{3(2J_1+1)(2J_1+2\rho_1^2+1)(2J_2+1)(2J_2+2\rho_2^2+1)} \n
&&\qquad\qquad\qquad\qquad
\times\left\{ \begin{array}{ccc}
                Q_1 & \tilde{Q}_1 &1 \\
                Q_2 & \tilde{Q}_2 &1 \\
                J   & J           &0
                \end{array}   \right\}
C^{Jm}_{Q_1m_1\; Q_2m_2}
C^{J\tilde{m}}_{\tilde{Q}_1\tilde{m}_1\;\tilde{Q}_2\tilde{m}_2}, \\
&&{\cal E}_{J_1M_1\rho_1\;J_2M_2\rho_2\;J_3M_3\rho_3} \n
&&=
\sqrt{6(2J_1+1)(2J_1+2\rho_1^2+1)(2J_2+1)(2J_2+2\rho_2^2+1)
     (2J_3+1)(2J_3+2\rho_3^2+1)} \n
&&\;\;\;\;\;\times (-1)^{-\frac{\rho_1+\rho_2+\rho_3+1}{2}}
\left\{ \begin{array}{ccc}
                Q_1 & \tilde{Q}_1 &1 \\
                Q_2 & \tilde{Q}_2 &1 \\
                Q_3 & \tilde{Q}_3 &1        
                \end{array}   \right\}
\left( \begin{array}{ccc}
        Q_1 & Q_2 & Q_3 \\
        m_1 & m_2 & m_3  
       \end{array}  \right)
\left( \begin{array}{ccc}
        \tilde{Q}_1 & \tilde{Q}_2 & \tilde{Q}_3 \\
        \tilde{m}_1 & \tilde{m}_2 & \tilde{m}_3  
       \end{array}  \right), \\
&&{\cal F}^{J_1M_1\kappa_1}_{J_2M_2\kappa_2\;JM}
=\sqrt{2(2J+1)^2(2J_2+1)(2J_2+2)} 
\left\{ \begin{array}{ccc}
                U_1 & \tilde{U}_1 &\frac{1}{2} \\
                U_2 & \tilde{U}_2 &\frac{1}{2} \\
                J   & J           &0        
                \end{array}   \right\}
C^{U_1m_1}_{U_2m_2\;Jm}
C^{\tilde{U}_1\tilde{m}_1}_{\tilde{U}_2\tilde{m}_2\;J\tilde{m}}, \\
&&{\cal G}^{J_1M_1\kappa_1}_{J_2M_2\kappa_2\;JM\rho}
=(-1)^{\frac{\rho}{2}}\sqrt{6(2J_2+1)(2J_2+2)(2J+1)(2J+2\rho^2+1)} \n
&&\qquad\qquad\qquad\;\;\;
\times\left\{ \begin{array}{ccc}
                U_1 & \tilde{U}_1 &\frac{1}{2} \\
                U_2 & \tilde{U}_2 &\frac{1}{2} \\
                Q   & \tilde{Q}   &1        
                \end{array}   \right\}
C^{U_1m_1}_{U_2m_2\;Qm}
C^{\tilde{U}_1\tilde{m}_1}_{\tilde{U}_2\tilde{m}_2\;\tilde{Q}\tilde{m}}.
\label{vertexfunctions}
\eeqa

\section{Spherical harmonics on $S^2$}
\setcounter{equation}{0}
\renewcommand{\theequation}{C.\arabic{equation}}
In this appendix, we summarize the definitions and the properties of 
the spherical harmonics on $S^2$. We set the radius of $S^2$ to $\mu^{-1}$.
Construction of the spherical harmonics on $S^2$ proceeds parallel to
that of the spherical harmonics on $S^3$. We again identify $S^2$ with
a coset space: $S^2=G/H=SO(3)/SO(2)$.
The generators of $G=SO(3)$ are $J_1$, $J_2$ $J_3$, and the generator of 
$H=SO(2)$ is $J_3$. The representative element of $G/H$ is
$\Upsilon'(\Omega')=e^{-i\varphi J_3}e^{-i\theta J_2}$, where 
$\Omega'=(\theta,\varphi)$ is the polar coordinates of $S^2$.
The spin $L$ spherical harmonics is defined by
\beqa
{\cal Y}^{Lq}_{Jm}=n_J^L\langle Jq | {\Upsilon'}^{-1}(\Omega') |Jm\rangle,
\eeqa
where $J$ takes $L, L+1, L+2,\cdots$ while $q$ takes $L$ or $-L$, and 
$n_J^L=\sqrt{\frac{2J+1}{2}}$ for $L\neq 0$ and $n_J^0=\sqrt{2J+1}$.
The spin $L$ spherical harmonics has the following properties.
\beqa
&&\int d\Omega' \sum_{q=\pm L}({\cal Y}^{Lq}_{J_1m_1})^{\ast}
{\cal Y}^{Lq}_{J_2m_2}
=\delta_{J_1J_2}\delta_{m_1m_2}, \n
&&\int d\Omega' ({\cal Y}^{L_1q_2+q_3}_{J_1m_1})^{\ast}
{\cal Y}^{L_2q_2}_{J_2m_2}{\cal Y}^{L_3q_3}_{J_3m_3}
=\frac{n_{J_1}^{L_1}n_{J_2}^{L_2}n_{J_3}^{L_3}}{2J_1+1}
C^{J_1q_2+q_3}_{J_2q_2\;J_3q_3}
C^{J_1m_1}_{J_2m_2\;J_3m_3}, \n
&&({\cal Y}^{Lq}_{Jm})^{\ast}=(-1)^{m-q}{\cal Y}^{L\:-q}_{J\:-m}, \n
&&\nabla_i{\cal Y}^{Lq}_{Jm}
=n_J^L\langle Jq| (-i\mu)J_i{\Upsilon'}^{-1}(\Omega') |Jm\rangle, \;\;\;
\mbox{for}\;\;i=1,2, \n
&&\nabla^2{\cal Y}^{Lq}_{Jm}=\mu^2(-J(J+1)+q^2){\cal Y}^{Lq}_{Jm}.
\label{calYjm}
\eeqa

The scalar spherical harmonics is
defined by $Y_{Jm}={\cal Y}^{00}_{Jm} \;\;(J=0,1,2,\cdots)$. 
The spinor spherical harmonics is defined by
$Y_{Jm\alpha}={\cal Y}^{\frac{1}{2}\alpha}_{Jm}\;\;(J=\frac{1}{2},\frac{3}{2},
\cdots)$. The transverse vector spherical harmonics is defined by
$Y^t_{Jmi=1}=\frac{1}{\sqrt{2}}(-{\cal Y}^{11}_{Jm}+{\cal Y}^{1-1}_{Jm})$ and
$Y^t_{Jmi=2}=-\frac{i}{\sqrt{2}}({\cal Y}^{11}_{Jm}+{\cal Y}^{1-1}_{Jm})$ 
$(J=1,2,\cdots)$
while the longitudinal vector spherical harmonics is defined by
$Y^l_{Jmi}=\epsilon_{ij}Y^t_{Jmj}\;\;(J=1,2,\cdots)$.
These spherical harmonics satisfy the following identities.
\beqa
&&\nabla^2 Y_{Jm}=-\mu^2J(J+1)Y_{Jm}, \n
&&\nabla^2 Y_{Jm\alpha}=-\mu^2(J(J+1)-\frac{1}{4})Y_{Jm\alpha}, \n
&&\nabla^2 Y^{t,l}_{Jmi}=-\mu^2(J(J+1)-1)Y^{t,l}_{Jmi}, \n
&&(\nabla_1\pm i\nabla_2)Y_{Jm\pm\frac{1}{2}}=-i\mu(J+\frac{1}{2})
Y_{Jm\mp\frac{1}{2}}, \n
&&\nabla_iY^t_{Jmi}=0, \n
&&\nabla_iY^l_{Jmi}=-\mu\sqrt{J(J+1)}Y_{Jm}, \n
&&Y^l_{Jmi}=\frac{1}{\mu\sqrt{J(J+1)}}\nabla_iY_{Jm}, \n
&&\epsilon_{ij}\nabla_iY^t_{Jmj}=-\mu\sqrt{J(J+1)}Y_{Jm}, \n
&&\epsilon_{ij}\nabla_iY^l_{Jmj}=0.
\label{calYjmS}
\eeqa

\section{1-loop divergences}
\setcounter{equation}{0}
\renewcommand{\theequation}{D.\arabic{equation}}
In this appendix, we give the 1-loop diagrams and the divergent part
of each diagram.
The nine diagrams
for the 1-loop self-energy of $A_i$ which is $(-i)$ times 
the 1-loop contribution to the 1PI part of the truncated
2-point function $\langle A_{JM\rho}(q)_{kl}A_{J'M'\rho'}(-q)_{k'l'}\rangle$
are shown in Fig. 7. 
The six diagrams for the 1-loop self-energy of 
$A_0$ which is $(-i)$ times the 1-loop contribution to 
the 1PI part of the truncated
2-point function$\langle B_{JM}(q)_{kl}B_{J'M'}(-q)_{k'l'}\rangle$ 
are shown in Fig. 8.
The diagram for the 1-loop self-energy of $c$
which is $(-i)$ times the 1-loop contribution to the 1PI part of the truncated
2-point function
$\langle c_{JM}(q)_{kl}\bar{c}_{J'M'}(-q)_{k'l'}\rangle$ 
are shown in Fig. 9.
The three diagram
for the 1-loop self-energy of $\psi^A$ 
which is $(-i)$ times the 1-loop contribution to the 1PI part of the truncated
2-point function
$\langle \psi_{JM\kappa}^A(q)_{kl}\psi_{J'M'\kappa'A'}^{\dagger}(q)_{kl'}
\rangle$ are shown in Fig. 10.
The two diagrams for the 1-loop correction to the ghost-ghost-gauge
interaction term which is $(-i)$ times the 1-loop contribution to the
1PI part of the truncated three point function
$\langle A_{JM\rho}(q)_{kl}c_{J'M'}(q')_{K'l'}\bar{c}_{J''M''}(q'')_{k''l''}
\rangle$ are shown in Fig. 11.
The five diagrams for the one-loop
correction to the Yukawa interaction term 
which is $(-i)$ times the 1-loop contribution to the
1PI part of the truncated three point function
$\langle (X_{AB}^{JM}(q))_{kl}\psi_{J'}^{A'}(q')_{k'l'}
\psi_{J''}^{B'}(q'')_{k''l''}\rangle$,
are shown in Fig. 12.

The 1-loop self-energy of $A_i$ takes the form
\beqa
g^2N\delta_{kl'}\delta_{lk'}
(-1)^{m-\tilde{m}+1}\delta_{JJ'}\delta_{M-M'}\delta_{\rho\rho'}
\Pi^A_{J}(q).
\eeqa
We list the the divergent part in the contribution of each diagram to
$\Pi^A_J(q)$.
\beqa
&&(A-a)=\sum_{J_1,J_2\neq0,M_1M_2}
\frac{2i\delta(0)}{4\sqrt{J_1(J_1+1)J_2(J_2+1)}} 
\mathcal{D}_{J_2M_2 \, J_1M_10 \, J-M\rho} 
\mathcal{D}_{J_1-M_1 \, J_2-M_20 \, J'-M'\rho'}, \n
&&(A-b)=\sum_{J_1,J_2\neq0,M_1M_2}\bigg[ -\frac{2i 
\delta(0)}{4\sqrt{J_1(J_1+1)J_2(J_2+1)}} 
\mathcal{D}_{J_2 M_2 \, J_1 M_1 0 \, J -M \rho} 
\mathcal{D}_{J_1-M_1 \, J_2 -M_2 0 \, J' -M' \rho'} \nonumber\\
&& \qquad\qquad\;\; + \frac{2i \delta(0)}{4J_2(J_2+1)} 
\mathcal{D}_{J_2M_2 \, J_1M_10 \, J-M\rho} 
\mathcal{D}_{J_2-M_2 \, J_1-M_10 \, J'-M'\rho'} \bigg], \n
&&(A-c)=\sum_{J_2\neq=0,J_1M_1M_2}\frac{-2i\delta(0)}{4J_2(J_2+1)} 
\bigg[
\mathcal{D}_{J_2 M_2 \, J-M\rho \, J_1M_10} 
\mathcal{D}_{J_2-M_2 \, J'-M'\rho' \, J_1-M_10} \nonumber\\
&&\qquad\qquad\;\; + \mathcal{D}_{J_2 M_2 \, J-M\rho \, J_1 M_1 \pm} 
\mathcal{D}_{J_2-M_2 \, J'-M'\rho' \, J_1 -M_1 \pm}
\bigg], \n
&&(A-d)=\sum_{J_2\neq=0,J_1M_1M_2}\frac{2i\delta(0)}{4 J_2(J_2+1)}
\mathcal{D}_{J_2 M_2 \, J_1 M_1\pm,J-M \rho} 
\mathcal{D}_{J_2-M_2 \, J'-M'-\rho'J_5M_5\pm} \nonumber\\
&&\qquad\qquad\;\; -\frac{4}{3}\Lambda_s^2 
-2\Lambda_s -\left[\frac{2}{3}q^2 +\frac{2}{5}(2J+2)^2+\frac{2}{5}
\right]\log(2\Lambda), \n
&&(A-e)=-\frac{8}{3} \Lambda_v^2 
-\frac{20}{3}\Lambda_v +\frac{4}{3}\log(2\Lambda), \n
&&(A-f)=\frac{4}{3} \Lambda_v^2 
+\frac{10}{3}\Lambda_v + \left[\frac{q^2}{6}+\frac{18}{5}(J+1)^2 
- \frac{14}{15}\right]\log(2\Lambda), \n
&&(A-g)=-12 \Lambda_s^2 -18 \Lambda_s, \n
&&(A-h)=4 \Lambda_s^2 
+ 6\Lambda_s + \frac{1}{2}\left[q^2 - (2J+2)^2 \right]
\log(2\Lambda),  \n
&&(A-i)=\frac{32}{3} \Lambda_f^2 
+ \frac{64}{3} \Lambda_f +\frac{4}{3} \left[ q^2 - (2J+2)^2 \right] 
\log(2\Lambda). 
\label{divergentpartsofselfenergyofAi}
\eeqa
Note that the terms proportional to $\delta(0)$ cancel among
$(A-a)\sim (A-d)$.

The 1-loop self-energy of $A_0$ takes the form
\beqa
g^2N\delta_{kl'}\delta_{lk'}(-1)^{m-\tilde{m}}\delta_{JJ'}\delta_{M-M'}
\Pi^B_{J}(q).
\eeqa
We list the the divergent part in the contribution of each diagram to
$\Pi^B_J(q)$.
\beqa
&&(B-a)=4 \Lambda_v^2 + 10 \Lambda_v -2 \log(2 \Lambda), \n
&&(B-b)=-4 \Lambda_v^2 
-10 \Lambda_v + \left[ 2 + \frac{10}{3}J(J+1)\right] \log(2\Lambda), \n
&&(B-c)=-\frac{32}{3}J(J+1) \log(2\Lambda), \n
&&(B-d)=12 \Lambda_s^2 + 18 \Lambda_s, \n
&&(B-e)=-12 \Lambda_s^2 -18 \Lambda_s + 2J(J+1) \log(2\Lambda), \n
&&(B-f)=\frac{16}{3} J(J+1) \log(2\Lambda).
\label{divergentpartsofselfenergyofA0}
\eeqa

The 1-loop self-energy of $c$ takes the form
\beqa
g^2N\delta_{kl'}\delta_{lk'}(-1)^{m-\tilde{m}}\delta_{JJ'}\delta_{M-M'}
\Pi^c_{J}(q).
\eeqa
The divergent part in the contribution of the diagram to
$\Pi^c_J(q)$ is
\beqa
(G-a)=4iJ(J+1) \left(-\frac{2}{3}\right) \log(2 \Lambda).
\label{divergentpartsofselfenergyofghosts}
\eeqa

The 1-loop self-energy of $\psi^A$ takes the form
\beqa
g^2N\delta_{kl'}\delta_{lk'}
\delta_{JJ'}\delta_{MM'}\delta_{\kappa\kappa'}\delta^A_{A'}
\Pi^{\psi}_{J}(q).
\eeqa
We list the the divergent part in the contribution of each diagram to
$\Pi^{\psi}_J(q)$.
\beqa
&&(F-a)=\left(\frac{1}{2} q 
- \frac{1}{6} \kappa (2J+\frac{3}{2}) \right) \log (2 \Lambda), \n
&&(F-b)=\left(\frac{2}{3} \kappa (2J+\frac{3}{2}) \right)
\log(2 \Lambda), \n
&&(F-c)=\frac{3}{2}\left( q + \kappa (2J+\frac{3}{2}) \right) \log(2 \Lambda).
\label{divergentpartsofselfenergyoffermions}
\eeqa

The two diagrams for the one-loop correction to the ghost-ghost-gauge
interaction term vanish:
\beqa
(GV-a)=0, \;\;\;\;\;
(GV-b)=0.
\label{divergentpartsofghostghostgaugeinteraction}
\eeqa

The 1-loop correction to the Yukawa interaction term takes the form
\beqa
&&2ig^3N\delta_{AB}^{A'B'}
\left(\delta_{kl'}\delta_{k'l''}\delta_{k''l}
(-1)^{m'-\tilde{m}'+\frac{\kappa'}{2}}
F^{J''M''\kappa''}_{J'-M'\kappa'\;J_3-M_3}
+\delta_{kl''}\delta_{k'l}\delta_{k''l'}
(-1)^{m''-\tilde{m}''+\frac{\kappa''}{2}}
F^{J'M'\kappa'}_{J''-M''\kappa''\;J_3-M_3} \right) \n
&&\times 2\pi\delta(q+q'+q'')\Gamma^{Y}_{JJ'J''}(q',q'').
\eeqa
We list the the divergent part in the contribution of each diagram to
$\Gamma^{Y}_{JJ'J''}(q',q'')$.
\beqa
&&(Y-a)=\frac{1}{2}\log(2\Lambda), \n
&&(Y-b)=\frac{1}{2}\log(2\Lambda), \n
&&(Y-c)=\frac{1}{2}\log(2\Lambda), \n
&&(Y-d)=\log(2\Lambda), \n
&&(Y-e)=0.
\label{divergentpartsofYukawainteraction}
\eeqa

\begin{figure}[htbp]
\begin{center}
 \begin{minipage}{0.2\hsize}
  \begin{center}
   \includegraphics[width=30mm]{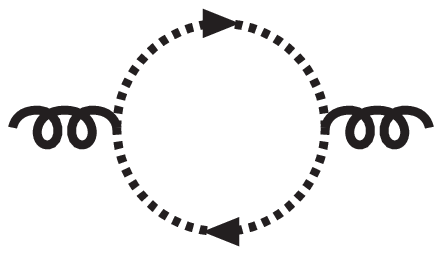}
\\
(A-a)
\label{fig:A7} 
  \end{center}
 \end{minipage}
 \begin{minipage}{0.2\hsize}
  \begin{center}
   {\includegraphics[width=30mm]{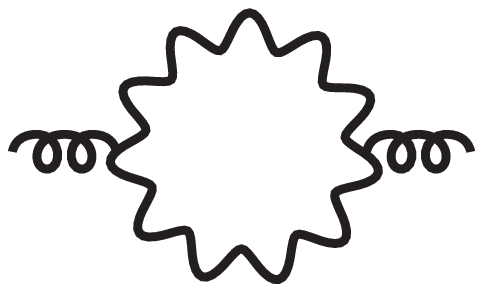}}
\\
(A-b)
\label{fig:A6}
  \end{center}
\end{minipage}
 \begin{minipage}{0.2\hsize}
  \begin{center}
\includegraphics[width=30mm]{{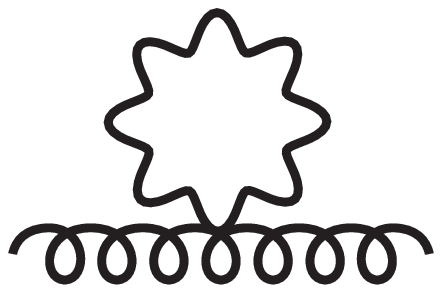}}
\\
(A-c)
\label{fig:A2}
\end{center}
 \end{minipage}
 \begin{minipage}{0.2\hsize}
  \begin{center}
  {\includegraphics[width=30mm]{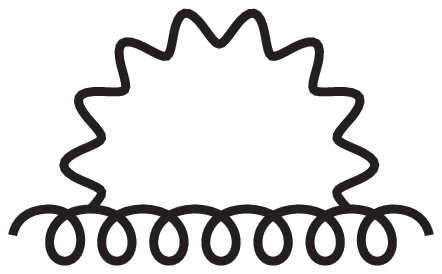}}
\\
(A-d)
\label{fig:A5}
\end{center}
 \end{minipage}
\\
 \begin{minipage}{0.2\hsize}
  \begin{center}
   \includegraphics[width=30mm]{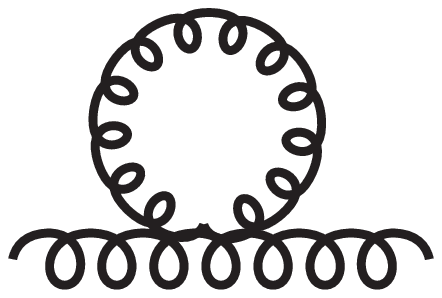}
   (A-e)
   \label{fig:A1}
  \end{center}
 \end{minipage}
 \begin{minipage}{0.2\hsize}
  \begin{center}
   \includegraphics[width=30mm]{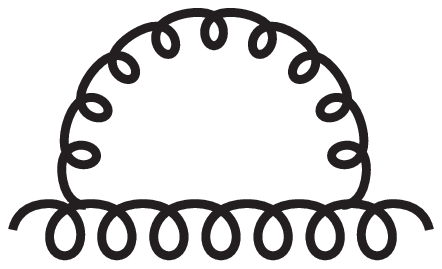}
\\
   (A-f)
   \label{fig:A4}
  \end{center}
 \end{minipage}
\\
 \begin{minipage}{0.2\hsize}
  \begin{center}
   \includegraphics[width=30mm]{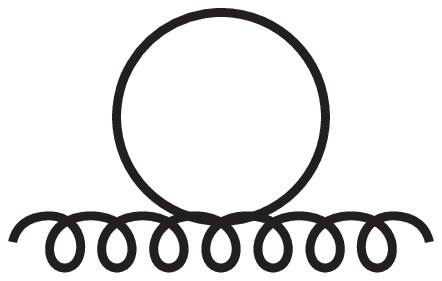}
   (A-g)
   \label{fig:A1}
\end{center}
 \end{minipage}
 \begin{minipage}{0.2\hsize}
  \begin{center}
   \vspace*{4mm}
   \includegraphics[width=30mm]{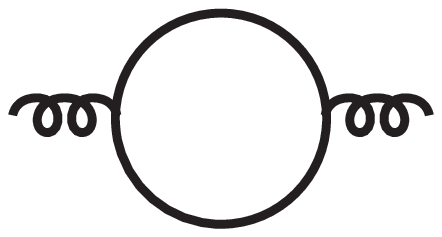}
   (A-h) 
   \label{fig:A5}
  \end{center}
 \end{minipage}
 \begin{minipage}{0.2\hsize}
  \begin{center}
   \includegraphics[width=30mm]{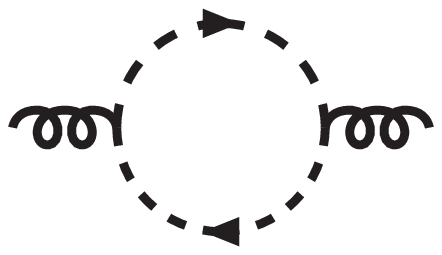}
   (A-i)
   \label{fig:A6}
  \end{center}
 \end{minipage}
\caption{Diagrams for the one-loop self energy of $A_i$. The curly line
represents the propagator of $A_i$. The wavy line represents the propagator
of $A_0$. The dotted line represents the propagator of the ghost.
The solid line represents the propagator of $X_{AB}$. The dashed line
represents the propagator of $\psi^A$.}
\label{self-energy of A}
\end{center}
\end{figure}

\begin{figure}[htbp]
\begin{center}
 \begin{minipage}{0.2\hsize}
  \begin{center}
   \includegraphics[width=30mm]{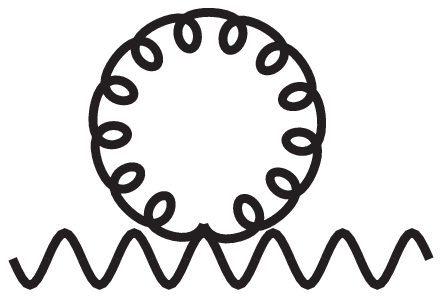}
   (B-a)
   \label{fig:B1}
  \end{center}
 \end{minipage}
 \begin{minipage}{0.2\hsize}
  \begin{center}
   \vspace*{10mm}
   \includegraphics[width=30mm]{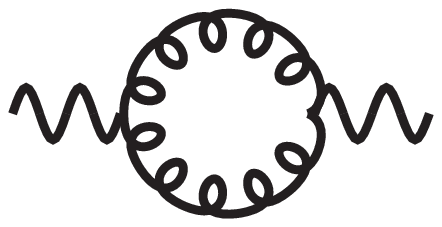}
   (B-b)
   \label{fig:B4}
  \end{center}
 \end{minipage}
 \begin{minipage}{0.2\hsize}
  \begin{center}
   \includegraphics[width=30mm]{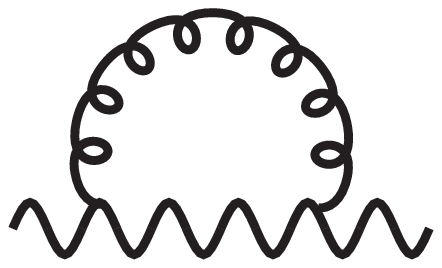}
   (B-c)
   \label{fig:B3}
  \end{center}
 \end{minipage}
\\
 \begin{minipage}{0.2\hsize}
  \begin{center}
   {\includegraphics[width=30mm]{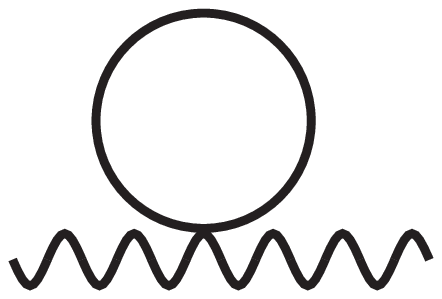}}
   (B-d)
   \label{fig:B2}
  \end{center}
 \end{minipage}
 \begin{minipage}{0.2\hsize}
  \begin{center}
   {\includegraphics[width=30mm]{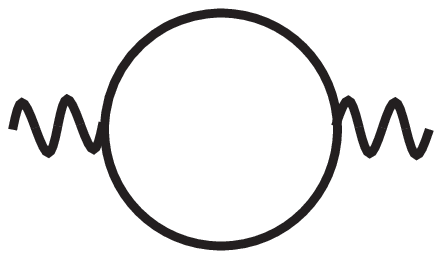}}
   (B-e)
   \label{fig:B5}
  \end{center}
 \end{minipage}
 \begin{minipage}{0.2\hsize}
  \begin{center}
   {\includegraphics[width=30mm]{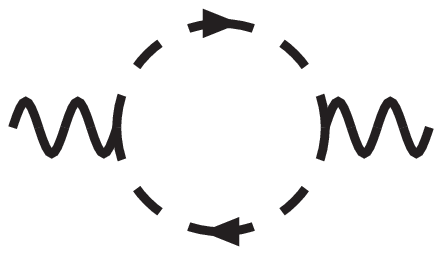}}
   (B-f)
   \label{fig:B6}
  \end{center}
 \end{minipage}
\caption{Diagrams for the one-loop self energy of $A_0$. The curly line
represents the propagator of $A_i$. 
The solid line represents the propagator of $X_{AB}$. The dashed line
represents the propagator of $\psi^A$.}
\end{center}
\end{figure}

\begin{figure}[htbp]
\begin{center}
\begin{minipage}{0.2\hsize}
  \begin{center}
   {\includegraphics[width=30mm]{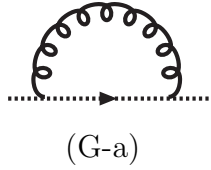}}
   (G-a)
  \label{G-aFig}
  \end{center}
 \end{minipage}
\end{center}
\caption{Diagram for the self-energy of the ghost. The curly line
represents the propagator of $A_i$. 
The dotted line represents the propagator of the ghost.}
\label{fig:A1}
\end{figure}

\begin{figure}[htbp]
\begin{center}
 \begin{minipage}{0.2\hsize}
  \begin{center}
   {\includegraphics[width=30mm]{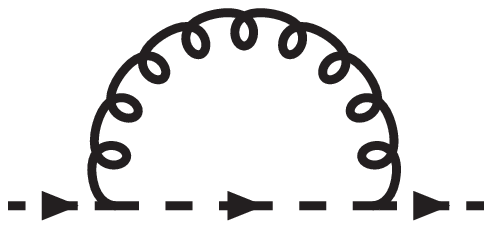}}
   (F-a)
   \label{fig:A1}
\end{center}

 \end{minipage}
 \begin{minipage}{0.2\hsize}
  \begin{center}
   {\includegraphics[width=30mm]{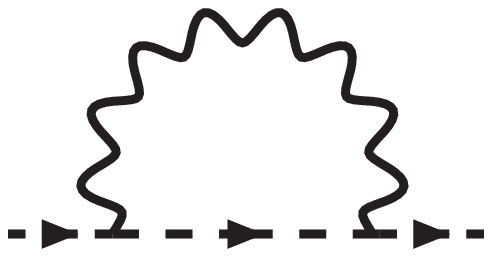}}
   (F-b)
   \label{fig:F2}
  \end{center}
 \end{minipage}
 \begin{minipage}{0.2\hsize}
  \begin{center}
   {\includegraphics[width=30mm]{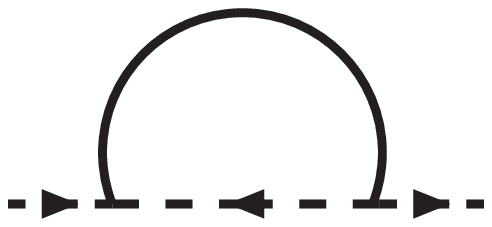}}
   (F-c)
   \label{fig:A6}
  \end{center}
 \end{minipage}
\end{center}
\caption{Diagrams for the one-loop self energy of $\psi^A$. The curly line
represents the propagator of $A_i$. The wavy line represents the propagator
of $A_0$. 
The solid line represents the propagator of $X_{AB}$. The dashed line
represents the propagator of $\psi^A$.}
\end{figure}

\begin{figure}[htbp]
\begin{center}
 \begin{minipage}{0.2\hsize}
  \begin{center}
   {\includegraphics[width=30mm]{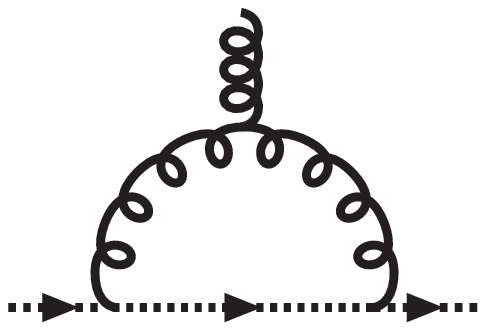}}
   (GV-a)
   \label{fig:A1}
\end{center}

 \end{minipage}
 \begin{minipage}{0.2\hsize}
  \begin{center}
{\includegraphics[width=30mm]{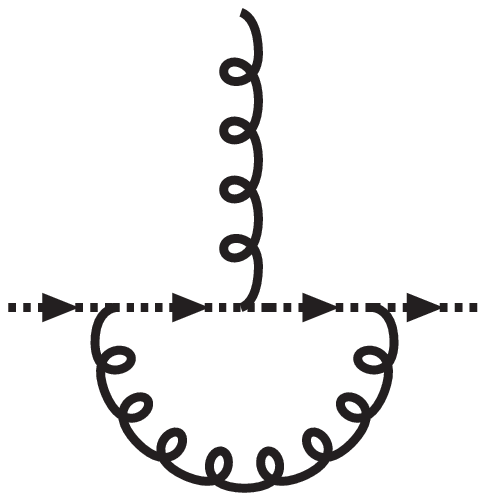}}
   (GV-b)
   \label{fig:A5}
  \end{center}
 \end{minipage}
\caption{Diagrams for the one-loop correction to the ghost-ghost-gauge
interaction vertex. The curly line
represents the propagator of $A_i$. 
The dotted line represents the propagator of the ghost.}
\end{center}
\end{figure}

\begin{figure}[htbp]
\begin{center}
 \begin{minipage}{0.2\hsize}
  \begin{center}
\includegraphics[width=30mm]{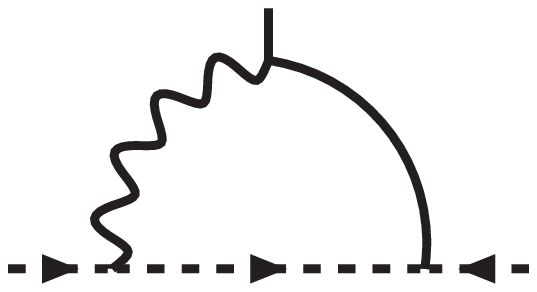}
  (Y-a)
  \label{fig:Y1}
\end{center}
 \end{minipage}
 \begin{minipage}{0.2\hsize}
  \begin{center}
  {\includegraphics[width=30mm]{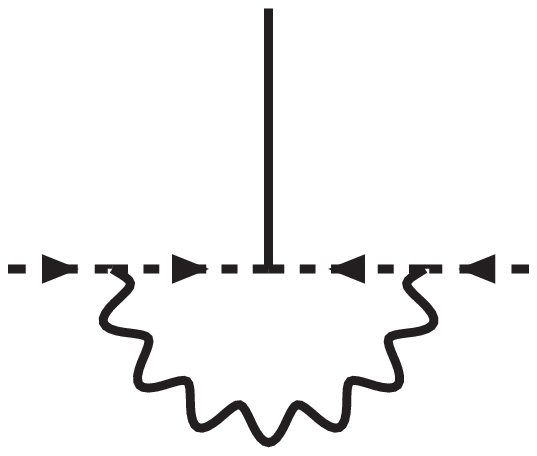}}
  (Y-b)
  \label{fig:Y2}
\end{center}
 \end{minipage}
 \begin{minipage}{0.2\hsize}
  \begin{center}
  {\includegraphics[width=30mm]{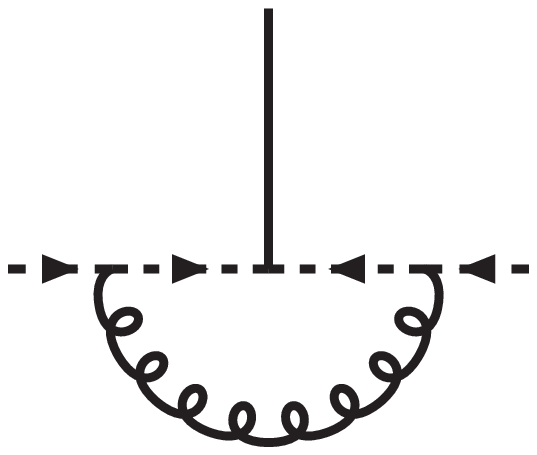}}
  (Y-c)
  \label{fig:Y3}
  \end{center}
\end{minipage}
 \begin{minipage}{0.2\hsize}
  \begin{center}
   \includegraphics[width=30mm]{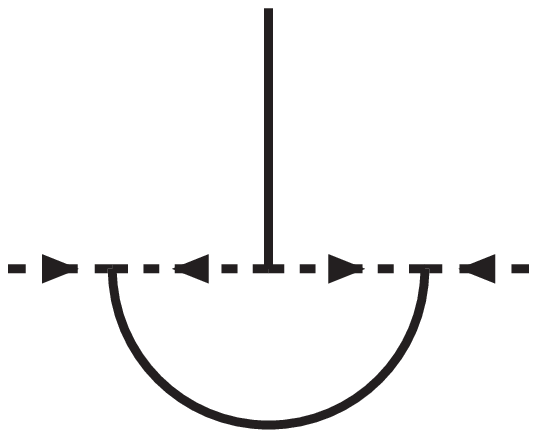}
   (Y-d)
   \label{fig:Y4} 
  \end{center}
 \end{minipage}
 \begin{minipage}{0.2\hsize}
  \begin{center}
   \includegraphics[width=30mm]{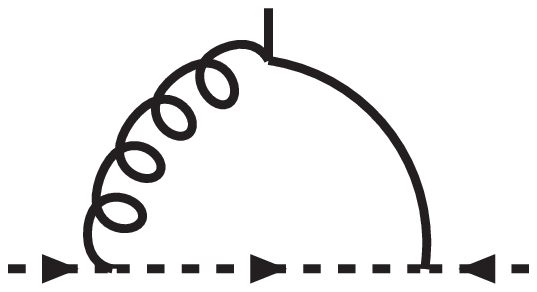}
   (Y-e)
   \label{Y-eFig} 
  \end{center}
 \end{minipage}
\caption{Diagrams for the one-loop correction to the Yukawa interaction.
The curly line
represents the propagator of $A_i$. 
The wavy line represents the propagator of $A_0$.
The solid line represents the propagator of $X_{AB}$. The dashed line
represents the propagator of $\psi^A$.}
\end{center}
\end{figure}

\newpage

\section{1-loop effective action in the truncated theories}
\setcounter{equation}{0}
\renewcommand{\theequation}{E.\arabic{equation}}
In this appendix, we give the expressions for the 1-loop effective action
around the time-dependent BPS solution in the truncated theories.
In the expressions, we omit the factor $g^2N\int dt$ to make them compact.

The 1-loop effective action in the plane-wave matrix model is
\beqa
&&\Gamma_{eff}^{Z0}=-\sqrt{4+l}, \n
&&\Gamma_{eff}^{Y}=-4\sqrt{1+l}, \n
&&\Gamma_{eff}^{A}=-3\sqrt{4+l}, \n
&&\Gamma_{eff}^{F}=4(\sqrt{4+l}+\sqrt{1+l}\:).
\label{1-loop effective action in plane wave matrix model}
\eeqa

The 1-loop effective action in ${\cal N}=4$ SYM on $R\times S^2$ is
\beqa
&&\Gamma_{eff}^{Z0}=-\sqrt{4+l}, \n
&&\Gamma_{eff}^{Z}=-\sum_{J \in \mathbf{Z}_{>0}}(2J+1)
                    (\sqrt{4J^2+l}+\sqrt{(2J+2)^2+l}\:), \n
&&\Gamma_{eff}^{Y}=-4\sum_{J \in \mathbf{Z}_{\ge 0}}
                     (2J+1)\sqrt{(2J+1)^2+l}, \n
&&\Gamma_{eff}^{A}=-\sum_{J \in \mathbf{Z}_{\ge 0}}
                    ((2J+3)\sqrt{(2J+2)^2+l}+(2J+1)\sqrt{(2J+2)^2+l}\:),\n
&&\Gamma_{eff}^{F}=2\sum_{J \in \mathbf{Z}_{\ge 0}}
                    (2J+2)(\sqrt{(2J+2)^2+l}+\sqrt{(2J+1)^2+l}\:) \n
&&\qquad\quad\;
                    +2\sum_{J \in \frac{1}{2}+\mathbf{Z}_{\ge 0}}
                    (2J+1)(\sqrt{(2J+2)^2+l}+\sqrt{(2J+1)^2+l}\:), \n
&&\Gamma_{eff}^{c.t.}=-\frac{g^2Nl}{2}\left(\Pi_{J=0}^X(1)+\frac{1}{2}\right).
\label{1-loop effective action in N=4 SYM on R times S^2}
\eeqa

The 1-loop effective action in
${\cal N}=4$ SYM on $R\times S^3/Z_k$ with $k$ even is
\beqa
&&\Gamma_{eff}^{Z0}=-\sqrt{4+l}, \n
&&\Gamma_{eff}^{Z}=-\left(\sum_{n \in \mathbf{Z}_{>0}}\sum_{v=0}^{\frac{k}{2}-1}
                       +\left.\sum_{v=1}^{\frac{k}{2}-1}\right|_{n=0}\right) \n
&&\qquad\qquad\;\;        (kn+2v+1)(2n+1)
                          (\sqrt{(kn+2v)^2+l}+\sqrt{(kn+2v+2)^2+l}\:), \n
&&\Gamma_{eff}^{Y}=-4\sum_{n \in \mathbf{Z}_{\ge 0}}\sum_{v=0}^{\frac{k}{2}-1}
                     (kn+2v+1)(2n+1)\sqrt{(kn+2v+1)^2+l}, \n
&&\Gamma_{eff}^{A}=-\sum_{n \in \mathbf{Z}_{\ge 0}}\sum_{v=0}^{\frac{k}{2}-1}
                     (kn+2v+3)(2n+1)\sqrt{(kn+2v+2)^2+l} \n
&&\qquad\quad\; 
-\left(\sum_{n \in \mathbf{Z}_{>0}}\sum_{v=0}^{\frac{k}{2}-1}
       +\left.\sum_{v=1}^{\frac{k}{2}-1}\right|_{n=0}\right)
(kn+2v-1)(2n+1)\sqrt{(kn+2v)^2+l}, \n
&&\Gamma_{eff}^{F}
=2\sum_{n \in \mathbf{Z}_{\ge 0}}\sum_{v=0}^{\frac{k}{2}-1}
(kn+2v+2)(2n+1)(\sqrt{(kn+2v+2)^2+l}+\sqrt{(kn+2v+1)^2+l}\:) \n
&&\qquad\;\;\;
+2\left(\sum_{n \in \mathbf{Z}_{>0}}\sum_{v=0}^{\frac{k}{2}-1}
       +\left.\sum_{v=1}^{\frac{k}{2}-1}\right|_{n=0}\right)
(kn+2v)(2n+1)(\sqrt{(kn+2v+1)^2+l}+\sqrt{(kn+2v)^2+l}\:), \n
&&\Gamma_{eff}^{c.t.}=-\frac{g^2Nl}{2}\left(\Pi_{J=0}^X(1)+\frac{1}{2}\right).
\label{1-loop effective action in N=4 SYM on R times S^3/Z_k with k even}
\eeqa

The 1-loop effective action in
${\cal N}=4$ SYM on $R\times S^3/Z_k$ with $k$ odd is
\beqa
&&\Gamma_{eff}^{Z0}=-\sqrt{4+l}, \n
&&\Gamma_{eff}^{Z}
=-\left(\sum_{n \in \mathbf{Z}_{>0}}\sum_{v=0}^{\frac{k}{2}-\frac{1}{2}}
+\left.\sum_{v=1}^{\frac{k}{2}-\frac{1}{2}}\right|_{n=0}\right) \n
&&\qquad\qquad\;\;
(kn+2v+1)(n+1)(\sqrt{(kn+2v)^2+l}+\sqrt{(kn+2v+2)^2+l}\:) \n
&&\qquad\quad
-\sum_{n\in\mathbf{Z}_{\ge 0}}\sum_{v=1}^{\frac{k}{2}-\frac{1}{2}}
(kn+2v)n(\sqrt{(kn+2v-1)^2+l}+\sqrt{(kn+2v+1)^2+l}\:), \n
&&\Gamma_{eff}^{Y}
=-4\sum_{n \in \mathbf{Z}_{\ge 0}}\sum_{v=0}^{\frac{k}{2}-\frac{1}{2}}
(kn+2v+1)(n+1)\sqrt{(kn+2v+1)^2+l} \n
&&\qquad\quad 
-4\sum_{n \in \mathbf{Z}_{\ge 0}}\sum_{v=1}^{\frac{k}{2}-\frac{1}{2}}
(kn+2v)n\sqrt{(kn+2v)^2+l}, \n
&&\Gamma_{eff}^{A}
=-\sum_{n \in \mathbf{Z}_{\ge 0}}\sum_{v=0}^{\frac{k}{2}-\frac{1}{2}}
(kn+2v+3)(n+1)\sqrt{(kn+2v+2)^2+l} \n
&&\qquad\quad
-\left(\sum_{n \in \mathbf{Z}_{>0}}\sum_{v=0}^{\frac{k}{2}-\frac{1}{2}}
       +\left.\sum_{v=1}^{\frac{k}{2}-\frac{1}{2}}\right|_{n=0}\right)
(kn+2v-1)(n+1)\sqrt{(kn+2v)^2+l} \n
&&\qquad\;\;\;
-\sum_{n\in\mathbf{Z}_{\ge 0}}\sum_{v=1}^{\frac{k}{2}-\frac{1}{2}}
((kn+2v+2)n\sqrt{(kn+2v+1)^2+l}+(kn+2v-2)n\sqrt{(kn+2v-1)^2+l}\:), \n
&&\Gamma_{eff}^{F}
=2\sum_{n \in \mathbf{Z}_{\ge 0}}\sum_{v=0}^{\frac{k}{2}-\frac{1}{2}}
(kn+2v+2)(n+1)(\sqrt{(kn+2v+2)^2+l}+\sqrt{(kn+2v+1)^2+l}\:) \n
&&\qquad\quad
+2\left(\sum_{n \in \mathbf{Z}_{>0}}\sum_{v=0}^{\frac{k}{2}-\frac{1}{2}}
       +\left.\sum_{v=1}^{\frac{k}{2}-\frac{1}{2}}\right|_{n=0}\right)
(kn+2v)(n+1)(\sqrt{(kn+2v+1)^2+l}+\sqrt{(kn+2v)^2+l}\:) \n
&&\qquad\quad
+2\sum_{n\in\mathbf{Z}_{\ge 0}}\sum_{v=1}^{\frac{k}{2}-\frac{1}{2}}
((kn+2v+1)n(\sqrt{(kn+2v+1)^2+l}+\sqrt{(kn+2v)^2+l}\:) \n
&&\qquad\qquad\qquad\qquad
+(kn+2v-1)n(\sqrt{(kn+2v)^2+l}+\sqrt{(kn+2v-1)^2+l}\:)), \n
&&\Gamma_{eff}^{c.t.}=-\frac{g^2Nl}{2}\left(\Pi_{J=0}^X(1)+\frac{1}{2}\right).
\label{1-loop effective action in N=4 SYM on R times S^3/Z_k with k odd}
\eeqa


\begin{thebibliography}{99}




\bibitem{Lin:2005nh}
  H.~Lin and J.~Maldacena,
  ``Fivebranes from gauge theory,''
  arXiv:hep-th/0509235.
  
\bibitem{LLM}H.~Lin, O.~Lunin and J.~Maldacena,
``Bubbling AdS space and 1/2 BPS geometries,''
 JHEP {\bf 0410} (2004) 025 [arXiv:hep-th/0409174].
  
\bibitem{Jevicki} 
S.~Corley, A.~Jevicki and S.~Ramgoolam,
``Exact correlators of giant gravitons from dual N = 4 SYM theory,''
  Adv.\ Theor.\ Math.\ Phys.\  {\bf 5} (2002) 809
  [arXiv:hep-th/0111222].  

\bibitem{Berenstein}
D.~Berenstein, 
``A toy model for the AdS/CFT correspondence,'' 
 JHEP {\bf 0407} (2004) 018 [arXiv:hep-th/0403110]. 

\bibitem{Berenstein:2002jq}
  D.~Berenstein, J.~M.~Maldacena and H.~Nastase,
  ``Strings in flat space and pp waves from N = 4 super Yang Mills,''
  JHEP {\bf 0204} (2002) 013
  [arXiv:hep-th/0202021].
  
\bibitem{Maldacena:2002rb}
  J.~Maldacena, M.~M.~Sheikh-Jabbari and M.~Van Raamsdonk,
  ``Transverse fivebranes in matrix theory,''
  JHEP {\bf 0301} (2003) 038
  [arXiv:hep-th/0211139].

\bibitem{Dasgupta:2002ru}
  K.~Dasgupta, M.~M.~Sheikh-Jabbari and M.~Van Raamsdonk,
  ``Protected multiplets of M-theory on a plane wave,''
  JHEP {\bf 0209} (2002) 021
  [arXiv:hep-th/0207050].




\bibitem{Dasgupta:2002hx}
  K.~Dasgupta, M.~M.~Sheikh-Jabbari and M.~Van Raamsdonk,
  ``Matrix perturbation theory for M-theory on a PP-wave,''
  JHEP {\bf 0205} (2002) 056
  [arXiv:hep-th/0205185].



\bibitem{Kim:2002if}
  N.~w.~Kim and J.~Plefka,
  ``On the spectrum of pp-wave matrix theory,''
  Nucl.\ Phys.\ B {\bf 643}, 31 (2002)
  [arXiv:hep-th/0207034].

\bibitem{Kim:2003rz}
  N.~w.~Kim, T.~Klose and J.~Plefka,
  ``Plane-wave matrix theory from N = 4 super Yang-Mills on R x S**3,''
  Nucl.\ Phys.\ B {\bf 671}, 359 (2003)
  [arXiv:hep-th/0306054].


\bibitem{Klose:2003qc}
  T.~Klose and J.~Plefka,
  ``On the integrability of large N plane-wave matrix theory,''
  Nucl.\ Phys.\ B {\bf 679}, 127 (2004)
  [arXiv:hep-th/0310232].

  
  


\bibitem{Fischbacher:2004iu}
  T.~Fischbacher, T.~Klose and J.~Plefka,
  ``Planar plane-wave matrix theory at the four loop order: Integrability without BMN scaling,''
  JHEP {\bf 0502}, 039 (2005)
  [arXiv:hep-th/0412331].
\bibitem{Beisert:2005wv}
  N.~Beisert and T.~Klose,
  ``Long-range gl(n) integrable spin chains and plane-wave matrix theory,''
  arXiv:hep-th/0510124.

\bibitem{Breitenlohner:1982jf}
  P.~Breitenlohner and D.~Z.~Freedman,
  ``Stability In Gauged Extended Supergravity,''
  Annals Phys.\  {\bf 144} (1982) 249.


\bibitem{Nicolai:1988ek}
  H.~Nicolai, E.~Sezgin and Y.~Tanii,
  ``Conformally Invariant Supersymmetric Field Theories On S**P X S**1 And Super P-Branes,''
  Nucl.\ Phys.\ B {\bf 305} (1988) 483.



\bibitem{Bergshoeff:1988jx}
  E.~Bergshoeff, A.~Salam, E.~Sezgin and Y.~Tanii,
  ``N=8 Supersingleton Quantum Field Theory,''
  Nucl.\ Phys.\ B {\bf 305} (1988) 497.

\bibitem{Okuyama:2002zn}
  K.~Okuyama,
  ``N = 4 SYM on R x S(3) and pp-wave,''
  JHEP {\bf 0211}, 043 (2002)
  [arXiv:hep-th/0207067].

\bibitem{Hashimoto}
A.~Hashimoto, S.~Hirano and N.~Itzhaki, 
``Large branes in AdS and their field theory dual,'' 
JHEP {\bf 0008} (2000) 051 [arXiv:hep-th/0008016]. 

\bibitem{Witten:1998zw}
  E.~Witten,
  ``Anti-de Sitter space, thermal phase transition, and confinement in  gauge
theories,''
  Adv.\ Theor.\ Math.\ Phys.\  {\bf 2} (1998) 505
  [arXiv:hep-th/9803131].
  
\bibitem{Sundborg}
  B.~Sundborg,
  ``The Hagedorn transition, deconfinement and N = 4 SYM theory,''
  Nucl.\ Phys.\ B {\bf 573} (2000) 349
  [arXiv:hep-th/9908001].


\bibitem{Aharony}
  O.~Aharony, J.~Marsano, S.~Minwalla, K.~Papadodimas and M.~Van Raamsdonk,
  ``The Hagedorn / deconfinement phase transition in weakly coupled large N
  gauge theories,''
  Adv.\ Theor.\ Math.\ Phys.\  {\bf 8} (2004) 603
  [arXiv:hep-th/0310285].
  
  
  
\bibitem{Hawking-Page}
S.W. Hawking and D.N. Page,
``Thermodynamics of black holes in anti-de Sitter Space,"
Comm. Math. Phys. {\bf 87} (1983) 577.

\bibitem{Cappelli:1988vw}
See for example: 
  A.~Cappelli and A.~Coste,
  ``On The Stress Tensor Of Conformal Field Theories In Higher Dimensions,''
  Nucl.\ Phys.\ B {\bf 314}, 707 (1989).
  
\bibitem{Bergshoeff:1987dh}
  E.~Bergshoeff, M.~J.~Duff, C.~N.~Pope and E.~Sezgin,
  ``Supersymmetric Supermembrane Vacua And Singletons,''
  Phys.\ Lett.\ B {\bf 199} (1987) 69.

\bibitem{Salam:1981xd}
  A.~Salam and J.~A.~Strathdee,
  ``On Kaluza-Klein Theory,''
  Ann. Phys.\  {\bf 141}, 316 (1982).
  
\bibitem{Cutkosky}
  R.~E.~Cutkosky,
  ``Harmonic Functions And Matrix Elements For Hyperspherical Quantum Field Models,''
  J.\ Math.\ Phys.\  {\bf 25} (1984) 939.

  
\bibitem{Hamada}
  K.~j.~Hamada and S.~Horata,
  ``Conformal algebra and physical states in non-critical 3-brane on R x S**3,''
  Prog.\ Theor.\ Phys.\  {\bf 110}, 1169 (2004)
  [arXiv:hep-th/0307008].
  
\bibitem{Aharony:2005bq}
  O.~Aharony, J.~Marsano, S.~Minwalla, K.~Papadodimas and M.~Van Raamsdonk,
  ``A first order deconfinement transition in large N Yang-Mills theory on a small S**3,''
  Phys.\ Rev.\ D {\bf 71}, 125018 (2005)
  [arXiv:hep-th/0502149].



\bibitem{Deger:1998nm}
  S.~Deger, A.~Kaya, E.~Sezgin and P.~Sundell,
  ``Spectrum of D = 6, N = 4b supergravity on AdS(3) x S(3),''
  Nucl.\ Phys.\ B {\bf 536} (1998) 110
  [arXiv:hep-th/9804166].

\bibitem{Minahan:2002ve}
  J.~A.~Minahan and K.~Zarembo,
  ``The Bethe-ansatz for N = 4 super Yang-Mills,''
  JHEP {\bf 0303}, 013 (2003)
  [arXiv:hep-th/0212208].


\bibitem{'tHooft}
G. 't Hooft,
``Renormalization of massless Yang-Mills fields,"
Nucl. Phys. {\bf B33} (1971) 173.


\bibitem{vmk}
D. Varshalovich, A. Moskalev and V. Khersonskii, {\it Quantum Theory of Angular Momentum} 
(World Scientific, Singapore, 1988). 

\bibitem{Beisert:2003tq}
  N.~Beisert, C.~Kristjansen and M.~Staudacher,
  Nucl.\ Phys.\ B {\bf 664} (2003) 131
  [arXiv:hep-th/0303060].

\bibitem{Shin:2003np}
  H.~Shin and K.~Yoshida,
  ``One-loop flatness of membrane fuzzy sphere interaction in plane-wave matrix model,''
  Nucl.\ Phys.\ B {\bf 679}, 99 (2004)
  [arXiv:hep-th/0309258].
  

\bibitem{Shin:2004az}
  H.~Shin and K.~Yoshida,
  ``Membrane fuzzy sphere dynamics in plane-wave matrix model,''
  Nucl.\ Phys.\ B {\bf 709}, 69 (2005)
  [arXiv:hep-th/0409045].
  
\bibitem{Shin:2005tb}
  H.~Shin and K.~Yoshida,
  ``Point-like graviton scattering in plane-wave matrix model,''
  arXiv:hep-th/0511072.
  

\bibitem{Kawahara:2006hs}
  N.~Kawahara, J.~Nishimura and K.~Yoshida,
  ``Dynamical aspects of the plane-wave matrix model at finite temperature,''
  arXiv:hep-th/0601170.
  
  
\bibitem{Yamada}
D. Yamada and L. G. Yaffe, 
``Phase Diagram of ${\cal N}=4$ Super-Yang-Mills Theory with $R$-Symmetry
Chemical Potentials," hep-th/0602074.


\bibitem{Wadia}
L.~Alvarez-Gaume, P.~Basu, M.~Marino and S.~R.~Wadia,
``Blackhole / string transition for the small Schwarzschild blackhole of
 $AdS(5) \times S^5$ and critical unitary matrix models,''
 hep-th/0605041.












%


%









  

  






















































  



%
%
%
%
%










\end{thebibliography}
\end{document}